\documentclass[11pt]{article}

  \usepackage{smoothing_paper}
\usepackage[section]{placeins}
\usepackage{tabularx,ragged2e,booktabs,caption}

\newcommand{\ie}{\emph{i.e.}}

\newcommand{\expt}[1]{\mathrm{E}\left[#1\right]}

\newcommand{\rset}{\mathbb{R}}

\newcommand{\PERIOD}{.}
\newcommand{\COMMA}{,}

\newcommand{\Ordo}[1]{{\mathcal{O}}\left(#1\right)}

\makeatletter
\def\BState{\State\hskip-\ALG@thistlm}
\makeatother

\title{Multilevel Monte Carlo  with Numerical Smoothing   for  Robust and Efficient   Computation of Probabilities and Densities}

\author{Christian Bayer\thanks{
		Weierstrass Institute for Applied Analysis and Stochastics (WIAS),
		Berlin, Germany.}
	\and Chiheb Ben Hammouda\thanks{Mathematical Institute, Utrecht University, Utrecht, The Netherlands ({\tt b.h.chiheb@uu.nl}).} 
	\and  Ra\'ul Tempone\thanks{King Abdullah University of Science and Technology (KAUST), Computer, Electrical and Mathematical Sciences \& Engineering Division (CEMSE),  Saudi Arabia.} \thanks{Alexander von Humboldt Professor in Mathematics for Uncertainty Quantification, RWTH Aachen University, Germany.}}

\begin{document}
	\date{}
\maketitle

\begin{abstract}
The multilevel Monte Carlo (MLMC) method   is  highly efficient  for estimating  expectations of a functional of a solution to a stochastic differential equation (SDE). However, MLMC estimators may be  unstable  and  have a poor (noncanonical) complexity in the case of low regularity of the functional.  To overcome this issue, we  extend our previously introduced  idea of numerical smoothing  in (Quantitative Finance, 23(2), 209-227, 2023),  in the context of deterministic quadrature methods to the MLMC setting. The numerical smoothing technique is based on root-finding methods  combined with  one-dimensional numerical integration with respect to a single well-chosen variable.   This study is motivated by the computation of probabilities of events, pricing options with  a discontinuous payoff, and density estimation problems for dynamics where the discretization of the underlying stochastic processes is necessary. The analysis and numerical experiments reveal that  the  numerical smoothing significantly   improves  the strong convergence,  and consequently,  the complexity and  robustness (by making the kurtosis at  deep levels  bounded)  of the MLMC method.  In particular,  we      show  that   numerical smoothing   enables  recovering the MLMC complexities obtained for Lipschitz functionals due to the optimal variance decay rate when using the Euler--Maruyama scheme.  For  the Milstein scheme,  numerical smoothing recovers the canonical MLMC complexity  even for the nonsmooth integrand mentioned above.    Finally, our approach efficiently estimates univariate and multivariate density functions.

\

\textbf{Keywords}  Multilevel Monte Carlo, numerical smoothing, probability estimation,  density estimation, robustness, complexity,  Monte Carlo,  option pricing

\textbf{2010 Mathematics Subject Classification} 	 62P05, 65C05,  65D30, 65Y20, 91G20, 91G60.
\end{abstract}

\section{Introduction}
In  several  applications such as probability computation,   distribution functions or density estimation,  digital/barrier option pricing,   sensitivity computation (particularly financial Greeks), and risk estimation,  one is  interested in efficiently computing the expectation of a functional $g$ of a solution,   $X$, to a stochastic differential equation (SDE):
	\begin{equation}\label{eq:QoI}
		\expt{g(X)},
\end{equation}
even when $g$ exhibits low regularity.

Monte Carlo (MC) methods (standard  and multilevel MC (MLMC) \cite{giles2015multilevel_2}) can be used to approximate the expectation in \eqref{eq:QoI}.   Although
the convergence rate of the standard MC method  is insensitive to  the input space dimensionality and  regularity of the functional $g$, the convergence is slow. In contrast, the MLMC method (based on a hierarchical representation of the expectation of interest and with a better   convergence speed than the standard MC method) is negatively affected by  the low regularity of $g$.  These adverse effects consist of (i) a nonoptimal variance decay rate that affects the complexity of the MLMC method (see \cite{avikainen2009irregular,giles2009analysing,giles2023mlmc} and  Sections~\ref{sec:Details of our approach} and \ref{sec:Numerical smoothing with MLMC})  and (ii) a high kurtosis at the deep levels of MLMC, which deteriorates the robustness and performance of the estimator (see  Sections~\ref{sec:Details of our approach} and \ref{sec:Numerical smoothing with MLMC}).  Furthermore, when estimating densities ($g$  in \eqref{eq:QoI} is a  Dirac delta function),  standard (without smoothing) or regularized MC and MLMC methods either fail due to   infinite variance or have an error that explodes with the dimensions (see Section~\ref{sec: MLMC for approximating densities}).

This work addresses the mentioned challenges for cases  where  analytic  (bias-free) smoothing   of the integrand cannot be performed.  We extend our  numerical smoothing idea introduced in  \cite{bayer2020numerical}  to the MLMC estimator to improve its robustness and complexity when  computing the  expected value of a discontinuous function,  particularly when computing probabilities, estimating densities or  pricing options with discontinuous payoffs. This technique,  previously introduced in \cite{bayer2020numerical} in  the context of deterministic quadrature methods,  is based on root-finding  methods combined with a one-dimensional (1D) numerical integration with respect to (w.r.t.) a single well-chosen variable. 

Previously,  the authors of \cite{avikainen2009irregular,giles2009analysing} used the MLMC method without smoothing for pricing options with discontinuous payoffs  and obtained  poor performance (worst-case complexity of the MLMC method). Afterward, various treatments \cite{Burgos2011TheCO,giles2013numerical,giles2015multilevel,madoc37052, krumscheid2018multilevel,haji2021adaptive} were proposed  to  deal with discontinuous functionals efficiently when using the MLMC method.   These methods can be classified as follows: (i)  methods  based on adaptivity and branching ideas \cite{haji2021adaptive, giles2022multilevel}, which require specialized design for specific problems, (ii) methods based on conditional smoothing with respect to the last Brownian motion increments as in \cite{giles2008improved, giles2013numerical}, where the smoothing effect vanishes as the time step size $\Delta t \rightarrow 0$, and (iii) methods based on parametric regularization and kernel smoothing ideas as in \cite{giles2015multilevel}, which may suffer from exponential error growth with respect to the dimension of the underlying process (as explained in Section \ref{sec: MLMC for approximating densities}). For instance, \cite{giles2013numerical} used implicit smoothing based on   conditional expectation tools. Although this technique improved the variance decay  rate and  complexity of MLMC when using the Milstein scheme,  it did not help with the Euler discretization.  Furthermore, in general cases, the  dynamics may make it difficult to derive an analytic expression for the conditional expectation of interest. Using the Milstein scheme may have major drawbacks: (i)  it is expensive to compute for high-dimensional dynamics due to the L\'evy areas terms; (ii) the  design of a suitable coupling strategy   is challenging; and  (iii) the kurtosis may explode at the deep levels.    The authors of    \cite{giles2015multilevel} suggested a different approach based on parametric smoothing.  They carefully constructed a regularized version of the functional, based on a  regularization parameter that   depends on the degree of smoothness of the function of interest.  Despite offering better performance than the standard (without smoothing) MLMC estimator and a clear setting for the theoretical analysis, this approach has a practical disadvantage regarding the difficulty of its generalization toward the cases where (i) no prior knowledge of the degree of smoothness of the function of interest exists and (ii) more challenging dynamics are considered than the  geometric Brownian motion (GBM).   We refer to \cite{giles2023mlmc} for a detailed review of the various MLMC ideas employed when computing an expected value of a discontinuous function.

We mention that other similar and  different smoothing techniques were previously proposed for deterministic quadrature techniques  (e.g., bias-free, conditional sampling, and preintegration)   to improve the performance of deterministic quadrature techniques (e.g., (adaptive) sparse grid quadrature and 	quasi-Monte Carlo (QMC)) for the applications of   option pricing with discontinuous payoffs  \cite{achtsis2013conditional,weng2017efficient,griewank2017high,bayersmoothing,bayer2018hierarchical,bayer2020numerical,bayer2022optimal} and estimating univariate  density of random variables  (rdvs) \cite{ben2021density,gilbert2021analysis,l2022monte}.  However, the focus of this work  is  to propose a different approach than in \cite{Burgos2011TheCO,giles2013numerical,giles2015multilevel,madoc37052, krumscheid2018multilevel,haji2021adaptive} to improve the performance of MLMC methods when computing an expected value of a discontinuous function with applications in probability computation, digital option pricing and univariate/multivariate density estimation.

The main contributions of  this work are summarized as follows:
		 	\begin{itemize}
		 		\item Compared with the case without smoothing, this analysis  reveals that the employed numerical smoothing technique  improves (i)   the convergence of the variance  of MLMC, (ii) the complexity of the MLMC estimator  owing to the  improvement in  the variance decay rate  and (ii)  the  robustness of the  estimator  by significantly reducing and better controlling the kurtosis at   deep levels (making it bounded).  In particular,  we   theoretically  (see Theorems~\ref{corrol: Lipschitz of the integrands} and \ref{corrol: Lipschitz of the integrands_density}) and numerically demonstrate that   numerical smoothing   enables  recovering the MLMC complexities obtained for Lipschitz functionals by proving that the optimal variance decay  rate is recovered  when using the Euler--Maruyama scheme.  		 	Using the Euler scheme, we obtain  rates of variance decay  and MLMC complexity similar to those reported in \cite{giles2008improved,giles2013numerical} without  employing  higher-order schemes, such as the Milstein scheme. For the  Milstein scheme, we  numerically illustrate that  numerical smoothing recovers   the canonical  MLMC complexity. 
		 		\item  The proposed approach  efficiently  estimates univariate and multivariate  density functions: a task that previous   MC-or MLMC-based methods either fail to achieve due to the infinite variance or have an error that explodes with the dimension when using parametric smoothing or kernel density ideas. Estimating  densities using the MLMC method in \cite{giles2015multilevel} results in  a mean squared error (MSE)  behavior  similar to that   obtained using    kernel density techniques, where the error  increases exponentially w.r.t.~the dimension of the underlying process. However, due to the exact conditional expectation w.r.t.~the Brownian bridge, the error of our approach is  restricted to the root-finding error when approximating the  discontinuity location, which does not increase  exponentially w.r.t.~the dimension  (see Section \ref{sec: MLMC for approximating densities} for further details). Although we provide pointwise density estimates, the proposed approach can be easily extended to approximate functions using similar ideas  as in  \cite{giles2015multilevel,krumscheid2018multilevel} using interpolation grids.  
		 		\item Unlike  \cite{ben2021density,gilbert2021analysis,l2022monte}, which only considered the problem of estimating univariate density of rdvs using the QMC method, the proposed approach is based on the MLMC idea and  is  designed for estimating univariate/multivariate densities for   dynamics where the discretization of the underlying stochastic processes is necessary (\ie,   solution to an  SDE). Moreover,  the methods in \cite{ben2021density,l2022monte}  are based on  kernel density techniques, which have the previously mentioned issue of an exponential increase of the error  w.r.t.~the dimension.
		 		\item  The conditioning/smoothing  in the MLMC estimators  in \cite{giles2008improved,giles2013numerical} is done w.r.t.~the Brownian increments, implying that the smoothing advantage vanishes as $\Delta t \rightarrow 0$.  Instead, in this work, we smooth w.r.t.~$\Ordo{1}$ random variable, ensuring that the smoothing effect does not vanish as $\Delta t \rightarrow 0$.  In  \cite{giles2008improved,giles2013numerical}  satisfactory results were only obtained for  the Milstein scheme but not for the Euler--Maruyama scheme.
		 		\item Our approach prioritizes smoothing, yielding variance reduction as a byproduct.  In contrast,  \cite{achtsis2013conditional},  which introduced the idea of conditional sampling  to improve  QMC performance  for option pricing with discontinuous payoffs,   focused primarily on variance reduction. This distinction is evidenced by the fact that our work, an extension of \cite{bayer2020numerical}, centers on achieving smoother integrands (refer to  \cite{bayer2020numerical} for more details about smoothness analysis), while \cite{achtsis2013conditional} seeks variance reduction with a smoother integrand as a secondary outcome.  	Additionally, our method  is based on the Brownian bridge construction for the path generation and we achieve the numerical smoothing w.r.t. Gaussian rdvs, whereas   \cite{achtsis2013conditional} used  the linear transformation method and the conditioning was done w.r.t. uniform rdvs.   In our  context, the Brownian bridge  construction is primarily used to locate the discontinuity in a small-dimensional manifold compared to the original dimension. 	 	Lastly, our theoretical results   related to variance decay, complexity rates, and estimator robustness draw from a unique toolkit  and  distinct analytical tools and  completely differ from the one in \cite{achtsis2013conditional}.  We emphasize that in  high-dimensional scenarios, our approach finds greater advantage in MLMC over QMC because the convergence of the latter  deteriorates as dimension increases, while the MLMC estimator complexity is dimensionally resilient.
		 \end{itemize}
The remainder of the paper is organized as follows.  Section~\ref{sec:General setting} introduces the problem setting and explains the numerical smoothing technique. Section~\ref{sec: num_smoothing_revisited} briefly revisits the idea  in \cite{bayer2020numerical}. Then, Section~\ref{sec: MLMC for approximating densities} extends this idea for the density estimation application. Section~\ref{sec:Details of our approach} explains and analyzes the proposed approach, combining the MLMC estimator with numerical smoothing. Next, Sections~\ref{sec:Error discussion in the context of MLMC method}, \ref{sec:Work discussion in the context of MLMC method}, and \ref{sec:robustness_MLMC} present the error, work, and robustness analysis, respectively. Finally, Section~\ref{sec:Numerical smoothing with MLMC} reports the results of the numerical experiments conducted for pricing digital options (equivalently computing probability) and estimating density under the GBM and Heston models. Further, it illustrates the advantages of the proposed approach over the standard MLMC estimator (without smoothing) for the Euler--Maruyama and Milstein schemes.

\section{Problem Setting and Numerical Smoothing Idea} \label{sec:General setting}
\subsection{Problem Setup}
To showcase the application of the proposed approach, we work mainly with two possible structures of functional $g$:
\begin{align}
  \text{(i)}\: g(\mathbf{x})&=\mathbf{1}_{(\phi(\mathbf{x}) \ge 0)}; \quad \mathbf{x} \in \rset^d, \label{eq:proba_comp}\\
     \text{(ii)}\:  g(\mathbf{x})&=\delta\left(\phi(\mathbf{x}) = 0\right), \quad \mathbf{x} \in \rset^d, \label{eq:density_comp} 
\end{align}
where  the function  $\phi:  \rset^d \mapsto \rset$ is assumed to be smooth. Case (i) applies to estimating probability or pricing financial digital options. Case (ii)  applies to estimating density, where $\delta(.)$ is the Dirac delta function.  Both cases can relate to computing sensitivities (particularly Greeks as financial applications). We refer to Remark~\ref{rem: Extending  numerical smoothing for computing sensitivities} for the connection to sensitivities.
\begin{notation}\label{notation_2}
	We introduce the notation $\mathbf{x}_{-j}$ to denote    a vector with length $d-1$, representing all  variables other than $x_j$ in $\mathbf{x} \in \rset^d$, $d\ge 1$.  Abusing the  notation,  we define  $\phi(\mathbf{x})=\phi(x_j,\mathbf{x}_{-j})$. Without loss of generality, in the following, we will  use $j=1$.
	\end{notation}
For  the ease of presentation, we assume that, for fixed $\mathbf{x}_{-1}$,  the function $\phi(x_1,\mathbf{x}_{-1})$  has a simple root or is positive for all $x_1 \in \rset$. This  assumption is guaranteed by the monotonicity condition \eqref{assump:Monotonicity condition} and    infinite growth condition \eqref{assump:Growth condition}
	\begin{align}
		\frac{\partial \phi}{\partial x_1}(\mathbf{x}) &>0,\: \forall \:  \mathbf{x} \in \rset^d \: \: \textbf{(Monotonicity condition)\footnotemark}  \label{assump:Monotonicity condition}\\
		\underset{x_1 \rightarrow +\infty}{\lim} \phi(\mathbf{x})&=\underset{x_1 \rightarrow +\infty}{\lim} \phi(x_1,\mathbf{x}_{-1})=+\infty, \: \forall \: \mathbf{x}_{-1} \in \rset^{d-1}\: \text{or} \:\: \frac{\partial^2 \phi} {\partial x_1^2}(\mathbf{x}) \ge 0,\: \forall \: \mathbf{x} \in \rset^{d}  \: \: \textbf{(Growth condition)}  \label{assump:Growth condition} \PERIOD
	\end{align}
\footnotetext{We present the monotonicity condition for an increasing function without  loss of generality.}
As  stated in  Remark 2.4 in \cite{bayer2020numerical},  the numerical smoothing idea and consequently the proposed  approach   can be easily extended to the case of finitely  many roots  (when the number of roots is known a priori). Furthermore, we   revisit this extension for the density estimation case in Section~\ref{sec: MLMC for approximating densities}.
\subsection{Revisiting Numerical Smoothing}\label{sec: num_smoothing_revisited}
This Section briefly  revisits the numerical smoothing idea that was introduced in \cite{bayer2020numerical} in the context of determinsitic quadrature methods when pricing financial options. We refer the reader to Sections~2.1 and 2.2  in  \cite{bayer2020numerical}  for additional details. We aim  to efficiently approximate $\expt{g(\mathbf{X}(T))}$ at final time $T$,  where $\mathbf{X}(t):=(X^{(1)}(t), \dots, X^{(d)}(t))$ solves the following SDE\footnote{We assume that the $\{W^{(j)}\}_{j=1}^d$ are uncorrelated and that the correlation terms are included  in the diffusion terms $b_{ij}$. Moreover,  without restriction, the diffusion coefficients $b_{ij}$ can be stochastic as well.}
	\begin{equation}\label{eq:SDE_interest}
		dX_{t}^{(i)}=a_i(\mathbf{X}_t) dt + \sum_{j=1}^d b_{ij}(\mathbf{X}_t) dW_{t}^{(j)} \PERIOD
\end{equation}
The use of the Brownian bridge construction for path simulation implies that   $\mathbf{W}:=(W^{(1)},\dots, W^{(d)})$ can be  represented hierarchically as
	\begin{equation}\label{eq:bridge_construction}
		W^{(j)}(t)= \frac{t}{T} W^{(j)}(T)+B^{(j)}(t)= \frac{t}{\sqrt{T}} Z_1^{(j)}+B^{(j)}(t), \: 1 \le j \le d \COMMA
	\end{equation}
where  $\mathbf{Z}_1:=(Z^{(1)}_1 , \dots, Z^{(d)}_1)$  are  independent and identically distributed standard Gaussian  rdvs, and  $\{B^{(j)}\}_{j=1}^d$ are independent Brownian bridges.

For $1 \le j\le d$, we denote  by $(Z_1^{(j)},\dots,Z_N^{(j)})$ $N$ standard Gaussian independent rdvs, where $N$ represents the number of time steps  in the discretization ($\Delta t=\frac{T}{N}$). In addition,  $\psi^{(j)}: (Z_2^{(j)},\dots,Z_N^{(j)}) \mapsto (B_1^{(j)},\dots,B_N^{(j)})$  denotes the mapping of the Brownian bridge construction, and  $\Phi: \left(\Delta t, \mathbf{Z}_1 ,\mathbf{B}\right) \mapsto \bar{\mathbf{X}}^{\Delta t}(T)$ denotes the mapping  of the time-stepping scheme, where  $\mathbf{B}:=\left(B^{(1)}_1,\dots,B^{(1)}_N,\dots, B^{(d)}_1,\dots,B^{(d)}_N\right)$ is the discretized noncorrelated  Brownian bridge\footnote{Without loss of generality,   the   correlated Brownian bridge can be obtained  via simple matrix multiplication.} and  $\bar{\mathbf{X}}^{\Delta t}(T):= \left(\bar{X}_T^{(1)}, \dots,\bar{X}_T^{(d)} \right)$. Then, the  quantity of interest  is expressed as
\begin{align}\label{eq: option price as integral_basket}
		\expt{g(\mathbf{X}(T))}&\approx	\expt{g\left(\bar{X}_T^{(1)}, \dots,\bar{X}_T^{(d)} \right)}=\expt{g(\bar{\mathbf{X}}^{\Delta t}(T))} \nonumber\\
		&=\expt{g\circ \Phi  \left(\Delta t, \mathbf{Z}_1 ,\mathbf{B} \right)} \nonumber\\
		&=\expt{g\circ\Phi  \left(\Delta t, \mathbf{Z}_1 ,\psi^{(1)}(Z_2^{(1)}, \dots, Z_N^{(1)}), \dots, \psi^{(d)}(Z_2^{(d)},\dots,Z^{(d)}_N)\right)} \nonumber\\
		&=:\int_{\rset^{d \times N}} G(z_1^{(1)}, \dots, z_N^{(1)}, \dots, z_1^{(d)},\dots,z^{(d)}_N)) \rho_{d \times N}(\mathbf{z}) dz_1^{(1)} \dots dz_N^{(1)} \dots z_1^{(d)} \dots dz^{(d)}_N \COMMA
	\end{align}
where  $\rho_{d \times N}$ represents the $d \times N$  multivariate Gaussian density.

Due to \eqref{eq:proba_comp} and \eqref{eq:density_comp},   the irregularity  is characterized  by $\phi(\bar{\mathbf{X}}^{\Delta t}(T;\mathbf{z}_{1},\mathbf{z}^{(1)}_{-1},\dots,\mathbf{z}^{(d)}_{-1} ))=0$.\footnote{The locations may differ depending on the considered functional.} A natural choice of smoothing directions is  $\mathbf{Z}_1$ for two reasons. First,  in this work, we consider functionals  depending on the terminal value (at the final time $T$) of the stochastic process. Second,  the Brownian bridge construction creates a hierarchy of importance for the  rdvs  such that  the coarsest factors $\mathbf{Z}_1$ tends to be  the most contributing to the information in $\bar{\mathbf{X}}^{\Delta t}$.    Depending on the structure of $\phi$, the root-finding problem can be  reduced to a lower dimension than $d$, potentially one, by adopting a  linear mapping, $\mathcal{A}$ ($d \times d $ matrix),   for the coarsest factors  $\mathbf{Z}_1$, that is
	\begin{equation}\label{eq:linear_transformation}
		\mathbf{Y}=\mathcal{A} \mathbf{Z}_1 \COMMA
	\end{equation}
where $\mathcal{A}$ is generally selected from a family of rotations. We refer to Remark~\ref{rem:exp_rotation} for an example of $\mathcal{A}$.
\begin{remark}[Example of the linear mapping $\mathcal{A}$]\label{rem:exp_rotation} If we consider an arithmetic basket call option, that is $\phi(\mathbf{x})=\sum_{i=1}^{d} w_i x_i$ where $\{w_i\}_{i=1}^d$ represent the  weights,  then   a sufficiently suitable selection of   $\mathcal{A}$ is  a rotation matrix, with the first  row  leading to $Y_1=\sum_{i=1}^d Z_1^{(i)}$ up to rescaling without any constraint for the remaining rows.  In practice, we construct $\mathcal{A}$ by fixing the first row to\footnote{Note that $\mathbf{1}_{1 \times d}$ denotes the row vector with dimension $d$,  where all its coordinates are $1$.}  $\frac{1}{\sqrt{d}} \mathbf{1}_{1 \times d}$,  and the remaining rows are obtained using the Gram--Schmidt procedure.
\end{remark}
 Then   for fixed $\mathbf{y}_{-1}$, $\mathbf{z}^{(1)}_{-1},\dots,\mathbf{z}^{(d)}_{-1} $ (see notation~\ref{notation_2}),   we determine the 1D   discontinuity location $y^{\ast}_1$ (first component of $\mathbf{y}$ in \eqref{eq:linear_transformation}) by   solving 
\begin{equation*}\label{eq:roots_function}
		\phi(\bar{\mathbf{X}}^{\Delta t}(T))=	\phi(\bar{\mathbf{X}}^{\Delta t}(T;y^\ast_1,   \mathbf{y}_{-1},\mathbf{z}^{(1)}_{-1},\dots,\mathbf{z}^{(d)}_{-1} ))=0.
	\end{equation*}
We employ the Newton iteration method to determine the approximated discontinuity location $\bar{y}^\ast_1$.

Based on  \eqref{eq: option price as integral_basket}, the second step  of the  numerical smoothing idea presented in \cite{bayer2020numerical} involves  performing the numerical preintegration,  as follows:
	\begin{align}\label{eq: pre_integration_step_wrt_y1_basket}
		\expt{g(\mathbf{X}(T))}\approx	\expt{g\left(\bar{X}_T^{(1)}, \dots,\bar{X}_T^{(d)} \right)} 
		&=:\expt{I\left(\mathbf{Y}_{-1}, \mathbf{Z}^{(1)}_{-1},\dots,\mathbf{Z}^{(d)}_{-1} \right)}\\ \nonumber
		&\approx\expt{\bar{I}\left(\mathbf{Y}_{-1}, \mathbf{Z}^{(1)}_{-1},\dots,\mathbf{Z}^{(d)}_{-1} \right)} \COMMA
	\end{align}
where
	\begin{align}\label{eq:smooth_function_after_pre_integration}
		I\left(\mathbf{y}_{-1},\mathbf{z}^{(1)}_{-1},\dots,\mathbf{z}^{(d)}_{-1}\right)&=\int_{\rset} G\left(y_1,\mathbf{y}_{-1},\mathbf{z}^{(1)}_{-1},\dots,\mathbf{z}^{(d)}_{-1} \right) \rho_{1}(y_1) dy_1 \nonumber\\
		&= \int_{-\infty}^{y^\ast_1} G\left(y_1,\mathbf{y}_{-1},\mathbf{z}^{(1)}_{-1},\dots,\mathbf{z}^{(d)}_{-1} \right) \rho_{1}(y_1) dy_1+ \int_{y_1^\ast}^{+\infty} G\left(y_1,\mathbf{y}_{-1},\mathbf{z}^{(1)}_{-1},\dots,\mathbf{z}^{(d)}_{-1} \right) \rho_{1}(y_1) dy_1,
	\end{align}
and $\bar{I}$ is the  approximation of $I$  obtained using the Newton iteration and  a two-sided Laguerre quadrature rule, which  is expressed as follows:
	\begin{equation}\label{eq:expression_h_bar}
		\bar{I}(\mathbf{y}_{-1},\mathbf{z}^{(1)}_{-1},\dots,\mathbf{z}^{(d)}_{-1}  ) \coloneqq \sum_{k=0}^{M_{\text{Lag}}} \eta_k \; G\left( \zeta_k\left(\bar{y}^{\ast}_1\right),
		\mathbf{y}_{-1},\mathbf{z}^{(1)}_{-1},\dots,\mathbf{z}^{(d)}_{-1}   \right),
	\end{equation}
where $M_{\text{Lag}}$ represents the number of Laguerre quadrature points  $\zeta_k \in \R$ with $\zeta_0 = \bar{y}^{\ast}_1$ and corresponding weights $\eta_k$\footnote{The points $\zeta_k$ must be selected  systematically  depending on $\bar{y}^{\ast}_1$.}.

Equations~\eqref{eq:smooth_function_after_pre_integration} and \eqref{eq:expression_h_bar} can be easily extended to the case  in which  finitely  many discontinuities exist. We refer to Remark~2.4 presented in \cite{bayer2020numerical} for this extension.
\subsection{Extending the Numerical Smoothing  Idea to Density Estimation}\label{sec: MLMC for approximating densities}
This Section extends the numerical smoothing idea to approximate the density at point $u$, $\rho_{X_T}(u)$, for the stochastic process $X$, at time $T$, whose dynamics are given by \eqref{eq:SDE_interest}:
	\begin{align}\label{eq:density_function}
		\rho_{X_T}(u)=\expt{\delta(X(T)-u)}.
	\end{align}
For the 1D case,  we let $\mathbf{Z}$ be the Gaussian random  vector used for Brownian bridge construction, then by conditioning w.r.t.~$ \mathbf{Z}_{-1}$, we obtain 
	\begin{align}\label{eq:density_estimation_MLMC}
		\rho_{X_T}(u)=\expt{\delta(X(T)-u)} \approx \expt{\delta(\bar{X}^{\Delta t}(T)-u)} &=\expt{ \expt{\delta(\bar{X}^{\Delta t}(T)-u) \mid  \mathbf{Z}_{-1}}}\nonumber\\
		&=\frac{1}{\sqrt{2 \pi}}\expt{\exp\left(-\left(Y^\ast(u)\right)^2/2\right) \left|\frac{d Y^\ast}{dx}(u)\right|}\nonumber\\
		&\approx\frac{1}{\sqrt{2 \pi}}\expt{\exp\left(-\left(\bar{Y}^\ast(u)\right)^2/2\right) \left|\frac{d \bar{Y}^\ast}{dx}(u)\right|}
\end{align}
where $Y^\ast(x)$ and    $\bar{Y}^\ast(x)$   are  the exact and approximate discontinuity locations,  respectively, and obtained numerically by solving: $\bar{X}^{\Delta t}(T; \bar{Y}^\ast(x), \mathbf{Z}_{-1})=x$.
		\begin{remark}[Extending  numerical smoothing for density estimation  to the case of  multiple roots]\label{rem: Generalization of numerical smoothing to the case of countable multiple roots}
		For the case in which there are  finitely  many discontinuities,   \eqref{eq:density_estimation_MLMC} can be extended to  \eqref{eq:density_estimation_MLMC_multiple roots}
		\begin{align}\label{eq:density_estimation_MLMC_multiple roots}
			\rho_{X_T}(u)=\expt{\delta(X(T)-u)} \approx \expt{\delta(\bar{X}^{\Delta t}(T)-u)} &=\frac{1}{\sqrt{2 \pi}}\expt{\sum_{i=1}^{R}\exp\left(-\frac{\left(Y_i^\ast(u)\right)^2}{2}\right) \left|\frac{d Y_i^\ast}{dx}(u)\right|}\nonumber\\
			&\approx  \frac{1}{\sqrt{2 \pi}} \sum_{i=1}^{R} \expt{\exp\left(\frac{-\left(\bar{Y}_i^\ast(u)\right)^2}{2}\right) \left|\frac{d \bar{Y}_i^\ast}{dx}(u)\right|},
		\end{align}
		where $\{Y_i^\ast(u)\}_{i=1}^{R}$  and  $\{\bar{Y}_i(u)\}_{i=1}^{R}$ are  the exact and approximated discontinuities,  respectively.
	\end{remark}
Equation \eqref{eq:density_estimation_MLMC} can be generalized to the multidimensional case, with the difference that  a root-finding procedure in the $d$-dimensional space characterized by the coarsest factor in each dimension must be  performed. Explicitly,  for $\mathbf{u} \in \rset^d$
\begin{align}\label{eq:density_estimation_MLMC_multidim}
			\rho_{\mathbf{X}_T}(\mathbf{u})=\expt{\delta(\mathbf{X}(T)-\mathbf{u})}\approx \expt{\delta(\bar{\mathbf{X}}^{\Delta t}(T)-\mathbf{u})}&= \expt{\rho_d\left(\mathbf{Y}^\ast(\mathbf{u})\right) \left|\det\left(\mathbf{J} \left(\mathbf{u}\right)\right)\right| }=: \expt{F\left(\mathbf{Y}_{-1}, \mathbf{Z}^{(1)}_{-1},\dots,\mathbf{Z}^{(d)}_{-1}; \mathbf{u} \right)} \nonumber\\
			&\approx \expt{\rho_d\left(\overline{\mathbf{Y}}^\ast(\mathbf{u})\right) \left|\det\left(\overline{\mathbf{J}} \left(\mathbf{u}\right)\right)\right|}\nonumber\\
			&=: \expt{\bar{F}\left(\mathbf{Y}_{-1}, \mathbf{Z}^{(1)}_{-1},\dots,\mathbf{Z}^{(d)}_{-1}; \mathbf{u} \right)}\COMMA
	\end{align}
where $\mathbf{Y}^\ast(\mathbf{x})$ and   $\overline{\mathbf{Y}}^\ast(\mathbf{x})$ are  the exact and approximate discontinuity locations,  respectively, and  obtained  by solving:  $ \bar{\mathbf{X}}^{\Delta t}(T; \overline{\mathbf{Y}}^\ast(\mathbf{x}), \mathbf{Y}_{-1}, \mathbf{Z}^{(1)}_{-1},\dots,\mathbf{Z}^{(d)}_{-1})=\mathbf{x}$.  In addition, $\mathbf{J}$ and $\bar{\mathbf{J}}$ are   the Jacobian matrices with $\mathbf{J}_{ij}=\frac{\partial y_i^\ast }{\partial x_j}$ and $\bar{\mathbf{J}}=\frac{\partial \bar{y}_i^\ast }{\partial x_j}$. Finally,  $\det \left( . \right)$  denotes the  determinant of a matrix.

The numerical smoothing  procedure  presented in \eqref{eq:density_estimation_MLMC} and \eqref{eq:density_estimation_MLMC_multidim} enables the MLMC estimator (see Section~\ref{sec:Details of our approach})  to  compute density functions.   We recall that the MLMC method without  smoothing   fails due to the infinite variance caused by the singularity of  the delta function.  Moreover, owing to the exact conditional expectation,  the only error  present in the proposed smoothing approach corresponds  to the root-finding procedure for finding the discontinuity location,  which does not depend exponentially on the dimension of the problem.   	The QMC method with kernel density estimation techniques, as in \cite{ben2021density, l2022monte},   or the MLMC  method combined with parametric smoothing approach, as    in \cite{giles2015multilevel}   can be used as  an alternative to our approach. However,  this class of   approaches has a pointwise  error that increases exponentially  w.r.t.~the dimension of the state vector $\mathbf{X}$ (or a   vector-valued  function that depends on the density of $\mathbf{X}$). For instance,  for a $d$-dimensional problem,  the kernel density estimator with a bandwidth matrix, $\mathcal{H}=\diag(h,\dots,h)$,  has an  MSE on  the order of  $c_1 M^{-1} h^{-d}+c_2 h^{4}$, where $M$ is the number of samples and $c_{1}$ and  $c_2$ are constants. 
	\begin{remark}[Extending  numerical smoothing for computing sensitivities]\label{rem: Extending  numerical smoothing for computing sensitivities}
	The proposed  approach can be extended to computing sensitivities, particularly financial Greeks  using efficient MLMC methods based on pathwise or likelihood ratio  approaches  (see \cite{glasserman2004monte}). These methods rely on the smoothness of the payoff function (or its derivative). For illustration, we  denote the payoff function by  $g_\theta(X_T)$,  where $\theta$ represents a parameter of interest. 	 The quantity of interest   can then be expressed as $u(\theta):= E[g_{\theta}(X_T)]$. The pathwise estimate, $u'(\theta)=E[\frac{d g_{\theta}(X_T)}{d \theta}]$, is unbiased  and applicable if enough smoothness conditions hold for $g$ and its deriavtive (see Section~7.2.2 in \cite{glasserman2004monte}).  As an alternative, in the  likelihood ratio method, we write  $u(\theta)= \int g(x_T)   \rho_\theta(x_T) dx_T $, where $\rho_\theta(x_T)$ is the density of $X_T$ depending on the parameter  $\theta$. Then, if the interchange of differentiation and expectation is justified, we obtain $u'(\theta)=E[g(X_T)\frac{d   \log(\rho_\theta(X_T))}{d \theta}]$. When $g(.)$ is discontinuous, the performance of the MLMC method  deteriorates, as explained earlier.  In future work, we intend to explore these directions further.
		\end{remark}
	\begin{remark}[Extending  numerical smoothing for inference problems]
	The proposed  approach can be    adapted to solving  inference problems  \cite{hoel2016multilevel,beskos2017multilevel,warne2018multilevel} using MLMC. Instead of   smoothing the observable  using kernel-based method as in \cite{warne2018multilevel}, we can adapt the  numerical smoothing idea as an alternative.
 	\end{remark}	
\section{MLMC Combined with Numerical Smoothing}\label{sec:Details of our approach}
Using  the MLMC method, as described in \cite{giles2008multilevel,giles2015multilevel_2}, our approach   aims to efficiently approximate the resulting expectation obtained after the numerical smoothing  step,    defined by \eqref{eq: pre_integration_step_wrt_y1_basket}-\eqref{eq:expression_h_bar}, when $g(\mathbf{x})=\mathbf{1}_{(\phi(\mathbf{x}) \ge 0)}$, or  \eqref{eq:density_estimation_MLMC} and \eqref{eq:density_estimation_MLMC_multidim}, when $g(\mathbf{x})=\delta\left(\phi(\mathbf{x}) = 0\right)$. 

We  construct our MLMC estimator as follows: First, we consider a hierarchy of nested meshes of the time interval $[0,T]$, indexed by $\ell=L_0, L_0+1, \dots, L$.  $\Delta t_{0}$ denotes the step size used at level $\ell=L_0$. The size of the subsequent time steps for levels $\ell \geq L_0+1$ is given by $ \Delta t_{\ell}=K^{-\ell} \Delta t_{0}$, where $K{>}1$ is a given constant integer. In this work, we take $K = 2$. Moreover,  $M_{\text{Lag},\ell}$ and  $\text{TOL}_{\text{Newton},\ell}$  denote the number of Laguerre quadrature points  and  the tolerance of the Newton method at the level $\ell$, respectively.  Hereafter,  to simplify notation, $\bar{I}_{\ell}$ corresponds to  $\bar{I}$  expressed in  \eqref{eq:expression_h_bar} (or $\bar{F}$ expressed in  \eqref{eq:density_estimation_MLMC_multidim}, when estimating densities) computed using  $ \Delta t_{\ell}$,  $\text{TOL}_{\text{Newton},\ell}$ and $M_{\text{Lag},\ell}$.\footnote{We do not need the Laguerre integration points when estimating densities.} Finally, we denote by  $M_{\ell}$  the number of samples at level $\ell$. 
 
Consider now the following telescoping decomposition of  $\expt{\bar{I}_{L}}$
\begin{align*}
 	\expt{\bar{I}_{L}}&=  \quad \expt{\bar{I}_{L_0}}  \quad   +  \quad  \sum_{\ell=L_0+1}^{L} \expt{\bar{I}_{\ell}- \bar{I}_{\ell-1}}
 \end{align*} 
Then, by defining 
\begin{equation}\label{eq:MLMC_terms}
 	\widehat{Q}_{L_0}:= \frac{1}{M_{L_0}} \sum\limits_{m_{L_0}=1}^{M_{L_0}} \bar{I}_{L_0,[m_{L_0}]}; \quad \widehat{Q}_{\ell}:= \frac{1}{M_{\ell}} \sum\limits_{m_{\ell}=1}^{M_{\ell}}  \left(   \bar{I}_{\ell,[m_{\ell}]}-\bar{I}_{\ell-1,[m_{\ell}]}  \right), \: L_0+1 \le \ell \le L,
\end{equation}
we arrive at the unbiased MLMC estimator, $\hat{Q}$, of  $\expt{\bar{I}_L}$
	\begin{equation*}\label{eq:MLMC_estimator}
		\hat{Q}:= \sum\limits_{\ell=L_0}^{L} \hat{Q}_{\ell}\PERIOD
\end{equation*}
Notably, the key point in constructing \eqref{eq:MLMC_terms} is that both $\overline{I}_{\ell,[m_{\ell}]}$ and $\overline{I}_{\ell-1,[m_{\ell}]}$ are sampled using different time discretizations but with the same generated randomness.

\subsection{Error    analysis}\label{sec:Error discussion in the context of MLMC method}
This Section  analyzes  the different error contributions in our approach that combines  the MLMC   estimator  with the numerical smoothing to approximate $\expt{g(\mathbf{X}(T))}$ with $g$ given by \eqref{eq:proba_comp} or \eqref{eq:density_comp}.    Following the notation of Sections \ref{sec:General setting} and \ref{sec:Details of our approach},  we obtain  the  following error decomposition
\begin{align}\label{eq: error decomposition MLMC}
\expt{g(\mathbf{X}(T))}-\hat{Q}&= \underset{\text{Error I: bias or weak error}}{\underbrace{\expt{g(\mathbf{X}(T))}- \expt{g(\bar{\mathbf{X}}^{\Delta t_{L}}(T))}}}\nonumber\\
&+\underset{\text{Error II:  numerical integration and root-finding error}} {\underbrace{\expt{I_L\left(\mathbf{Y}_{-1}, \mathbf{Z}^{(1)}_{-1},\dots,\mathbf{Z}^{(d)}_{-1}\right)}- \expt{\bar{I}_L\left(\mathbf{Y}_{-1}, \mathbf{Z}^{(1)}_{-1},\dots,\mathbf{Z}^{(d)}_{-1} \right)}}}\nonumber\\
&+ \underset{\text{Error III: MLMC  statistical error}}{\underbrace{\expt{\bar{I}_L\left(\mathbf{Y}_{-1}, \mathbf{Z}^{(1)}_{-1},\dots,\mathbf{Z}^{(d)}_{-1}\right)}- \hat{Q}}},
\end{align} 
where $I_L$ corresponds to  $I$   in  \eqref{eq:smooth_function_after_pre_integration} (or $F$ in  \eqref{eq:density_estimation_MLMC_multidim}) computed with  $ \Delta t_{L}$.

Because we simulate the dynamics of $X$ using  Euler--Maruyama or Milstein  schemes,    we obtain
\begin{equation}\label{eq:Error I order MLMC}
\text{Error I}=\Ordo{\Delta t_L}.
\end{equation}
\text{Error II}  in  \eqref{eq: error decomposition MLMC} was analyzed in  \cite{bayer2020numerical}, and for  the  case $g(\mathbf{x})=\mathbf{1}_{(\phi(\mathbf{x}) \ge 0)}$, is  expressed as  
	\begin{align}\label{eq:Error II order}
		\text{Error II}&:=\expt{I_L\left(\mathbf{Y}_{-1}, \mathbf{Z}^{(1)}_{-1},\dots,\mathbf{Z}^{(d)}_{-1}\right)}- \expt{\bar{I}_L\left(\mathbf{Y}_{-1}, \mathbf{Z}^{(1)}_{-1},\dots,\mathbf{Z}^{(d)}_{-1} \right)}\nonumber\\
		&= \Ordo{M_{\text{Lag},L}^{-s/2}}+ \Ordo{\text{TOL}_{\text{Newton},L}},
	\end{align}
where   $s>0$ is related to the degree of regularity of the integrand, $G$, w.r.t.~$y_1$.\footnote{For the parts of the domain separated by the discontinuity location, the derivatives of $G$ w.r.t.~$y_1$ are bounded up to order $s$.}
	
For the density estimation case,  we  obtain $	\text{Error II}=\Ordo{\text{TOL}_{\text{Newton},L}}$  because we do not perform any numerical pre-integration.

  \text{Error III}  presents the corresponding statistical error. From the standard multilevel analysis (see \cite{giles2008multilevel,giles2015multilevel_2}),  we obtain
	\begin{equation}\label{eq:Error III, MLMC, order_1}
		\text{Error III} \propto \sqrt{\sum_{\ell=L_0}^L {(M^{\ast}_{\ell})}^{-1} V_{\ell}}= \sqrt{\sum_{\ell=L_0}^L \sqrt{C_{\ell} V_{\ell}}},
	\end{equation}
where $M^{\ast}_{\ell}$ is the optimal number of samples per level, 
\begin{equation*}
	V_{L_0}:=\text{Var}\left[\bar{I}_{L_0}\right], \:   V_{\ell}:=\text{Var}\left[\bar{I}_{\ell}-\bar{I}_{\ell-1}\right], \: L_0+1 \le \ell \le L,
\end{equation*}
and  $C_{\ell}$ is the cost per sample per level, given by\footnote{For the  case $g(\mathbf{x})=\delta(\phi(\mathbf{x}) = 0)$,  we do not have the term $M_{\text{Lag},\ell}$ in  $C_{\ell}$.}
\begin{equation}\label{eq:cost_estimates}
C_{\ell} \propto (\Delta t_{\ell})^{-1} (M_{\text{Lag},\ell}+N_{\text{iter},\ell})\propto (\Delta t_{\ell})^{-1} \left(M_{\text{Lag},\ell}+\log \left(\text{TOL}_{\text{Newton},\ell}^{-1}\right)\right), \: L_0 \le \ell \le L,
\end{equation}
where $N_{\text{iter},\ell}$ is the number of the Newton iterations at level $\ell$.\footnote{Under some mild conditions and using Taylor expansion, we can  show that Newton iteration has a second order convergence and conclude that $N_{\text{iter},\ell} \propto \log \left(\text{TOL}_{\text{Newton},\ell}^{-1}\right)$.}

Theorems \ref{corrol: Lipschitz of the integrands} and \ref{corrol: Lipschitz of the integrands_density} derive  estimates of the variances  $\{V_{\ell}\}_{\ell=L_0+1}^{L}$,   and show that $V_{\ell}=\Ordo{\Delta t_{\ell}}$ when using the  Euler-Maruyama scheme. The analysis when combining our approach with the Milstein scheme is left for future work.
	
Finally, using  \eqref{eq: error decomposition MLMC}, \eqref{eq:Error I order MLMC}, \eqref{eq:Error II order} and \eqref{eq:Error III, MLMC, order_1}, the total error estimate of our approach is
	\begin{align}\label{eq:total_error_estimate MLMC}
		\mathcal{E}_{\text{total}}&:=\expt{g(\mathbf{X}(T))}-\hat{Q}\nonumber\\
		&= \Ordo{\Delta t_{L}}+\Ordo{\sqrt{\sum_{\ell=L_0}^L {(M^{\ast}_{\ell})}^{-1} V_{\ell}}}+\Ordo{M_{\text{Lag},L} ^{-s/2}}+ \Ordo{\text{TOL}_{\text{Newton},L}}.
	\end{align}
\subsection{Strong Convergence results for MLMC with numerical smoothing}\label{sec:Strong Convergence results for MLMC with numerical smoothing}	
Before stating the main theorems and their proofs, we introduce  some needed notations and Assumptions.  For ease of notation, we show the proofs of Theorems \ref{corrol: Lipschitz of the integrands} and \ref{corrol: Lipschitz of the integrands_density} for the 1D case. 

We extend the approximate process $\bar{X}$ (using the Euler-Maruyama scheme  and defined on the time grid  $0=t_0<t_1<\ldots < t_N =T$)  of  $X$  in \eqref{eq:SDE_interest}  to all $t \in [0,T]$, and write
\begin{equation}\label{eq: Euler SDE}
	\bar{X} (t) = \bar{X} (0) + \int_{0}^t a(\bar{X}([s])) ds + \int_{0}^t b(\bar{X}([s])) dW(s),
\end{equation}
 where $[s]$ represents $s$ rounded down to the nearest discrete time $t_n$ ($0 \le n \le N$) on the given time mesh.

Moreover, we denote by $\bar{X}_{\ell} , \bar{X}_{\ell-1}$ the coupled paths of the  approximate process $\bar{X}$, simulated with time step sizes $\Delta t_{\ell}$  and  $\Delta t_{\ell-1}$, respectively. Then, using  \eqref{eq:bridge_construction},  we define    $\tilde{e}_{\ell}(t;  W_{\ell})$ and  $e_{\ell}(t; Y, B_{\ell})$ as
\begin{small}
\begin{align}\label{eq:error_eq}
	&(\bar{X}_{\ell}- \bar{X}_{\ell-1}) (t)= \int_0^t \left( a(\bar{X}_{\ell}  ([s]_{\ell}))-a(\bar{X}_{\ell-1} ([s]_{\ell-1}))\right) ds+ \int_0^t \left(b(\bar{X}_{\ell}  ([s]_{\ell}))-b(\bar{X}_{\ell-1} ([s]_{\ell-1}))\right)  dW_{\ell}(s)=: \tilde{e}_{\ell}(t;  W_{\ell})\nonumber\\
	&= \int_0^t \left( a(\bar{X}_{\ell}  ([s]_{\ell}))-a(\bar{X}_{\ell-1} ([s]_{\ell-1}))\right) ds+ \int_0^t \left(b(\bar{X}_{\ell}  ([s]_{\ell}))-b(\bar{X}_{\ell-1} ([s]_{\ell-1}))\right)  \frac{Y}{\sqrt{T}} ds\nonumber\\
	&+\int_0^t \left(b(\bar{X}_{\ell}  ([s]_{\ell}))-b(\bar{X}_{\ell-1} ([s]_{\ell-1}))\right)  dB_{\ell}(s)\nonumber\\
	&=: e_{\ell}(t; Y, B_{\ell}),
\end{align}
\end{small}
 where $W_{\ell}$ and  $B_{\ell}$ correspond to  the coupling Wiener process and related   Brownian bridge process at levels $\ell$ and $\ell-1$ in the MLMC estimator, respectively.

Finally,  for $\delta>0$, $\tilde{g}_{\delta}$  denotes  a  $C^\infty$ mollified  version of $g$ (\ie, obtained by convoluting $g$ with a mollifier).
\begin{notation}\label{notation}
	For sequences of rdvs $F_N$, we write  that
	$F_N = \mathcal{O}(1)$ if there exists a rdv $C$ with finite moments
	of all orders,  such that for all $N$, we have $\abs{F_N} \le C $ a.s.
\end{notation}
\begin{assumption}[Global Lipschitz continuity  of drift and diffusion coefficients]\label{ass: globally Lipschitz_drift_diffusion}\
	
	The drift and diffusion terms  in \eqref{eq:SDE_interest} ($a(\cdot)$ and  $b(\cdot)$) are  globally Lipschitz, that is, $\forall x,y \in \rset$, there exists $C>0$ such that
	\begin{equation*}
		\label{eq:ass_lipcoeff}
		\hspace{-4mm}\max{\left\{|a(x)-a(y)|, |b(x)-b(y)|\right\}\le C|x-y|}.
	\end{equation*}
\end{assumption}
	\begin{assumption}[Additional conditions for Theorem \ref{corrol: Lipschitz of the integrands}:  smoothness of drift and diffusion coefficients and  uniform boundedness of first order derivatives]\label{ass:  uniform boundedness of first order derivatives}\
		
		  The functions $a(\cdot)$ and  $b(\cdot)$ are   of class $C^2(\rset,\rset)$ with $a'(\cdot)$ and $b'(\cdot)$ being uniformly bounded. \label{assum2}
\end{assumption}
	\begin{assumption}[Additional conditions for Theorem \ref{corrol: Lipschitz of the integrands_density}:  smoothness of drift and diffusion coefficients  and uniform boundedness of first  and second order derivatives]   \label{ass:  uniform boundedness of first second order derivatives}\
		
	  The functions 	$a(\cdot)$ and  $b(\cdot)$  are   of class $C^3(\rset,\rset)$ with $a'(\cdot), b'(\cdot),a''(\cdot), b''(\cdot)$ being uniformly bounded. 
\end{assumption}
	\begin{assumption}[Conditions for Proposition \ref{lemma: boundary_condition_error growth}: uniform boundedness of the drift and diffusion coefficients]   \label{ass:  uniform boundedness of the drift and diffusion coefficients}\
		
		  The functions  $a(\cdot)$ and  $b(\cdot)$  are   uniformly bounded. 
\end{assumption}
\begin{remark}[On the relaxation of  Assumption \ref{ass:  uniform boundedness of the drift and diffusion coefficients}]
Assumption \ref{ass:  uniform boundedness of the drift and diffusion coefficients} is used for the proof of  Proposition \ref{lemma: boundary_condition_error growth}, needed in both Theorems \ref{corrol: Lipschitz of the integrands} and \ref{corrol: Lipschitz of the integrands_density}. However, it can be relaxed by proving   Proposition \ref{lemma: boundary_condition_error growth} differently  using instead  Assumption \ref{ass: globally Lipschitz_drift_diffusion}. We refer to Remark \ref{rem:ass_relaxation} for more details. Note that  even though the examples that we consider in Section \ref{sec:Numerical smoothing with MLMC} (the GBM and Heston models) do not satisfy some of the Assumptions \ref{ass: globally Lipschitz_drift_diffusion}--\ref{ass:  uniform boundedness of the drift and diffusion coefficients}, we still obtain the same estimates stated in Theorems  \ref{corrol: Lipschitz of the integrands} and \ref{corrol: Lipschitz of the integrands_density}, i.e., these assumptions are sufficient but not necessary.
	\end{remark}
\begin{theorem}[Variance estimates for  probabilities computation] \label{corrol: Lipschitz of the integrands} 	Let  the function $g$ as in \eqref{eq:proba_comp}. Then under  Assumptions \ref{ass: globally Lipschitz_drift_diffusion},  \ref{ass:  uniform boundedness of first order derivatives}, \ref{ass:  uniform boundedness of first second order derivatives}, \ref{ass:  uniform boundedness of the drift and diffusion coefficients},   \ref{ass:boundedness-derivative} and  \ref{ass:boundedness-inverse}, we obtain 
	\begin{equation}\label{eq:variance_estimates_prob}
	V_{\ell}=\Ordo{\Delta t_{\ell}}.
	\end{equation}
\end{theorem}
\begin{theorem}[Variance estimates for  densities estimation] \label{corrol: Lipschitz of the integrands_density}
	Let  the function $g$ as in  \eqref{eq:density_comp}. Then under  Assumption \ref{ass: globally Lipschitz_drift_diffusion}, \ref{ass:  uniform boundedness of first order derivatives}, \ref{ass:  uniform boundedness of first second order derivatives}, \ref{ass:  uniform boundedness of first second order derivatives}, \ref{ass:  uniform boundedness of the drift and diffusion coefficients}, \ref{ass:boundedness-derivative} and  \ref{ass:boundedness-inverse}, we obtain 
	\begin{equation}\label{eq:variance_estimates_density}
		V_{\ell}=\Ordo{\Delta t_{\ell}}.
	\end{equation}
\end{theorem}
\begin{proof}[Proof of Theorem \ref{corrol: Lipschitz of the integrands}]\label{proof: Lipschitz of the integrands_revised}
We want to show  $V_{\ell}  := \text{Var}\left[\bar{I}_{\ell}-\bar{I}_{\ell-1}\right] \le  \expt{\left(\bar{I}_{\ell}-\bar{I}_{\ell-1}\right)^2} =\Ordo{\Delta t_{\ell}}$.  
For $\delta>0$, we have  
\begin{small}
\begin{align}\label{eq:proof_result1}
	&\Delta I^{\delta}_{\ell}( B_{\ell}):=(\bar{I}^{\delta}_{\ell}-\bar{I}^{\delta}_{\ell-1})( B_{\ell})\nonumber\\
	&:= \int_\rset \left(  \tilde{g}_{\delta} (\bar{X}_{\ell} (T; y, B_{\ell})) - \tilde{g}_{\delta} (\bar{X}_{\ell-1} (T; y, B_{\ell}))  \right) \rho_1(y) dy\nonumber\\
	&= \int_\rset \left[ \int_{0}^{1} \tilde{g}_{\delta}'  \left(\underset{:=z(\theta; y,B_{\ell} )}{\underbrace{\bar{X}_{\ell-1}(T; y, B_{\ell})+\theta   e_{\ell}(T; y, B_{\ell}}})\right) d \theta  \right]  e_{\ell}(T; y, B_{\ell}) \; \rho_1(y) dy, \: \theta \in (0,1) \nonumber\\
	&= \int_\rset \left[ \int_{0}^{1}  \partial_y \tilde{g}_{\delta} (z(\theta;y,B_{\ell}) )  \left(\partial_y z(\theta;y,B_{\ell}) \right)^{-1} d \theta  \right]  e_{\ell}(T; y, B_{\ell}) \; \rho_1(y) dy  \quad (\text{using } \partial_y \tilde{g}_{\delta} = \tilde{g}_{\delta}'   \partial_y z) \nonumber\\
		&=  \int_{0}^{1} \left[ \int_\rset   \partial_y \tilde{g}_{\delta} (z(\theta;y,B_{\ell}) )  \left(\partial_y z(\theta;y,B_{\ell}) \right)^{-1}    e_{\ell}(T; y, B_{\ell}) \; \rho_1(y) dy\right]  d\theta  \:(\text{using Fubini's theorem})\nonumber\\
		&=-  \int_{0}^{1}  \left[  \int_\rset \tilde{g}_{\delta} (z(\theta;y,B_{\ell}) )\;  \partial_y \left(\left(\partial_y z(\theta;y,B_{\ell} ) \right)^{-1}     e_{\ell}(T; y, B_{\ell}) \; \rho_1(y)  \right)  dy \right] d \theta\:(\text{boundary terms vanish due to Proposition \ref{lemma: boundary_condition_error growth}})\nonumber\\
		&=  -\int_{0}^{1}  \left[  \int_\rset \tilde{g}_{\delta}(z(\theta;y,B_{\ell}))  \left(   e_{\ell}(T; y, B_{\ell})  \; \partial_y \left(\left(\partial_y z(\theta;y,B_{\ell}) \right)^{-1}  \rho_1(y) \right)  +  \left(\partial_y z(\theta;y,B_{\ell}) \right)^{-1}    \rho_1(y)\; \partial_y e_{\ell}(T; y, B_{\ell})  \right)  dy \right] d \theta\nonumber\\
			&=  -\int_{0}^{1}  \left[  \int_\rset e_{\ell}(T; y, B_{\ell}) \tilde{g}_{\delta}(z(\theta;y,B_{\ell}))   \left(   \partial_y \left(\left(\partial_y z(\theta;y,B_{\ell}) \right)^{-1}     \right) - y  \left(\partial_y z(\theta;y,B_{\ell}) \right)^{-1}\right)   \rho_1(y)dy \right] d \theta\nonumber\\
				&- \int_{0}^{1}  \left[  \int_\rset  \partial_y e_{\ell}(T; y, B_{\ell}) \tilde{g}_{\delta}(z(\theta;y,B_{\ell}))  \left(\partial_y z(\theta;y,B_{\ell}) \right)^{-1}     \rho_1(y) dy \right] d \theta \PERIOD
	\end{align}
\end{small}
Taking $\delta\rightarrow 0$ and applying  the dominated convergence theorem to   \eqref{eq:proof_result1}, we  obtain
\begin{small}
	\begin{align}\label{eq:proof_result2}
		\Delta I_{\ell}( B_{\ell})&:=(\bar{I}_{\ell}-\bar{I}_{\ell-1})( B_{\ell})\nonumber\\
		&=  \underset{(I)}{\underbrace{{ -\int_{0}^{1}  \left[  \int_\rset e_{\ell}(T; y, B_{\ell}) g(z(\theta;y,B_{\ell}))   \left(   \partial_y \left(\left(\partial_y z(\theta;y,B_{\ell}) \right)^{-1}     \right) - y  \left(\partial_y z(\theta;y,B_{\ell}) \right)^{-1}\right)  \rho_1(y) dy \right] d \theta}}}\nonumber\\
		& \underset{(II)}{\underbrace{{- \int_{0}^{1}  \left[  \int_\rset  \partial_y e_{\ell}(T; y, B_{\ell}) g(z(\theta;y,B_{\ell}))  \left(\partial_y z(\theta;y,B_{\ell}) \right)^{-1}       \rho_1(y) dy \right] d \theta}}} \PERIOD
	\end{align}
\end{small}
Using \eqref{eq:error_eq},  we recall that for Euler--Maruyama scheme and  $p\ge 1$, under Assumption \ref{ass: globally Lipschitz_drift_diffusion},  we have \cite{kloeden1992stochastic}
\begin{small}
	\begin{equation}
			\expt{\tilde{e}_{\ell}^{2p} (T)}=	\expt{e_{\ell}^{2p}(T)}=\Ordo{\Delta t_{\ell}^{p}}. \label{eq:L_p_moments_estimate} \\
	\end{equation}
\end{small}
Moreover,   Lemma   \ref{lemma: boundness of y-derivatives of the weak error} implies  that for any $p\ge1$ 
\begin{small}
\begin{equation}
	\expt{(\partial_y e_{\ell})^{2p}(T)}=\Ordo{\Delta t_{\ell}^{p}}. \label{eq:Y-derivativeL_p_moments_estimate} 
\end{equation}
\end{small}
For the   term (I) in   \eqref{eq:proof_result2},  taking expectation with respect to the Brownian bridge   and using H\"older's inequality twice ($p,q, p_1,q_1 \in (1,+\infty)$,  $\frac{1}{p}+\frac{1}{q}=1$ and $\frac{1}{p_1}+\frac{1}{q_1}=1$), result in
\begin{small}
	\begin{align}\label{eq:proof_result2_way2}
		E\left[\left( I\right)^2\right]	
		&\le  E\left[\left | \left|   g(z(\cdot; \cdot, B_{\ell})) \left(   \partial_y \left(\left(\partial_y z(\cdot;\cdot,B_{\ell}) \right)^{-1}     \right) - Y  \left(\partial_y z(\cdot;\cdot,B_{\ell}) \right)^{-1}\right)\right|\right|^2_{L_{\rho_1}^q ([0,1]\times \rset )}   \times \left|\left|   e_{\ell}(T; \cdot, B_{\ell})\right|\right|^2_{L_{\rho_1}^p (\rset)} \right]	\nonumber\\
		&\le  \left(E\left[\left | \left|   g(z(\cdot; \cdot,B_{\ell}))  \left(   \partial_y \left(\left(\partial_y z(\cdot;\cdot,B_{\ell}) \right)^{-1}     \right) - Y  \left(\partial_y z(\cdot;\cdot,B_{\ell}) \right)^{-1}\right)\right|\right|^{2q_1}_{L_{\rho_1}^q ([0,1]  \times \rset )}\right] \right)^{1/q_1}\\\nonumber
		&  \times  \left(E\left[\left|\left|    e_{\ell}(T;\cdot, B_{\ell})\right|\right|^{2p_1}_{L_{\rho_1}^p (\rset)} \right]\right)^{1/p_1}\PERIOD
	\end{align}
\end{small}
Choosing $p$ and $p_1$ such that $\frac{2p_1}{p}\le1$, and applying Jensen's inequality for  the second term in  the right-hand side  of \eqref{eq:proof_result2_way2}, we obtain
\begin{small}
\begin{align}\label{eq:L_p_moments_estimate_derivation}
	\left(E\left[\left|\left|    e_{\ell}(T; \cdot, B_{\ell})\right|\right|^{2p_1}_{L_{\rho_1}^p (\rset)} \right]\right)^{1/p_1} &= 	\left(E\left[\left(\int_\rset  \left| e^p_{\ell}(T; y, B_{\ell}) \right| \rho_1 dy\right)^{\frac{2p_1}{p}} \right]\right)^{1/p_1} 	\nonumber\\
	&\le	\left(E\left[\int_\rset   \left|e^p_{\ell}(T; y, B_{\ell})\right| \rho_1 dy\right]\right)^{\frac{2}{p}} 	\nonumber\\
	&= \Ordo{ \Delta t_{\ell}}  \:(\text{using Fubini's theorem and  \eqref{eq:L_p_moments_estimate}})\PERIOD
\end{align}
\end{small}
The first term in  the right-hand side  of \eqref{eq:proof_result2_way2} is bounded. In fact,  observe that  
\begin{small}
\begin{align}\label{eq:aux_results}
(\partial_y z(\theta;y,B_{\ell}))^{-1}&=   \left(\partial_y \bar{X}_{\ell-1}(T)\right)^{-1}  \left((1- \theta) + \theta \frac{\partial_y \bar{X}_{\ell}(T)}{\partial_y \bar{X}_{\ell-1}(T)}\right)^{-1},\\\nonumber
\partial_y \left((\partial_y z(\theta;y,B_{\ell}))^{-1}\right)&=   -\partial^2_y z(\theta;y,B_{\ell})(\partial_y z(\theta;y,B_{\ell}))^{-2},\\\nonumber
&= -\left((1- \theta) \partial^2_y \bar{X}_{\ell-1}(T)+ \theta \partial^2_y \bar{X}_{\ell}(T)\right) \left(\partial_y \bar{X}_{\ell-1}(T)\right)^{-2}  \left((1- \theta) + \theta \frac{\partial_y \bar{X}_{\ell}(T)}{\partial_y \bar{X}_{\ell-1}(T)}\right)^{-2}.
\end{align}
\end{small}
 Using Assumption \ref{ass:boundedness-inverse}, we obtain that $\left(\partial_y \bar{X}_{\ell-1}(T)\right)^{-1} $ and $\left(\partial_y \bar{X}_{\ell-1}(T)\right)^{-2}$ are bounded in moments, \ie,  $\Ordo{1}$ in the sense of notation \ref{notation}. Moreover, using Assumption \ref{ass:boundedness-derivative} and Lemma \ref{lem:dXdZ}, we obtain that  $\partial^2_y \bar{X}_{\ell-1}(T)$ and  $ \partial^2_y \bar{X}_{\ell}(T)$ are bounded in moments.  These results with \eqref{eq:aux_results} imply    $$\left(E\left[\left | \left|   g(z(\cdot;\cdot,B_{\ell} )) \left(   \partial_y \left(\left(\partial_y z(\cdot;\cdot,B_{\ell}) \right)^{-1}     \right) - Y  \left(\partial_y z(\cdot;\cdot,B_{\ell}) \right)^{-1}\right) \right|\right|^{2q_1}_{L_{\rho_1}^q (  [0,1]  \times \rset)}\right] \right)^{1/q_1}  <\infty,$$
 and consequently, we conclude that \eqref{eq:proof_result2_way2} $=\Ordo{\Delta t_{\ell}}$.
 
For the   term (II) in   \eqref{eq:proof_result2},  taking expectation with respect to the Brownian bridge   and using H\"older's  inequality twice ($p,q, p_1,q_1 \in (1,+\infty)$,  $\frac{1}{p}+\frac{1}{q}=1$ and $\frac{1}{p_1}+\frac{1}{q_1}=1$), result in
\begin{small}
	\begin{align}\label{eq:proof_result2_way2_termII}
		E\left[\left( II\right)^2\right]	
		&\le  E\left[\left | \left|   g(z(\cdot; \cdot, B_{\ell})) \left(\partial_y z(\cdot;\cdot,B_{\ell}) \right)^{-1}      \right|\right|^2_{L_{\rho_1}^q ( [0,1] \times  \rset)}   \times \left|\left|   \partial_y e_{\ell}(T; \cdot, B_{\ell})\right|\right|^2_{L_{\rho_1}^p (\rset)} \right]	\nonumber\\
		&\le  \left(E\left[\left | \left|   g(z(\cdot;\cdot, B_{\ell})) \left(\partial_y z(\cdot;\cdot,B_{\ell}) \right)^{-1}\right|\right|^{2q_1}_{L_{\rho_1}^q ([0,1] \times \rset)}\right] \right)^{1/q_1} \times  \left(E\left[\left|\left|    \partial_y e_{\ell}(T;\cdot, B_{\ell})\right|\right|^{2p_1}_{L_{\rho_1}^p (\rset)} \right]\right)^{1/p_1}\PERIOD
	\end{align}
\end{small}
Similarly to \eqref{eq:L_p_moments_estimate_derivation} and using \eqref{eq:Y-derivativeL_p_moments_estimate}, we obtain that  $ \left(E_{B_{\ell}}\left[\left|\left|  \partial_y   e_{\ell}(T;\cdot,B_{\ell})\right|\right|^{2p_1}_{L_{\rho_1}^p (\rset)} \right]\right)^{1/p_1}= \Ordo{\Delta t_{\ell}}$.  Moreover, as explained earlier and using \eqref{eq:aux_results}, we get the first term in  the right-hand side  of \eqref{eq:proof_result2_way2_termII} to be bounded. This concludes that \eqref{eq:proof_result2_way2_termII}$=\Ordo{\Delta t_{\ell}}$, 
and consequently finishes  the  proof. 
 \end{proof}
\begin{proof}[Proof of Theorem \ref{corrol: Lipschitz of the integrands_density}]\label{proof: Lipschitz of the integrands_revised_density}
We have
	\begin{small}
		\begin{align}\label{eq:proof_result1_density}
			&\Delta I^{\delta}_{\ell}( B_{\ell}):=(\bar{I}^{\delta}_{\ell}-\bar{I}^{\delta}_{\ell-1})( B_{\ell})\nonumber\\
		&= \int_\rset \left(  \tilde{g}_{\delta} (\bar{X}_{\ell} (T; y, B_{\ell})) - \tilde{g}_{\delta} (\bar{X}_{\ell-1} (T; y, B_{\ell}))  \right) \rho_1(y) dy\nonumber\\
		&= \int_\rset \left[ \int_{0}^{1} \tilde{g}_{\delta}'  \left(\underset{:=z(\theta; y,B_{\ell} )}{\underbrace{\bar{X}_{\ell-1}(T; y, B_{\ell})+\theta   e_{\ell}(T; y, B_{\ell}}})\right) d \theta  \right]  e_{\ell}(T; y, B_{\ell}) \; \rho_1(y) dy,  \nonumber\\
		&= \int_\rset \left[ \int_{0}^{1}  \partial_y \tilde{g}_{\delta} (z(\theta;y,B_{\ell}) )  \left(\partial_y z(\theta;y,B_{\ell}) \right)^{-1} d \theta  \right]  e_{\ell}(T; y, B_{\ell}) \; \rho_1(y) dy  \quad (\text{using } \partial_y \tilde{g}_{\delta} = \tilde{g}_{\delta}'   \partial_y z) \nonumber\\
		&=  \int_{0}^{1} \left[ \int_\rset   \partial_y \tilde{g}_{\delta} (z(\theta;y,B_{\ell}) )  \left(\partial_y z(\theta;y,B_{\ell}) \right)^{-1}    e_{\ell}(T; y, B_{\ell}) \; \rho_1(y) dy\right]  d\theta  \:(\text{using Fubini's theorem})\nonumber\\
		&=-  \int_{0}^{1}  \left[  \int_\rset \tilde{g}_{\delta} (z(\theta;y,B_{\ell}) )\;  \partial_y \left(\left(\partial_y z(\theta;y,B_{\ell} ) \right)^{-1}     e_{\ell}(T; y, B_{\ell}) \; \rho_1(y)  \right)  dy \right] d \theta\:(\text{boundary terms vanish due to Proposition \ref{lemma: boundary_condition_error growth}})\nonumber\\
			&= - \int_{0}^{1}  \left[  \int_\rset \tilde{G}_{\delta}'(z(\theta;y,B_{\ell}))  \left(   e_{\ell}(T;y, B_{\ell})  \; \partial_y \left(\left(\partial_y z(\theta;y,B_{\ell}) \right)^{-1}    \rho_1(y) \right)  +  \left(\partial_y z(\theta;y,B_{\ell}) \right)^{-1}    \rho_1(y)\; \partial_y e_{\ell}(T;y, B_{\ell})  \right)  dy \right] d \theta\nonumber\\
			&=  -\int_{0}^{1}  \left[  \int_\rset \partial y\tilde{G}_{\delta}(z(\theta; y))  \left(\partial_y z(\theta;y) \right)^{-1}  \left(   e_{\ell}(T;y, B_{\ell})  \; \partial_y \left(\left(\partial_y z(\theta;y) \right)^{-1}    \rho_1(y) \right)  +  \left(\partial_y z(\theta;y) \right)^{-1}    \rho_1(y)\; \partial_y e_{\ell}(T;y, B_{\ell})  \right)  dy \right] d \theta\nonumber\\
			&=  \int_{0}^{1}  \left[  \int_\rset \tilde{G}_{\delta}(z(\theta, y))   \partial y \left( \left(\partial_y z(\theta,y) \right)^{-1}  \left(   e_{\ell}(T;y, B_{\ell})  \; \partial_y \left(\left(\partial_y z(\theta;y) \right)^{-1}    \rho_1(y) \right)  +  \left(\partial_y z(\theta;y) \right)^{-1}    \rho_1(y)\; \partial_y e_{\ell}(T;y, B_{\ell})  \right)\right)  dy \right] d \theta\nonumber\\
			&=  \int_{0}^{1}  \left[  \int_\rset e_{\ell}(T;y, B_{\ell}) \tilde{G}_{\delta}(z(\theta;y,B_{\ell}))   \left( A_1(\theta;y,B_{\ell})\right)   \rho_1(y) dy \right] d \theta\nonumber\\
			&+ \int_{0}^{1}  \left[  \int_\rset \partial_y e_{\ell}(T;y, B_{\ell}) \tilde{G}_{\delta}(z(\theta; y,B_{\ell}))   \left( A_2(\theta;y,B_{\ell})\right)    \rho_1(y) dy \right] d \theta\nonumber\\
			&+\int_{0}^{1}  \left[  \int_\rset \partial_y^2e_{\ell}(T;y, B_{\ell}) \tilde{G}_{\delta}(z(\theta; y,B_{\ell}))   \left( A_3(\theta;y,B_{\ell})\right)    \rho_1(y) dy \right] d \theta\COMMA
		\end{align}
	\end{small}
	where 
	\begin{align*}
		A_1(\theta;y, B_{\ell})&=\left(\partial_y\left(\left(\partial_y z(\theta;y, B_{\ell}) \right)^{-1} \right)\right)^2+  \left(\partial_y z(\theta;y, B_{\ell}) \right)^{-2} \left(y^2-1\right)\\
		& + \left(\partial_y z(\theta;y, B_{\ell}) \right)^{-1} \left( \partial^2_y\left(\left(\partial_y z(\theta;y,B_{\ell}) \right)^{-1} \right)-3 y \partial_y\left(\left(\partial_y z(\theta;y, B_{\ell}) \right)^{-1} \right)  \right) \left(\partial_y z(\theta;y, B_{\ell}) \right)^{-1}
	\end{align*}
	\begin{align*}
		A_2(\theta;y,B_{\ell})&= \partial_y\left(\left(\partial_y z(\theta;y,B_{\ell}) \right)^{-1} \right) \left(\left(\partial_y z(\theta;y, B_{\ell}) \right)^{-1}+1 \right)  - y  \left( \left(\partial_y z(\theta;y, B_{\ell}) \right)^{-2} +  \left(\partial_y z(\theta;y, B_{\ell}) \right)^{-1}\right)
	\end{align*}
	\begin{align*}
		A_3(\theta;y, B_{\ell})&= (\left(\partial_y z(\theta;y, B_{\ell}) \right)^{-1}.
	\end{align*}
Taking $\delta\rightarrow 0$ and applying  the dominated convergence theorem to   \eqref{eq:proof_result1_density}, we  obtain
\begin{small}
	\begin{align}\label{eq:proof_result2_density}
		\Delta I_{\ell}( B_{\ell})&:=(\bar{I}_{\ell}-\bar{I}_{\ell-1})( B_{\ell})\nonumber\\
&=	\int_{0}^{1}  \left[  \int_\rset e_{\ell}(T;y, B_{\ell}) G(z(\theta;y, B_{\ell}))   \left( A_1(\theta;y, B_{\ell})\right)   \rho_1(y) dy \right] d \theta\nonumber\\
	&+ \int_{0}^{1}  \left[  \int_\rset \partial_y e_{\ell}(T;y, B_{\ell}) G(z(\theta; y, B_{\ell}))   \left( A_2(\theta;y, B_{\ell})\right)    \rho_1(y) dy \right] d \theta\nonumber\\
	&+\int_{0}^{1}  \left[  \int_\rset \partial_y^2e_{\ell}(T;y, B_{\ell})G(z(\theta;y, B_{\ell}))   \left( A_3(\theta;y, B_{\ell})\right)    \rho_1(y) dy \right] d \theta\COMMA
	\end{align}
\end{small}
	To derive the desired result for the density,  we redo the same steps \eqref{eq:proof_result2_way2}, \eqref{eq:L_p_moments_estimate_derivation} and \eqref{eq:proof_result2_way2_termII} in the proof of Theorem \ref{corrol: Lipschitz of the integrands}.  In addition to \eqref{eq:L_p_moments_estimate} and \ref{eq:Y-derivativeL_p_moments_estimate},  we  need  that, for $p\ge 1$,
	\begin{small}
		\begin{align}\label{eq:L_p_moments_estimate_2}
			\expt{(\partial^2_y e_{\ell})^{2p}}&=\Ordo{\Delta t_{\ell}^{p}}.
		\end{align}
	\end{small}
which can be proved in a similar way as  in the proof of Lemma \ref{lemma: boundness of y-derivatives of the weak error}  with further assuming that $a(\cdot)$ and  $b(\cdot)$ in \eqref{eq:SDE_interest} are   of class $C^3(\rset, \rset)$ (see Remark \ref{remark: extending lemma of  boundness of y-derivatives of the weak error}).

Finally, to conclude the proof, using Assumptions \ref{ass:boundedness-derivative} and \ref{ass:boundedness-inverse}, we get bounds on the terms  depending  on 	$A_1(\theta;y, B_{\ell})$, 	$A_2(\theta;y, B_{\ell})$ and 	$A_3(\theta;y, B_{\ell})$ in a similar way as in the proof of Theorem \ref{corrol: Lipschitz of the integrands} by deriving similar relations to \eqref{eq:aux_results} for   $\left(\partial_y\left(\left(\partial_y z(\theta;y, B_{\ell}) \right)^{-1} \right)\right)$, $\left(\partial_y z(\theta;y, B_{\ell}) \right)^{-2}$ and $\partial^2_y\left(\left(\partial_y z(\theta;y, B_{\ell}) \right)^{-1} \right)$.
\end{proof}


\subsection{Work   and Complexity  Analysis}\label{sec:Work discussion in the context of MLMC method}
From the MLMC analysis presented in \cite{giles2015multilevel_2} and from  \eqref{eq:cost_estimates} and  Theorems \ref{corrol: Lipschitz of the integrands} and \ref{corrol: Lipschitz of the integrands_density}, we obtain an estimate of the work of our approach as follows:
\begin{align}\label{eq:MLMC_work}
\text{Work} \left( L,L_0,\{M_{\text{Lag},\ell}\}_{\ell=L_0}^L , \{\text{TOL}_{\text{Newton},\ell}\}_{\ell=L_0}^L \right) & \propto   \sum_{\ell=L_0}^L M_{\ell}^\ast C_{\ell} \propto  \sum_{\ell=L_0}^L \sqrt{C_{\ell} V_{\ell}}  \nonumber\\
& \propto   \sum_{\ell=L_0}^L   \sqrt{M_{\text{Lag},\ell}+ \log\left(\text{TOL}_{\text{Newton},\ell}^{-1}\right)   }.
\end{align}
To achieve a certain error tolerance, $\text{TOL}$, with an optimal performance of our approach, one needs to solve  \eqref{eq:opt_MLMC_work} using  \eqref{eq:total_error_estimate MLMC} and \eqref{eq:MLMC_work}
	\begin{align}\label{eq:opt_MLMC_work}
		\begin{cases} 
			\underset{\left(L,L_0,\{M_{\text{Lag},\ell}\}_{\ell=L_0}^L , \{\text{TOL}_{\text{Newton},\ell}\}_{\ell=L_0}^L\right)}{\operatorname{min}} \: \text{Work} \left( L,L_0,\{M_{\text{Lag},\ell}\}_{\ell=L_0}^L , \{\text{TOL}_{\text{Newton},\ell}\}_{\ell=L_0}^L \right)  \\
			s.t. \:   \mathcal{E}_{\text{total}}=\text{TOL}
		\end{cases}
	\end{align}
In this work, we do not solve \eqref{eq:opt_MLMC_work}; however, we select the different parameters   heuristically\footnote{In our numerical experiments, we select $L_0$ such that $\text{Var}\left[\bar{I}_{L_0+1}{-}\bar{I}_{L_0}\right] \ll \text{Var}\left[\bar{I}_{L_0}\right]$, in order to  ensure the stability of the variance of the coupled paths of our MLMC estimator.}. A further investigation of optimizing  \eqref{eq:opt_MLMC_work} is left for a future study.

In Corollary \ref{corrol: Complexity of MLMC_smoothing},  we state the complexity of our approach, MLMC combined with numerical smoothing, compared with MLMC without smoothing.
\begin{corollary}[Complexity of MLMC with numerical smoothing]\label{corrol: Complexity of MLMC_smoothing}
 Under  the Assumptions of  Theorems  \ref{corrol: Lipschitz of the integrands} and \ref{corrol: Lipschitz of the integrands_density},	the complexity of MLMC  with numerical smoothing  using Euler--Maruyama  when computing probabilities  is $\Ordo{\text{TOL}^{-2-2/s} \left(\log(\text{TOL})\right)^2}$ (where  generally  $s \gg1$)  compared with $\Ordo{\text{TOL}^{-2.5}}$ for MLMC without smoothing. For the density estimation, the complexity of MLMC  with numerical smoothing  using Euler--Maruyama is $\Ordo{\text{TOL}^{-2} \left(\log(\text{TOL})\right)^2}$.
\end{corollary}
\begin{proof}\label{proof: complexity}
			Theorem 1   in \cite{giles2015multilevel_2} (see also Theorems 3.1 in \cite{giles2008multilevel} and  Theorem 1  in \cite{cliffe2011multilevel}) derives the computational complexity of the MLMC estimator     under different scenarios, depending on the values of $\alpha$ (weak convergence rate), $\beta$ (variance decay rate), and $\gamma$ (work rate). For  the Euler--Maruyama scheme, and  for  scenarios  with or without  numerical smoothing, we have $\gamma=1$. For non-Lipschitz functionals and without smoothing,  $V_{\ell}=\Ordo{\Delta t_{\ell}^{1/2}}$  (see \cite{giles2009analysing,avikainen2009irregular,giles2015multilevel_2}) (i.e.,  $\beta=1/2$).   Thus,  we obtain the worst-case MLMC complexity (i.e.,  $\Ordo{\text{TOL}^{-2.5}}$). 
			
			For our approach based on  the numerical smoothing idea, and by  Theorems  \ref{corrol: Lipschitz of the integrands} and \ref{corrol: Lipschitz of the integrands_density}, we recover  $V_{\ell}=\Ordo{\Delta t_{\ell}^{1}}$, \ie, $\beta=\gamma=1$. Recall that we require  an overall accuracy of order  $\text{TOL}$, \ie,  we desire to bound \eqref{eq: error decomposition MLMC} (equivalently \eqref{eq:total_error_estimate MLMC}) by $\text{TOL}$. Using  similar derivation and arguments  as   in the proof of Theorem 3.1  in \cite{giles2008multilevel} and Theorem 1  in \cite{cliffe2011multilevel},   to have Error III of order $\Ordo{\text{TOL}}$, we  choose 
			 	\begin{equation}\label{eq:number_of_samples_bound}
			 M_{\ell}^\ast  \le C\; \text{TOL}^{-2} (L-L_0+1) \Delta t_{\ell}+1,  \quad (C \: \text{ is a constant})
			 \end{equation}
			 		and  for Error I to be of   $\Ordo{\text{TOL}}$,   we obtain  
			 \begin{align}
			 	\sum_{\ell=L_0}^L  (\Delta t_{\ell})^{-1}  &= \Ordo{\text{TOL}^{-2}}, \: \text{and} \\
			 	L-L_0+1&= \Ordo{ \log(\text{TOL}^{-1})}.
			 \end{align}
			 Moreover, to bound Error II by  $\text{TOL}$, and using \eqref{eq:Error II order}, we obtain 
			 \begin{align}
			 M_{\text{Lag},L}&= \Ordo{\text{TOL}^{-2/s}}\\
			 N_{\text{iter},L}&= \log\left(\text{TOL}_{\text{Newton},L}^{-1}\right) =\Ordo{\log\left(\text{TOL}^{-1}\right)}
			 	\end{align}
		For simplification, we assume	that  on all levels ($L_0 \le \ell \le L$)   $M_{Lag,\ell}=M_{Lag,L}$ and $\text{TOL}_{\text{Newton},\ell}=\text{TOL}_{\text{Newton},L}$. Then, Using \eqref{eq:cost_estimates} and \eqref{eq:number_of_samples_bound}, we have the computational complexity of our MLMC estimator with numerical smoothing is  
		\begin{small}
				\begin{align}
				\sum_{\ell=L_0}^L M_{\ell}^\ast C_{\ell}  &\propto  	\sum_{\ell=L_0}^L M_{\ell}^\ast (\Delta t_{\ell})^{-1} \left(M_{\text{Lag},\ell}+\log \left(\text{TOL}_{\text{Newton},\ell}^{-1}\right)\right) \nonumber\\
				& \le  C \: \text{TOL}^{-2} (L-L_0+1)^2  \left(M_{\text{Lag},L}+\log \left(\text{TOL}_{\text{Newton,L}}^{-1}\right)\right) +   \left(M_{\text{Lag},L}+\log \left(\text{TOL}_{\text{Newton,L}}^{-1}\right)\right)  \sum_{\ell=L_0}^L  (\Delta t_{\ell})^{-1} \nonumber\\
				& =    \Ordo{\text{TOL}^{-2-(2/s)} \left(\log(\text{TOL})\right)^2} 
			\end{align}
			\end{small}
			When computing densities, the complexity of our MLMC estimator simplifies to   $\Ordo{\text{TOL}^{-2} \left(\log(\text{TOL})\right)^2}$.
	\end{proof}
\begin{remark}[About high-order schemes]  For non-Lipschitz observables, high-order schemes, such as the Milstein scheme, can improve the variance decay  rate \cite{giles2013numerical,giles2015multilevel_2} as compared with the Euler--Maruyama  scheme, thus improving  the MLMC estimator's complexity without the need for a smoothing procedure (see Section \ref{sec:Numerical smoothing with MLMC} for illustration). However,  this possibility  comes with some disadvantages compared to our approach: (i) for high-dimensional dynamics, coupling issues may arise and the scheme becomes computationally expensive and (ii) the deterioration of the robustness of the MLMC estimator because as $\Delta t$ decreases,   the kurtosis  explodes with   order $\Ordo{\Delta t_{\ell}^{-1}}$  compared to $\Ordo{\Delta t_{\ell}^{-1/2}}$ for Euler--Maruyama  without smoothing \cite{giles2015multilevel} and $\Ordo{1}$ for  our approach  (see Sections~\ref{sec:robustness_MLMC} and  \ref{sec:Numerical smoothing with MLMC}).
\end{remark}

\subsection{Robustness Analysis}\label{sec:robustness_MLMC}
 When approximating the expectation of  nonsmooth (non-Lipschitz) functionals, the standard MLMC  estimator (without smoothing) suffers from poor  robustness and performance owing to high kurtosis at deep levels (small $\Delta t_{\ell}$).  To explain this undesirable feature, we let $g$ denote a rdv and  $g_{\ell}$ denote the corresponding level $\ell$ numerical approximation. Further, we define $Y_{\ell}:= g_{\ell}-g_{\ell-1}$. The standard deviation of the sample variance for the rdv $Y_{\ell}$ is given by 
	\begin{equation}\label{eq:approx_var}
		\sigma_{\mathcal{S}^2(Y_{\ell})} =\frac{\text{Var}[Y_{\ell}]}{\sqrt{M_{\ell}}}  \sqrt{(\bar{\kappa}_{\ell}-1)+\frac{2}{M_{\ell}-1}},\quad L_0+1 \le \ell \le L,
	\end{equation}
	where $\bar{\kappa}_{\ell}$ is the kurtosis  at level $\ell$, given by
	\begin{equation}\label{eq: kurtosis_estimate}
	\bar{\kappa}_{\ell}=	\frac{\expt{\left(Y_{\ell}-\expt{Y_{\ell}}\right)^4}}{\left(\text{Var}\left[Y_{\ell}\right]\right)^2},\quad L_0+1 \le \ell \le L.
	\end{equation}
We recall that 	in the MLMC setting,  accurate estimates of $\bar{V}_{\ell}=\text{Var}[Y_{\ell}]$ are required because  the optimal number of samples per level, $M^{
		\ast}_{\ell} $, for the multilevel estimator is given by 
	\begin{equation*}\label{eq:optimal_number_samples}
		M^{
			\ast}_{\ell} \propto \sqrt{\bar{V}_{\ell} \bar{C}^{-1}_{\ell}  } \sum_{\ell=L_0}^L \sqrt{\bar{V}_{\ell} \bar{C}_{\ell} },\quad L_0+1 \le \ell \le L,
	\end{equation*}
	where  $\bar{C}_{\ell}$ is the cost per sample path per level.

From \eqref{eq:approx_var},  $\Ordo{\bar{\kappa}_{\ell}}$ samples are required to obtain a reasonable estimate of the variance  $\bar{V}_{\ell}$. Two possible consequences of the high kurtosis may occur, thus deteriorating the robustness and  performance of the MLMC estimator
\begin{itemize}
	\item The sample variance, $\bar{V}_{\ell}$, is underestimated (unreliable). Then, the required    confidence interval is not faithfully attained owing to $\sigma_{\mathcal{S}^2(Y_{\ell})}$ given by \eqref{eq:approx_var}.
	\item The sample variance, $\bar{V}_{\ell}$, is overestimated. In this case, too many sample paths are generated, and the algorithm takes substantially more time to run.
\end{itemize}
When using the Euler--Maruyama  scheme, the kurtosis at level $\ell$  for the MLMC method without numerical smoothing is on the order of $\Ordo{\Delta t_{\ell}^{-1/2}}$ \cite{giles2015multilevel}.   However, due to the numerical smoothing idea, the kurtosis at level $\ell$ for the proposed  approach is on the order of  $\Ordo{1}$, as indicated in Corollary~\ref{corrol: robustness of MLMC_smoothing} (see Section~\ref{sec:Numerical smoothing with MLMC} for more numerical illustrations of these behaviors).
\begin{corollary}[Bounded Kurtosis for MLMC with numerical smoothing]\label{corrol: robustness of MLMC_smoothing}
We let $\kappa_{\ell}$ be the kurtosis of the random variable  $Y_{\ell}:=\bar{I}_{\ell}-\bar{I}_{\ell-1}$ ($\bar{I}_{\ell} $ is defined in Section~\ref{sec:Details of our approach} using the Euler--Maruyama scheme). Then, under the assumptions of   Theorems~\ref{corrol: Lipschitz of the integrands} and \ref{corrol: Lipschitz of the integrands_density}, we obtain
		\begin{equation}\label{eq:kurtosis_estimates_order}
		\kappa_{\ell}=\Ordo{1}.
	\end{equation}
\end{corollary}
\begin{proof}\label{proof: kurtosis mlmc}
Using Theorem~\ref{corrol: Lipschitz of the integrands} and \ref{corrol: Lipschitz of the integrands_density}, we obtain 	$(\text{Var}[Y_{\ell}])^2=\Ordo{\Delta t_{\ell}^2}$.  Moreover, assuming  the global Lipschitz conditions for the drift and diffusion in Assumption \ref{ass: globally Lipschitz_drift_diffusion}, we obtain the $L_p$ moment estimate result  from \cite{bouleau1994numerical} (see also \cite{ben2015central}),  and that $\expt{\left(Y_{\ell}-\expt{Y_{\ell}}\right)^4}=\Ordo{\Delta t_{\ell}^2}$. Therefore, using   \eqref{eq: kurtosis_estimate}, we achieve the desired result  presented in \eqref{eq:kurtosis_estimates_order}.
\end{proof}
\begin{remark}
	We emphasize that some previous studies  \cite{giles2015multilevel,gou2016estimating,hammouda2017multilevel,ben2020importance} have  reported the problem of high kurtosis when using the MLMC estimator for different applications. In this work, we focus on probability computation and density estimation tasks where  high kurtosis is due to the low regularity of the functional. We illustrate how the numerical smoothing idea enables  overcoming this undesirable feature in the estimator.
	\end{remark}
\section{Numerical Experiments}\label{sec:Numerical smoothing with MLMC}
This section numerically illustrates  the advantages of combining the numerical smoothing idea with  MLMC when (i) computing  probability or  equivalently the price of a digital option (see Section~\ref{sec: MLMC for digital options}) and (ii)  approximating the density  of stochastic (assets) dynamics (see Section~\ref{sec: numerical MLMC for approximating densities and greeks}). We perform  tests for  Examples~\ref{exp: GBM} and \ref{exp: Heston}
 \begin{example}[The GBM discretized model]\label{exp: GBM}
Under this model, the  dynamics  are given by
		\begin{equation}\label{eq:dynamics_GBM}
		dX_t=\mu X_t dt+ \sigma X_t dW_t,
	\end{equation}
where  $\sigma$  indicates the  volatility;  $\mu$ denotes the drift and $W_{t}$ represents  a  Wiener process.
 \end{example}

\begin{example}[The Heston model    \cite{heston1993closed,broadie2006exact,kahl2006fast,andersen2007efficient}]\label{exp: Heston}
Under this model,	the  dynamics are given by
	\begin{align}\label{eq:dynamics Heston}
		dX_t&=\mu X_t dt+ \rho\sqrt{v_t}X_t dW_t^v+ \sqrt{1-\rho^2} \sqrt{v_t}X_t dW_t \nonumber\\
		dv_t&=\zeta (\theta-v_t)dt+\xi \sqrt{v_t} dW_t^v\COMMA
\end{align}
where $v_t$ represents the instantaneous variance; $\left(W_{t}^{S},W_{t}^{v}\right)$ are the correlated Wiener processes with correlation $\rho$; $\mu$  represents the asset's rate of return; $\theta$ denotes  the mean  variance; $\zeta$ indicates the rate at which $v_t$ reverts to $\theta$; and $\xi$ denotes the volatility of the volatility.
 \end{example}
We use the  Euler--Maruyama scheme and  a higher-order scheme (\ie,  the Milstein scheme)  to simulate  the GBM dynamics. To  simulate the Heston model, we use   the full truncation (FT) scheme \cite{lord2010comparison}, combined with the  Euler--Maruyama. In the  examples, we compare  (i) the standard MLMC estimator (without smoothing) and (ii) the proposed MLMC estimator combined with numerical smoothing (as explained in Sections~\ref{sec:General setting} and \ref{sec:Details of our approach}).  In    Figures~\ref{fig:euler_digital_GBM}, \ref{fig:milstein_digital_GBM}, \ref{fig:euler_digital_heston}, \ref{fig:density_GBM} and \ref{fig:density_heston}, $P_{\ell}$   denotes the numerical approximation of the quantity of interest at level $\ell$ of the MLMC estimator. In particular, $P_{\ell}= \bar{I}_{\ell}$ when using numerical smoothing. Moroever, in this section  we denote by   $\kappa_{L}$  the kurtosis at the finest level, $L$, and  by $(\alpha,\beta,\gamma)$  the  numerical estimates of  weak, variance decay, and work rates of MLMC, respectively. In addition, $\text{TOL}$ is the user-selected    tolerance. The  experiments were produced using  MATLAB (v. R2022a)  on an $8$-Core Intel Xeon W architecture.
\subsection{Pricing Digital Options/Computing Probability}\label{sec: MLMC for digital options}
We aim to approximate  the price of  digital options (equivalently a probability), expressed by
\begin{align}\label{eq:digital_option_price}
\expt{g(X(T))}=\expt{\mathbf{1}_{X(T)>K}},
\end{align}
where $X(T)$ is the asset price at the maturity $T$ and $K$ is the strike price.
\subsubsection{Pricing Digital Option/Computing Probability  under the GBM Model}\label{sec:Digital options under the GBM model}
We consider the  GBM model (Example \ref{exp: GBM}),  with parameters $X_0=K=100$, $T=1$, and   $\sigma=0.2$.  Table~\ref{table:Summary of our numerical results digital GBM.}  summarizes  the  results for approximating the probability/digital option price in  \eqref{eq:digital_option_price}. The reference value in this case is $0.460172$.
\begin{small}
	\begin{table}[h!]
		\centering
		\begin{tabular}{l*{4}{c}r}
			\toprule[1.5pt]
			Method      &   $\kappa_{L}$  & $\alpha$  & $\beta$  &  $\gamma$   & Numerical complexity \\
			\hline
			MLMC without smoothing (Euler--Maruyama) & $709$& $1$ & $1/2$ & $1$&  $\Ordo{\text{TOL}^{-2.5}}$ \\	
			\hline
			MLMC with numerical smoothing   (Euler--Maruyama)   & $3$& $1$ & $1$ & $1$&  $\Ordo{\text{TOL}^{-2} \left(\log(\text{TOL})\right)^2}$ \\
			\hline
			MLMC without smoothing (Milstein ) & $  116009$& $1$ & $1$ & $1$&  $\Ordo{\text{TOL}^{-2} \left(\log(\text{TOL})\right)^2}$ \\	
			\hline
			MLMC with numerical smoothing   (Milstein)   & $3$& $1$ & $2$ & $1$&   $\Ordo{\text{TOL}^{-2}}$ \\ 
			\bottomrule[1.25pt]
		\end{tabular}
		\caption{Digital option under the GBM model: Summary of the MLMC  results, which correspond to Figures \ref{fig:euler_digital_GBM},   \ref{fig:milstein_digital_GBM}, and \ref{fig:euler_digital_complexity} respectively.}
		\label{table:Summary of our numerical results digital GBM.}
	\end{table}
\end{small}

\begin{figure}[h!]
\centering
	\begin{subfigure}{0.45\textwidth}
\includegraphics[width=1\linewidth]{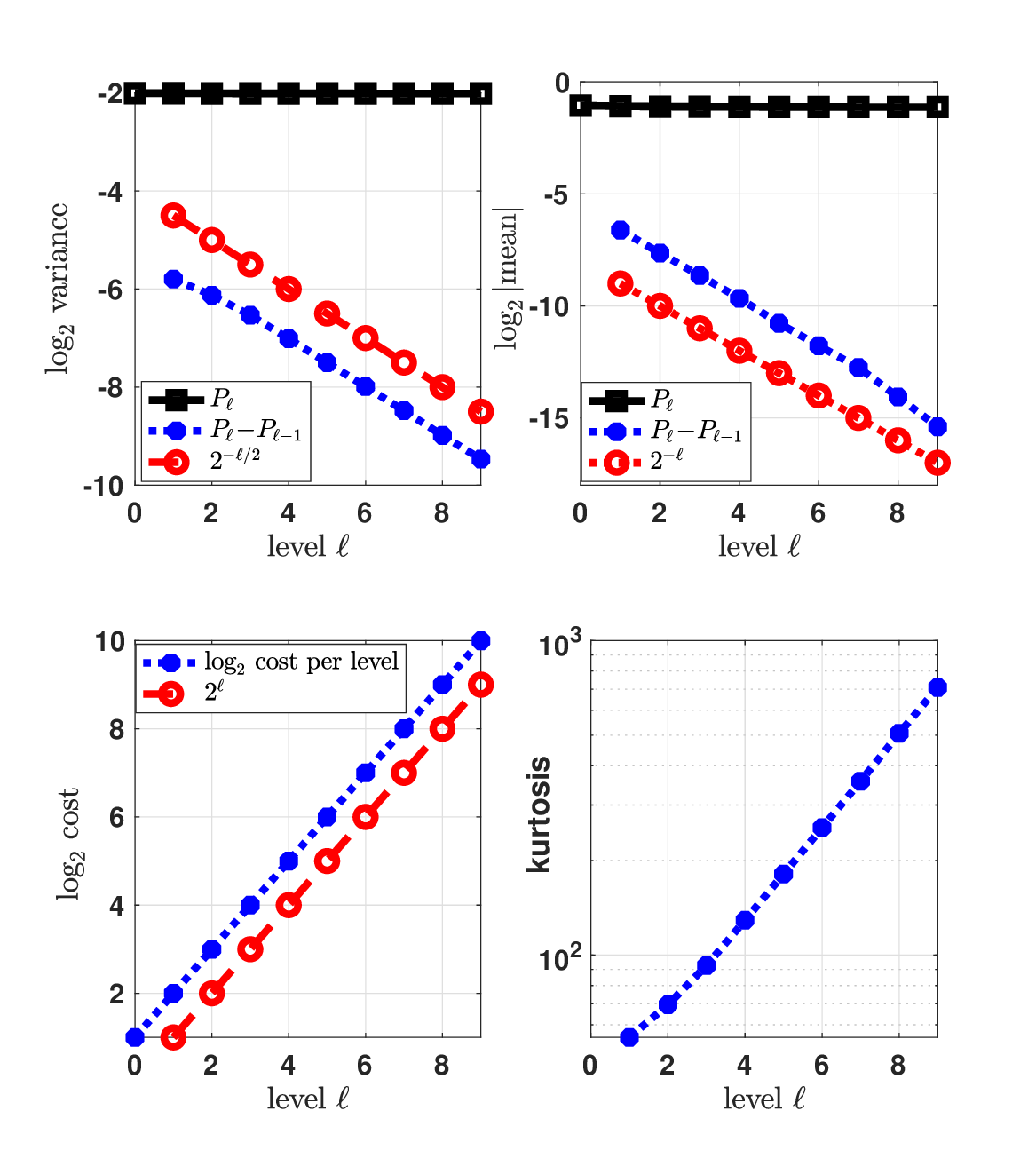}
\vspace{0.05mm}
	\caption{Without  smoothing}
	\label{fig:euler_digital_without_smoothing}
\end{subfigure}%
	\begin{subfigure}{0.45\textwidth}
		\includegraphics[width=1\linewidth]{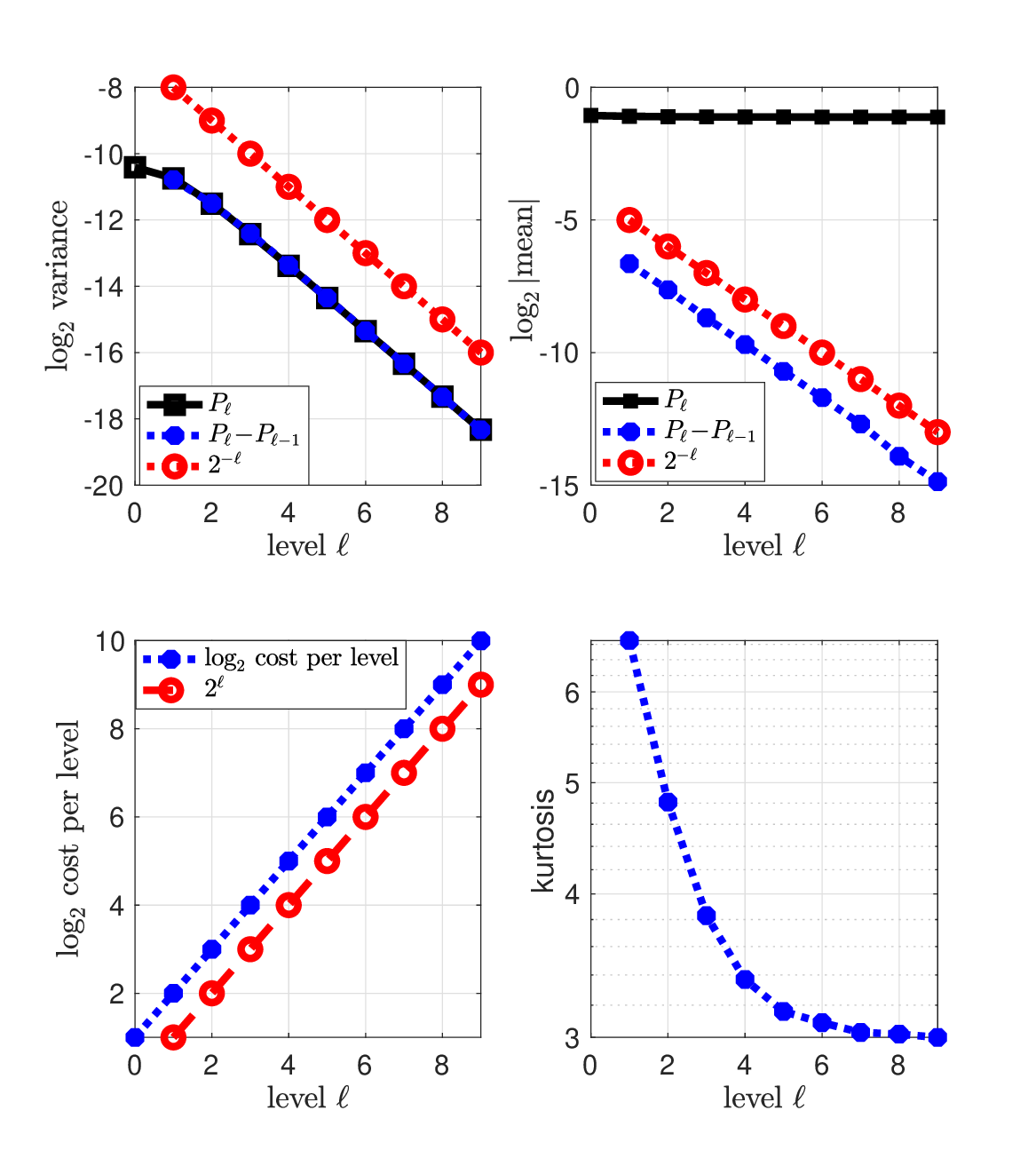}
			\caption{With numerical  smoothing $(\text{TOL}_{\text{Newton},\ell}=10^{-4}, M_{\text{Lag}, \ell}=8)$}
			\label{fig:euler_digital_with_smoothing_non_analytic}
	\end{subfigure}%
\caption{Probability/Digital option under GBM: Convergence plots for MLMC  combined with
 the Euler--Maruyama scheme. The bottom left plot corresponds to a scaled expected cost per level, \ie,  $\Ordo{2^{\gamma \ell}}$ (without tracking the constant), with $\gamma$ being the work growth rate.}
\label{fig:euler_digital_GBM}
\end{figure}

\begin{figure}[h!]
	\centering
	\begin{subfigure}{0.45\textwidth}
		\includegraphics[width=1\linewidth]{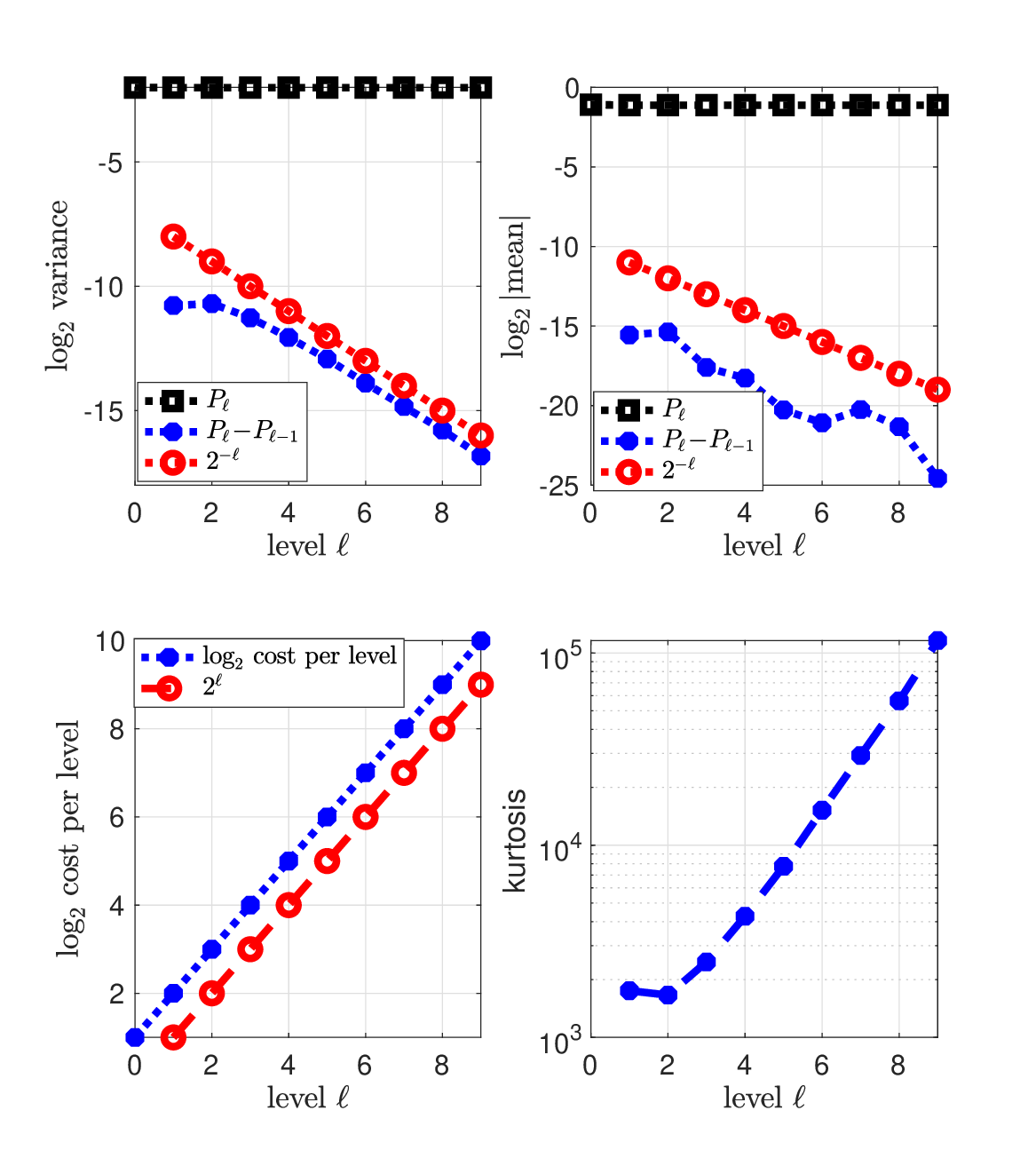}
		\vspace{0.05mm}
		\caption{Without  smoothing}
		\label{fig:milstein_digital_without_smoothing}
	\end{subfigure}%
	\begin{subfigure}{0.45\textwidth}
		\includegraphics[width=1\linewidth]{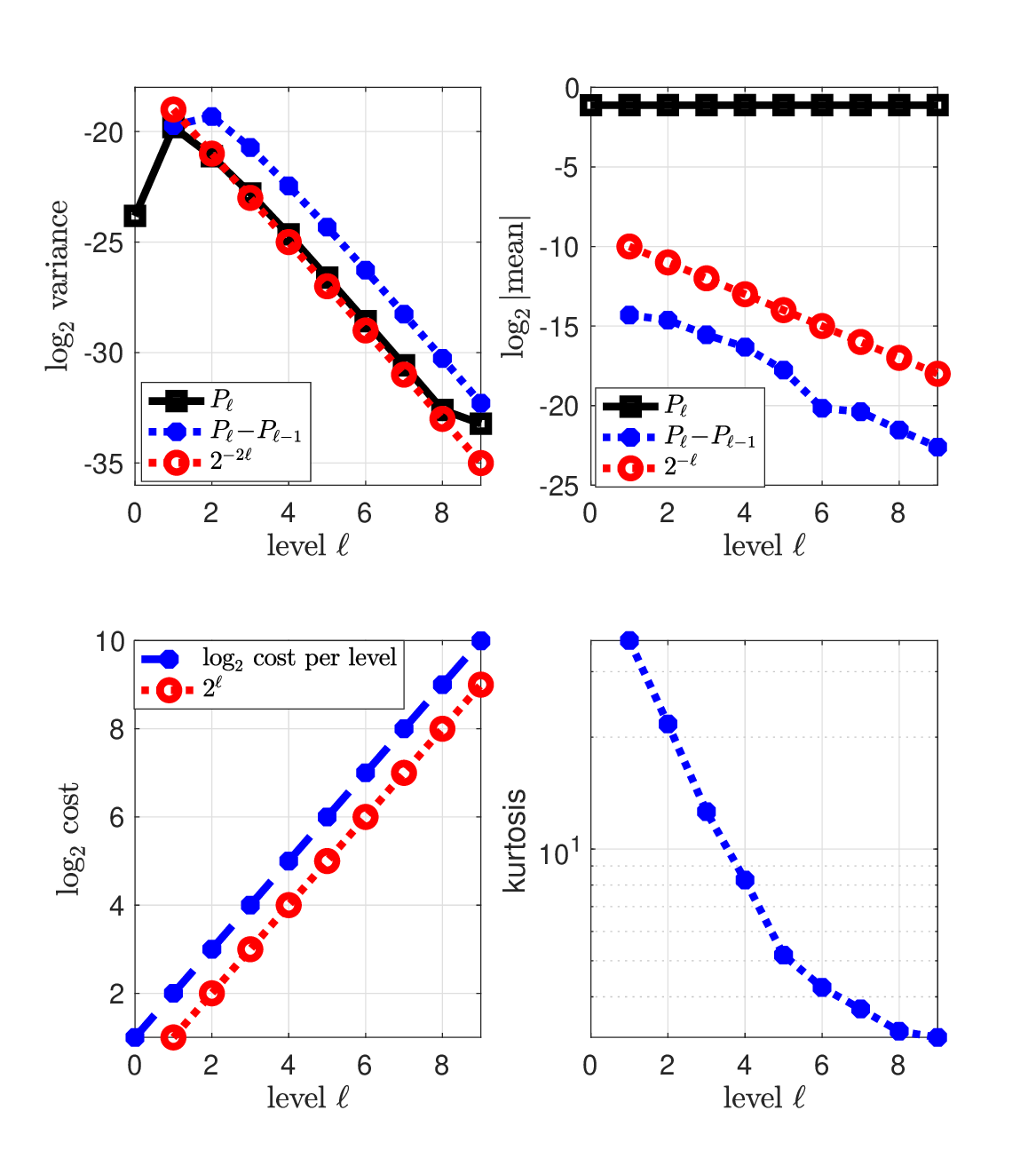}
		\caption{With numerical  smoothing $(\text{TOL}_{\text{Newton},\ell}=10^{-4}, M_{\text{Lag}, \ell}=8)$}
		\label{fig:milstein_digital_with_smoothing_non_analytic}
	\end{subfigure}%
	\caption{Probability/Digital option under GBM: Convergence plots for MLMC  combined with the Milstein  scheme.}
	\label{fig:milstein_digital_GBM}
\end{figure}

\begin{figure}[h!]
	\centering
	\includegraphics[width=0.53\linewidth]{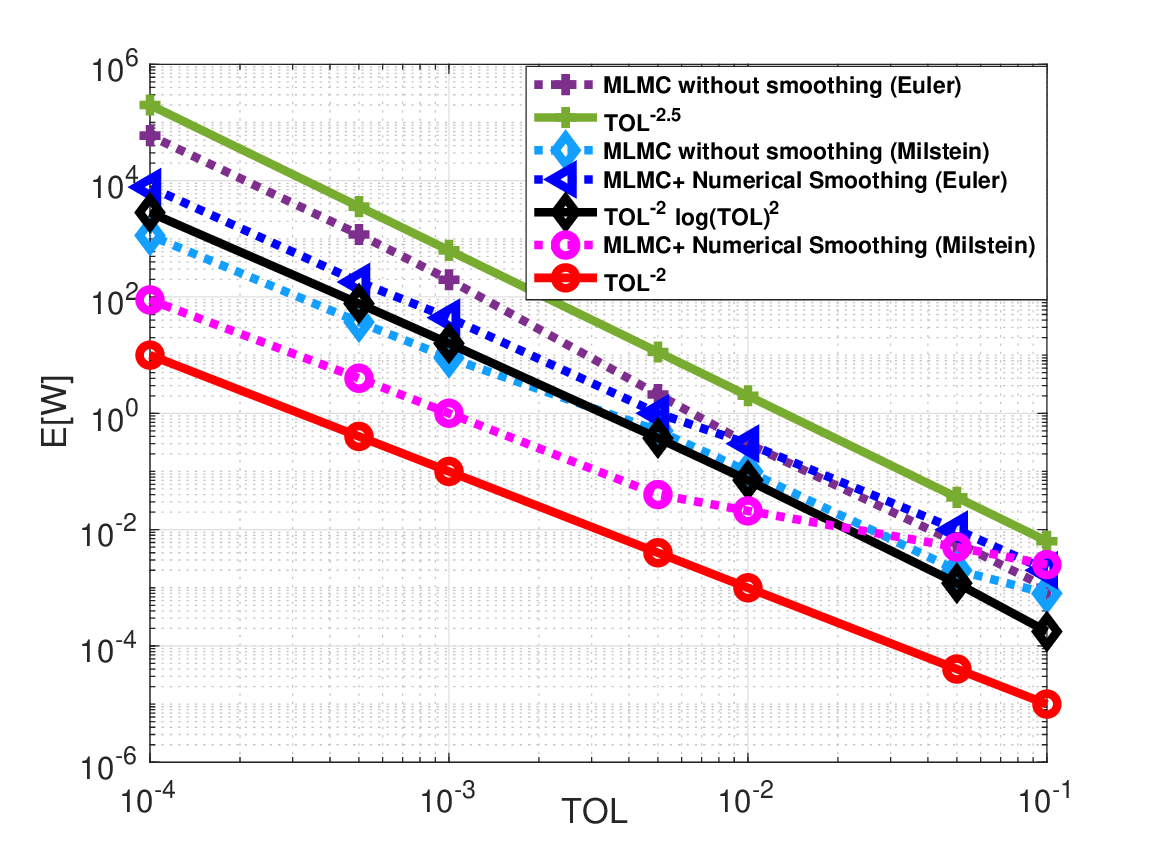}
	\caption{Digital option under the GBM model: Comparison of the numerical complexity (expected work  (in seconds), $\expt{W}$, vs tolerance, $\text{TOL}$, in a log--log scale)
		of  standard MLMC, and MLMC with numerical smoothing, combined with the Euler or Milstein schemes. When both  Euler and Milstein schemes, MLMC combined with numerical smoothing outperforms  standard MLMC, and   achieves a better numerical complexity rate. The  canonical MLMC complexity (\ie,  $\Ordo{\text{TOL}^{-2}}$) is obtained when using the Milstein scheme with our approach.}
	\label{fig:euler_digital_complexity}
\end{figure}

More details are illustrated in Figures~\ref{fig:euler_digital_GBM}, \ref{fig:milstein_digital_GBM}, and \ref{fig:euler_digital_complexity}. From  these figures and Table  \ref{table:Summary of our numerical results digital GBM.}, we obtain the following results:
\begin{enumerate}
\item The kurtosis is substantially reduced at  the finest level, $\kappa_{L}$, of the  MLMC algorithm  using  numerical smoothing for both Euler--Maruyama and Milstein schemes. The kurtosis becomes bounded and is reduced by a factor of $236$  for Euler--Maruyama (compare  the bottom right plots presented  in  Figures \ref{fig:euler_digital_without_smoothing} and \ref{fig:euler_digital_with_smoothing_non_analytic}), and more significantly by a factor of $ 38670$ for  the Milstein scheme (compare  the bottom right plots presented  in  Figures \ref{fig:milstein_digital_without_smoothing} and \ref{fig:milstein_digital_with_smoothing_non_analytic}). We emphasize that this is a crucial improvement regarding  the robustness and performance of the MLMC estimator, as explained in Section~\ref{sec:robustness_MLMC}.
\item  The numerical smoothing   considerably reduces the variance of the   coupled levels in MLMC and  improves the  variance decay rate,   from $\beta=1/2$ to $\beta=1$ for Euler-Maruyama (compare the  top left plots in Figures \ref{fig:euler_digital_without_smoothing} and \ref{fig:euler_digital_with_smoothing_non_analytic}), and to $\beta=2$ for  the Milstein scheme (compare the  top left plots in Figures \ref{fig:milstein_digital_without_smoothing} and \ref{fig:milstein_digital_with_smoothing_non_analytic}). This improvement results  in a reduction in  the order of MLMC numerical complexity from $\Ordo{\text{TOL}^{-2.5}}$ to $\Ordo{\text{TOL}^{-2} \left(\log(\text{TOL})\right)^2}$ for Euler--Maruyama  and  to the canonical complexity, \ie, $\Ordo{\text{TOL}^{-2}}$ for  the Milstein scheme (see Figure \ref{fig:euler_digital_complexity}).   Figure \ref{fig:euler_digital_complexity} indicates that  MLMC combined with numerical smoothing considerably outperforms standard MLMC  in computational work, especially for small tolerances.
\item For the proposed  MLMC estimator combined with numerical smoothing,  the variance of the level $0$ estimator is very small. The  numerical smoothing  can be seen as applying a  conditional expectation  w.r.t~ the terminal value. There is no path simulation at level $\ell= 0$, where there would usually be one timestep. Similar behavior was observed in  \cite{giles2008improved}.
\end{enumerate}
\begin{remark}\label{rem: MC_smoothing_GBM}
	Notably,  for the particular case of the GBM dynamics,  a decaying variance of $P_{\ell}$ in the top left plots presented in Figures \ref{fig:euler_digital_with_smoothing_non_analytic} and  \ref{fig:milstein_digital_with_smoothing_non_analytic} is expected  because we use a Brownian bridge for path construction. Additionally,  the integrand only depends on the terminal value of the Brownian bridge, which has a variance scale of the order $\Delta t$. Therefore, for this particular case, we expect the numerical complexity of the MC method with smoothing to be on the order of $\Ordo{\text{TOL}^{-2}}$. This feature does not hold anymore for the Heston model, as demonstrated later.
\end{remark}
\subsubsection{Pricing Digital Option/Computing Probability under  the Heston Model}\label{sec:Digital option under the Heston model}
\begin{figure}[h!]
	\centering
	\begin{subfigure}{0.45\textwidth}
		\includegraphics[width=1\linewidth]{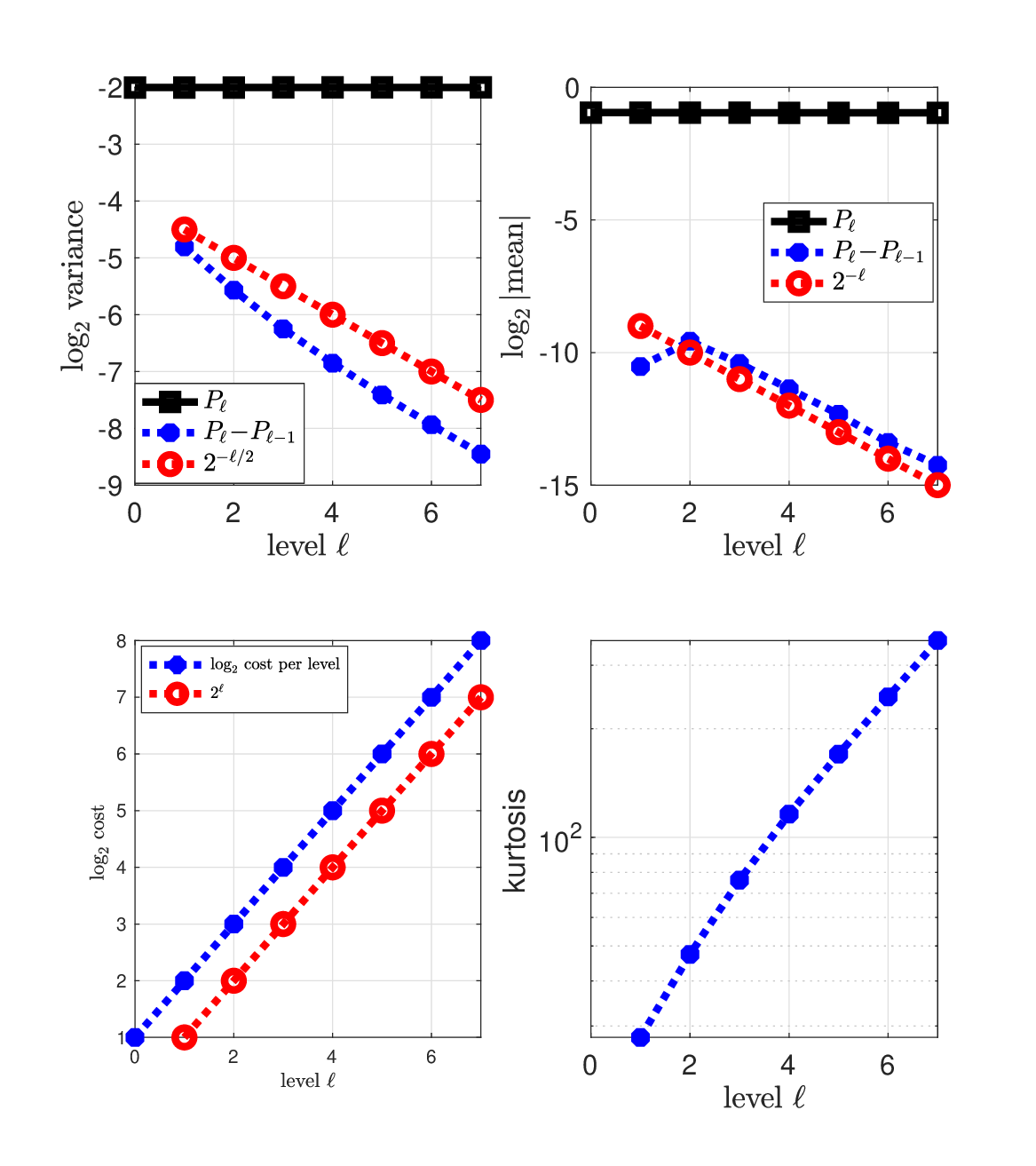}
		\vspace{0.05mm}
		\caption{Without  smoothing.}
		\label{fig: digital_Heston_without_smoothing}
	\end{subfigure}%
	\begin{subfigure}{0.45\textwidth}
		\includegraphics[width=1\linewidth]{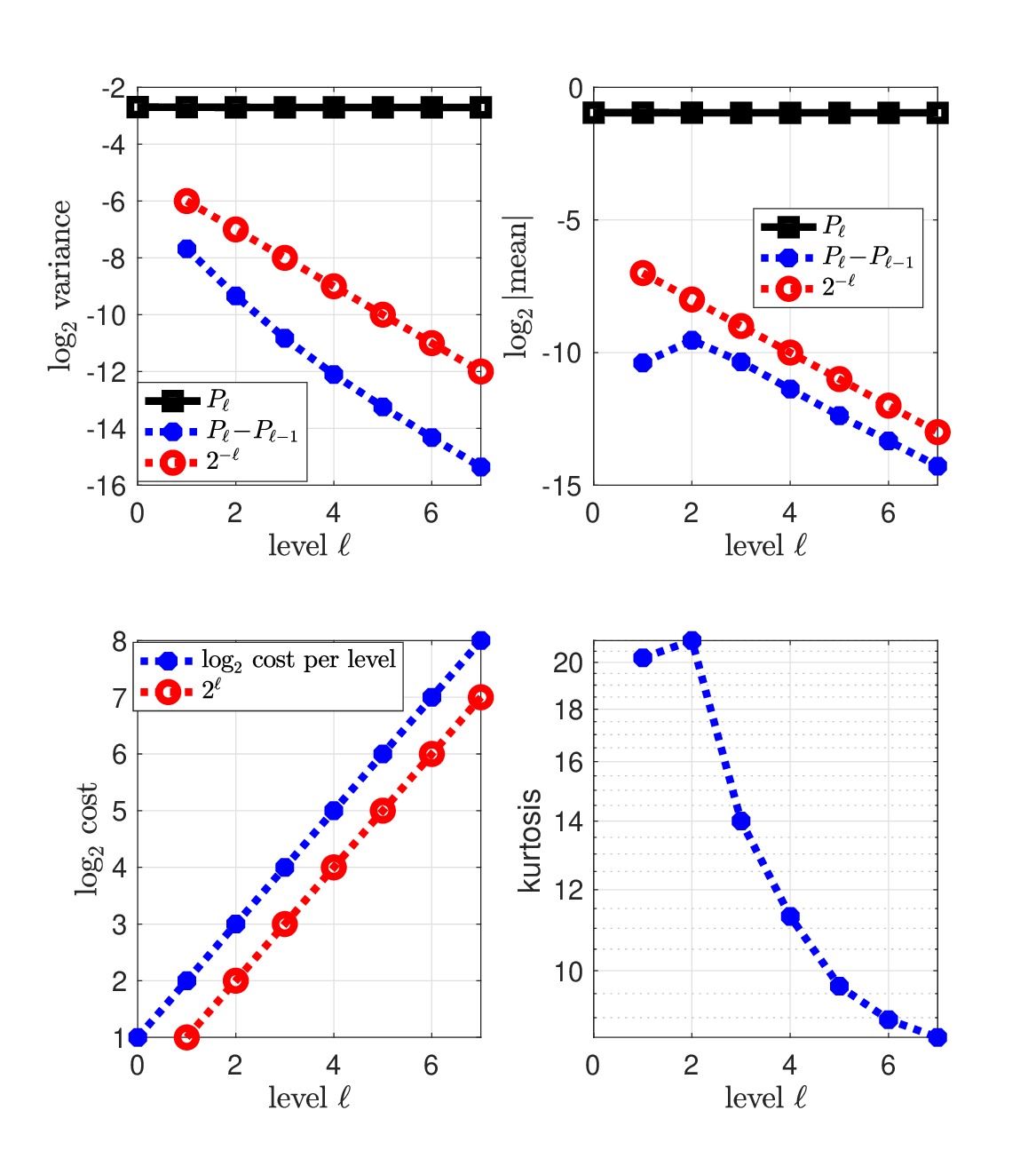}
		\caption{With numerical  smoothing $(\text{TOL}_{\text{Newton},\ell}=10^{-3}, M_{\text{Lag}, \ell}=32)$}
		\label{fig:digital_Heston_with_smoothing_FT_non_analytic}
	\end{subfigure}%
	\caption{Probability/Digital option under Heston: Convergence plots for MLMC  combined with
		the Euler--Maruyama scheme and FT.}
	\label{fig:euler_digital_heston}
\end{figure}

\begin{figure}[h!]
	\centering 
	\includegraphics[width=0.53\linewidth]{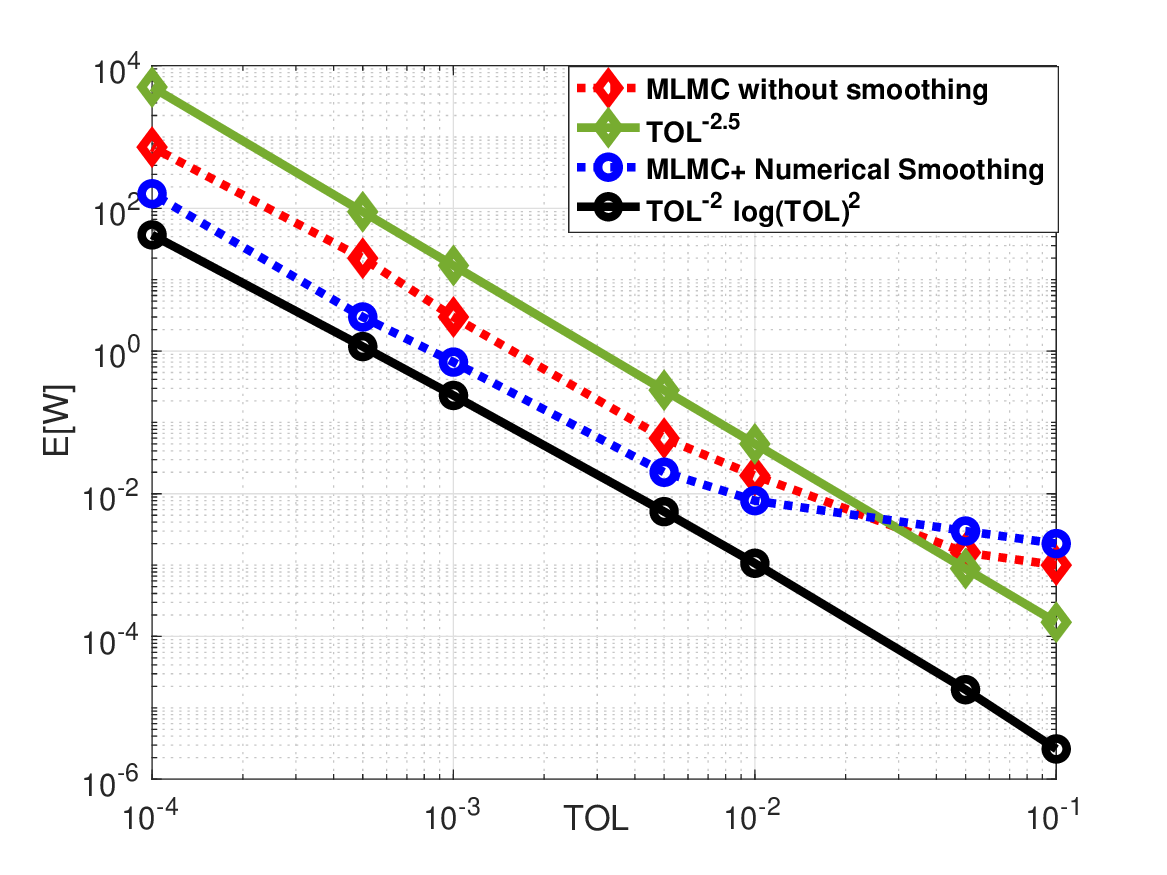}
	\caption{Digital option/probability under the Heston model: Comparison of the numerical complexity (expected work  (in seconds), $\expt{W}$, vs tolerance, $\text{TOL}$, in a log--log scale)
		of the   standard MLMC and   MLMC with  numerical smoothing.  MLMC combined with numerical smoothing  outperforms  standard MLMC and    achieves a better numerical complexity rate.} 
	\label{fig:Heston_digital_complexity}
\end{figure}

We consider the Heston model \eqref{eq:dynamics Heston},  with the parameters: $T=1$, $X_0=K=100$, $v_0=0.04$, $\mu=0$,  $\rho=-0.9$, $\zeta=1$, $\xi=0.1$,  and $\theta=0.0025$ (these parameters  do not satisfy the Feller condition, \ie, $\xi^2< 4 \zeta \theta$).    A reference solution, equal to $0.5146$,  was    obtained  by  the MC method. 
 Table \ref{table:Summary of our numerical results digital Heston.} summarizes  the  results for approximating the  probability/digital option price defined by  \eqref{eq:digital_option_price} using Euler--Maruyama scheme. Figures \ref{fig:euler_digital_heston} and \ref{fig:Heston_digital_complexity}  present more  details. From  these figures and Table \ref{table:Summary of our numerical results digital Heston.}, we obtain the following results:
\begin{enumerate}
\item The  kurtosis  substantially reduces at the finest level, $\kappa_{L}$, of MLMC when using numerical smoothing. The kurtosis is bounded and reduced by a factor of  $>27$.
\item Numerical smoothing  considerably  reduces  the variance of coupled levels in MLMC. Further, it improves the variance decay rate  from $\beta=1/2$ to $\beta=1$, implying  an improvement in the  MLMC numerical complexity from $\Ordo{\text{TOL}^{-2.5}}$ to $\Ordo{\text{TOL}^{-2} \left(\log(\text{TOL})\right)^2}$.
\end{enumerate}
	\begin{small}
	\begin{table}[!h]
		\centering
		\begin{tabular}{l*{4}{c}r}
			\toprule[1.5pt]
			Method      &     $\kappa_{L}$ & $\alpha$   &  $\beta$  &  $\gamma$   & Numerical complexity \\
			\hline
			MLMC without smoothing  (FT Euler--Maruyama) & $350$ & $1$  &  $1/2$&  $1$&  $\Ordo{\text{TOL}^{-2.5}}$\\	
			
			\hline
			MLMC with numerical smoothing (FT Euler--Maruyama)  & $9$ & $1$  &  $1$&  $1$ &  $\Ordo{\text{TOL}^{-2} \log(\text{TOL})^2}$\\ 
			
			\bottomrule[1.25pt]
		\end{tabular}
		\caption{Digital option/probability  under the Heston model: Summary of the MLMC numerical results, which correspond to Figures  \ref{fig:euler_digital_heston} and \ref{fig:Heston_digital_complexity}.}
		\label{table:Summary of our numerical results digital Heston.}
	\end{table}
\end{small}
\begin{remark}[On Milstein scheme for the Heston model]
In this work, we present the results of using the Milstein scheme with MLMC  for a scalar SDE in the context of the GBM example. However, when applied to multidimensional problems such as the Heston model, the Milstein scheme requires the simulation of expensive iterated It\^o integrals, known as the L\'evy areas. In future work, we plan to explore the potential of combining our numerical smoothing idea with the antithetic MLMC estimator proposed in \cite{giles2014antithetic}  to address this issue and improve overall performance.
\end{remark}
\subsection{Density Approximation}\label{sec: numerical MLMC for approximating densities and greeks}
\subsubsection{Density Approximation under the GBM Model}\label{sec:Approximating density under the GBM model}
We compute the density $\rho_{X(T)}$, given by  \eqref{eq:density_function},   at $u=1$ for the GBM  example with the parameters: $X_0=1$, $T=1$, and $\sigma=0.2$. In this case,  $X(T)$ is lognormally distributed with parameters $-\sigma^2/2$ and $\sigma$.  Table \ref{table:Summary of our numerical results density GBM.} summarizes   the  results of MLMC combined with numerical smoothing.  
\begin{small}
	\begin{table}[!h]
		\centering
		\begin{tabular}{l*{4}{c}r}
			\toprule[1.5pt]
			Method      &   $\kappa_{L}$ & $\alpha$   &  $\beta$  &  $\gamma$   & Numerical complexity \\
			\hline
			MLMC combined with numerical smoothing (Euler)  & $3$ & $1$  &  $1$&  $1$&  $\Ordo{\text{TOL}^{-2} \left(\log(\text{TOL})\right)^2} $\\
			\hline
			MLMC combined with numerical smoothing (Milstein)  & $3$ & $1$  &  $2$&  $1$&  $\Ordo{\text{TOL}^{-2} } $\\
			\bottomrule[1.25pt]
		\end{tabular}
		\caption{Density of GBM: Summary of the MLMC  results  for  computing the density $\rho_{X(T)}$ at  $u=1$,  where $X$ follows the GBM dynamics. These results correspond to Figures \ref{fig:euler_density_MLMC_with_smoothing} and  \ref{fig:euler_density_complexity}.}
		\label{table:Summary of our numerical results density GBM.}
	\end{table}
\end{small}

\begin{figure}[h!]
	\centering
	\begin{subfigure}{0.45\textwidth}
		\includegraphics[width=1\linewidth]{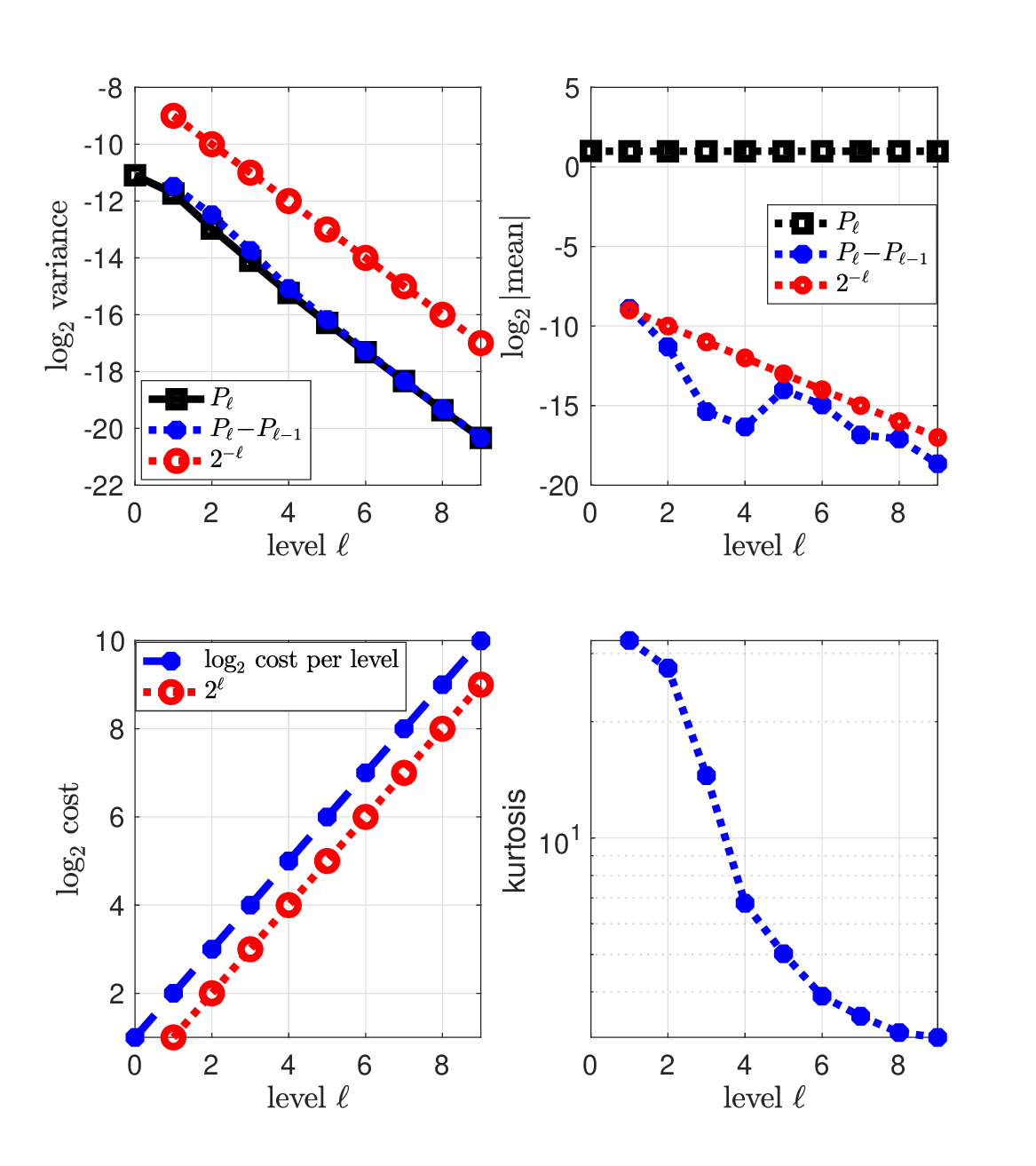}
		\caption{Euler--Maruyama}
		\label{fig:euler_density_MLMC_with_smoothing}
	\end{subfigure}%
	\begin{subfigure}{0.45\textwidth}
		\includegraphics[width=1\linewidth]{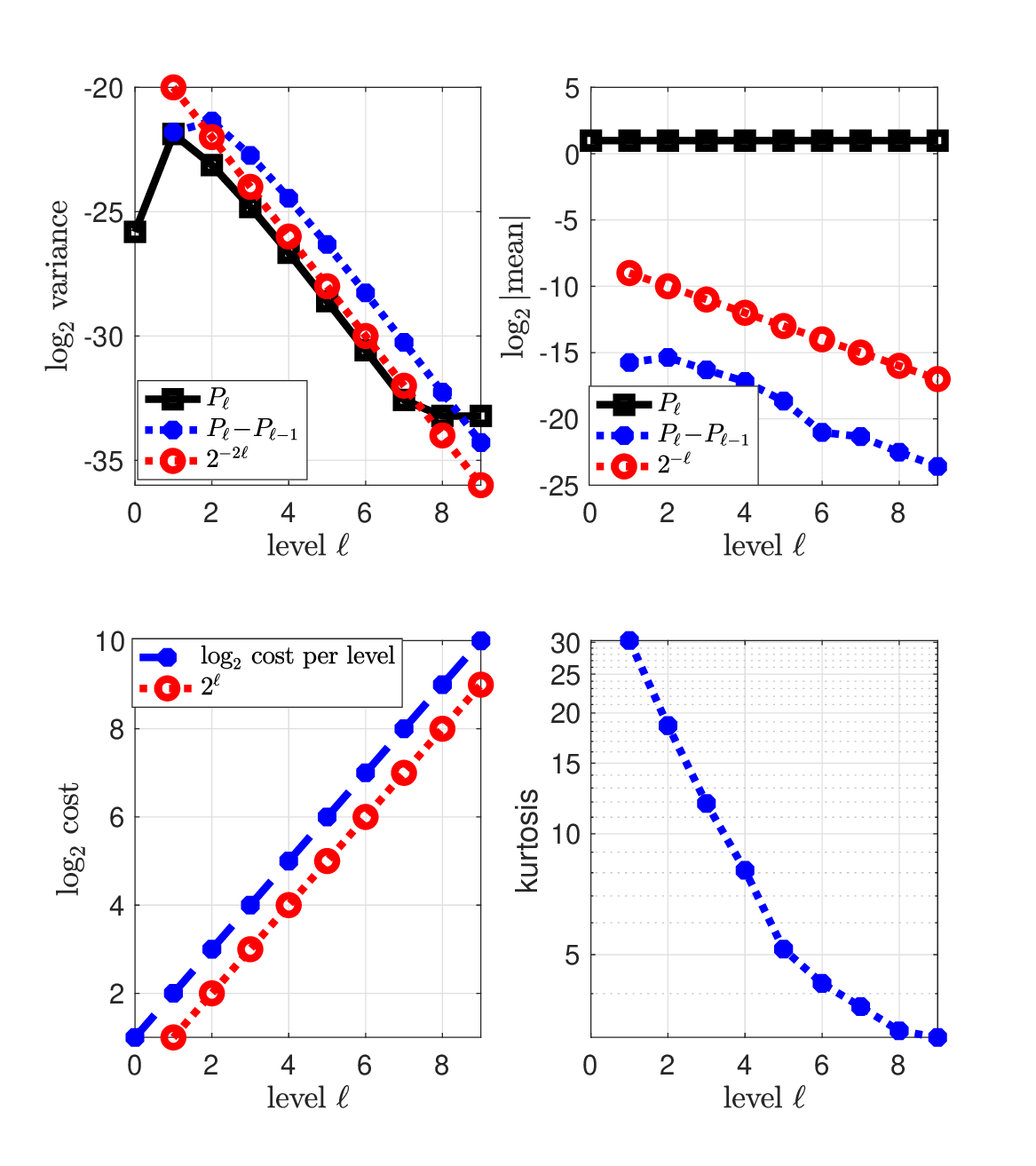}
		\caption{Milstein}
		\label{fig:Milstein_density_MLMC_with_smoothing}
	\end{subfigure}%
	\caption{Density of GBM: Convergence plots for MLMC with numerical smoothing  $(\text{TOL}_{\text{Newton},\ell}=10^{-4})$ for computing the density $\rho_{X(T)}$ at $u=1$, where $X$ follows the GBM dynamics. Remark \ref{rem: MC_smoothing_GBM}  also holds in this  example.}
	\label{fig:density_GBM}
\end{figure}

\begin{figure}[h!]
\centering
\includegraphics[width=0.53\linewidth]{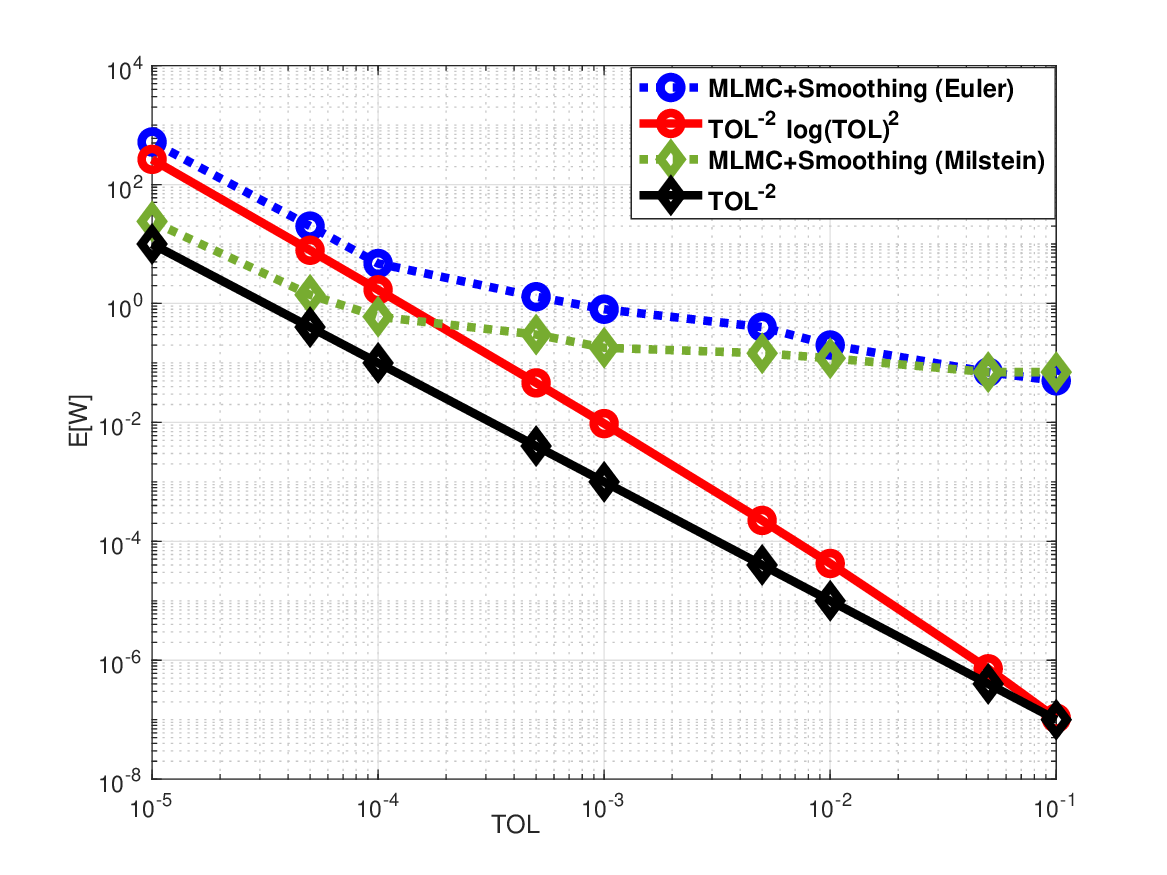}
\caption{Density of GBM: Numerical complexity  of MLMC with numerical smoothing for computing  the density $\rho_{X(T)}$ at $u=1$, where $X$ follows  the GBM dynamics.  The  canonical MLMC complexity (\ie,  $\Ordo{\text{TOL}^{-2}}$) is obtained when using the Milstein scheme with our approach.}
\label{fig:euler_density_complexity}
\end{figure}

Figure \ref{fig:euler_density_MLMC_with_smoothing} depicts  the detailed convergence, where we  verify that the kurtosis is bounded and that the variance decay rate is on order $1$ for the  Euler--Maruyama and $2$ for the Milstein scheme,  resulting in a numerical complexity of the MLMC estimator to be on the order  of  $\Ordo{\text{TOL}^{-2} \left(\log(\text{TOL})\right)^2} $ for Euler--Maruyama and $\Ordo{\text{TOL}^{-2}} $ for Milstein scheme, as confirmed in Figure \ref{fig:euler_density_complexity}.

\subsubsection{Asset Price and Joint Densities Approximation  under the Heston Model}\label{sec:Approximating density under the Heston model}
\begin{figure}[h!]
	\centering
	\begin{subfigure}{0.45\textwidth}
		\includegraphics[width=1\linewidth]{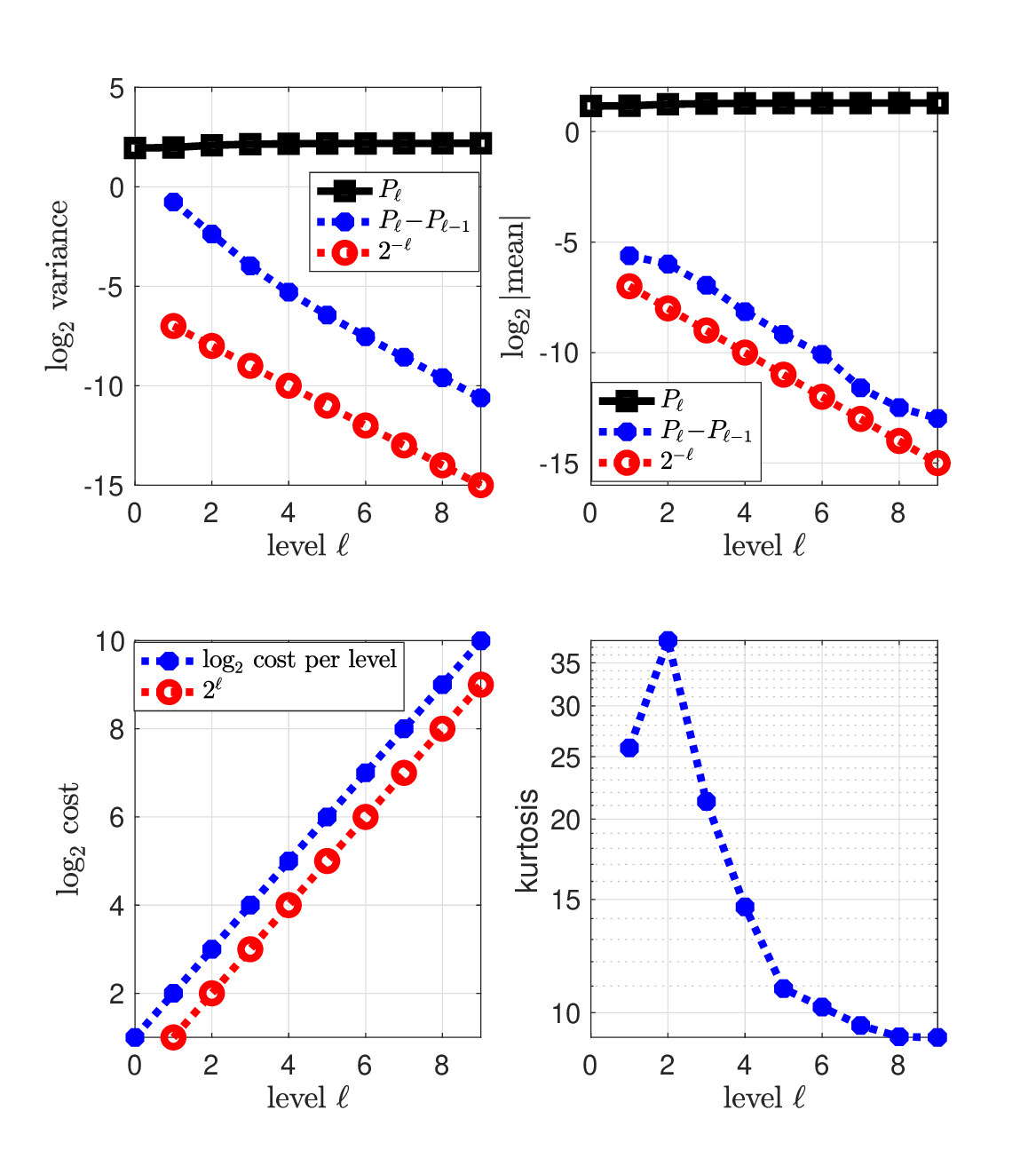}
		\caption{Asset price density}
		\label{fig:Heston_density_MLMC_with_smoothing_FT_euler}
	\end{subfigure}%
	\begin{subfigure}{0.45\textwidth}
		\includegraphics[width=1\linewidth]{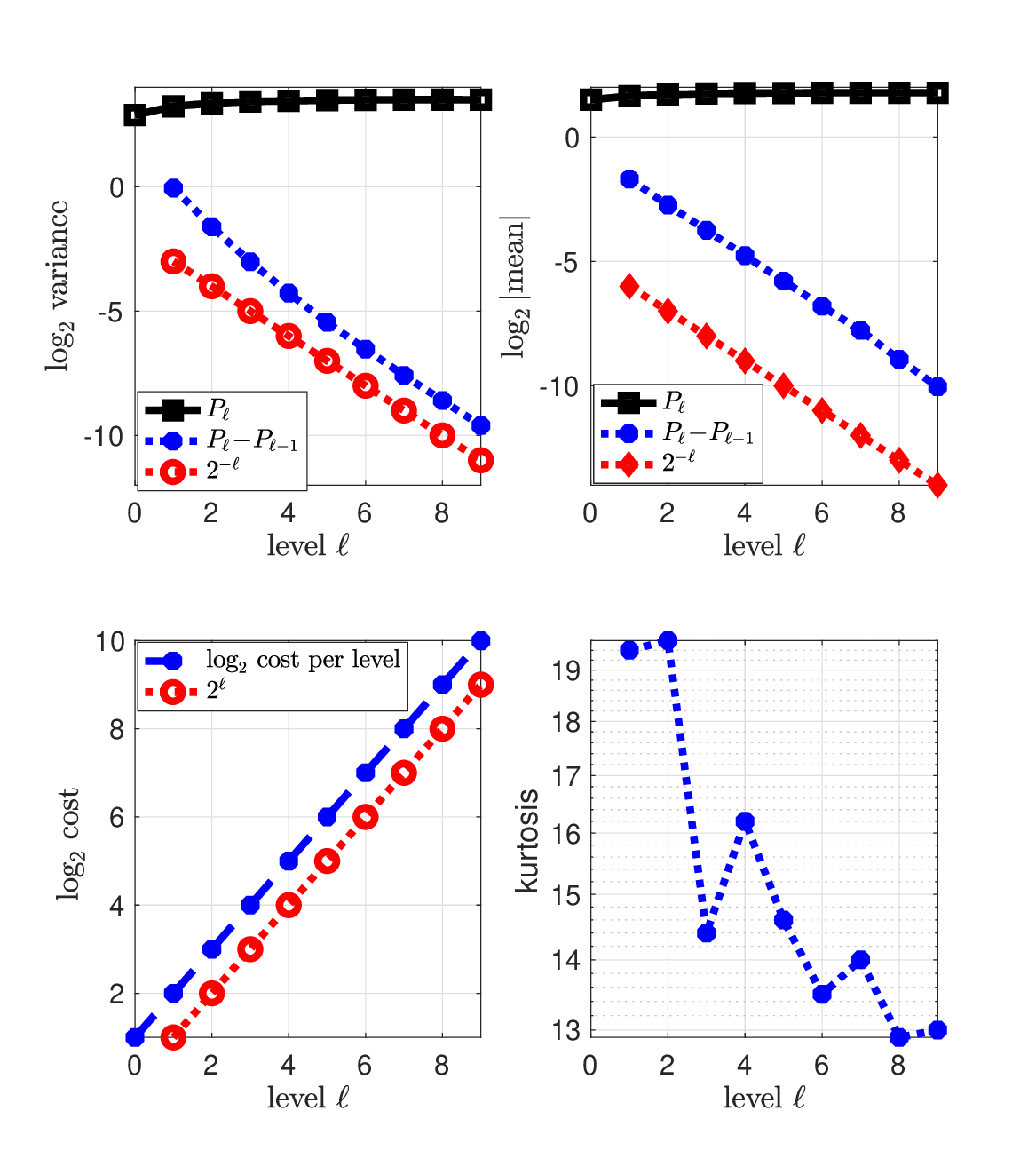}
		\caption{Joint density}
		\label{fig:joint_density_heston}
	\end{subfigure}%
	\caption{Density of Heston: Convergence plots for MLMC with numerical smoothing $(\text{TOL}_{\text{Newton},\ell}=10^{-2})$  combined with  the  FT   scheme, for computing the asset price  density $\rho_{X(T)}$ at $u=1$ and the  joint density $\rho_{X(T), v(T)}$ at $u=1$ and $v=0.04$.}
	\label{fig:density_heston}
\end{figure}

\begin{figure}[h!]
\centering
\includegraphics[width=0.53\linewidth]{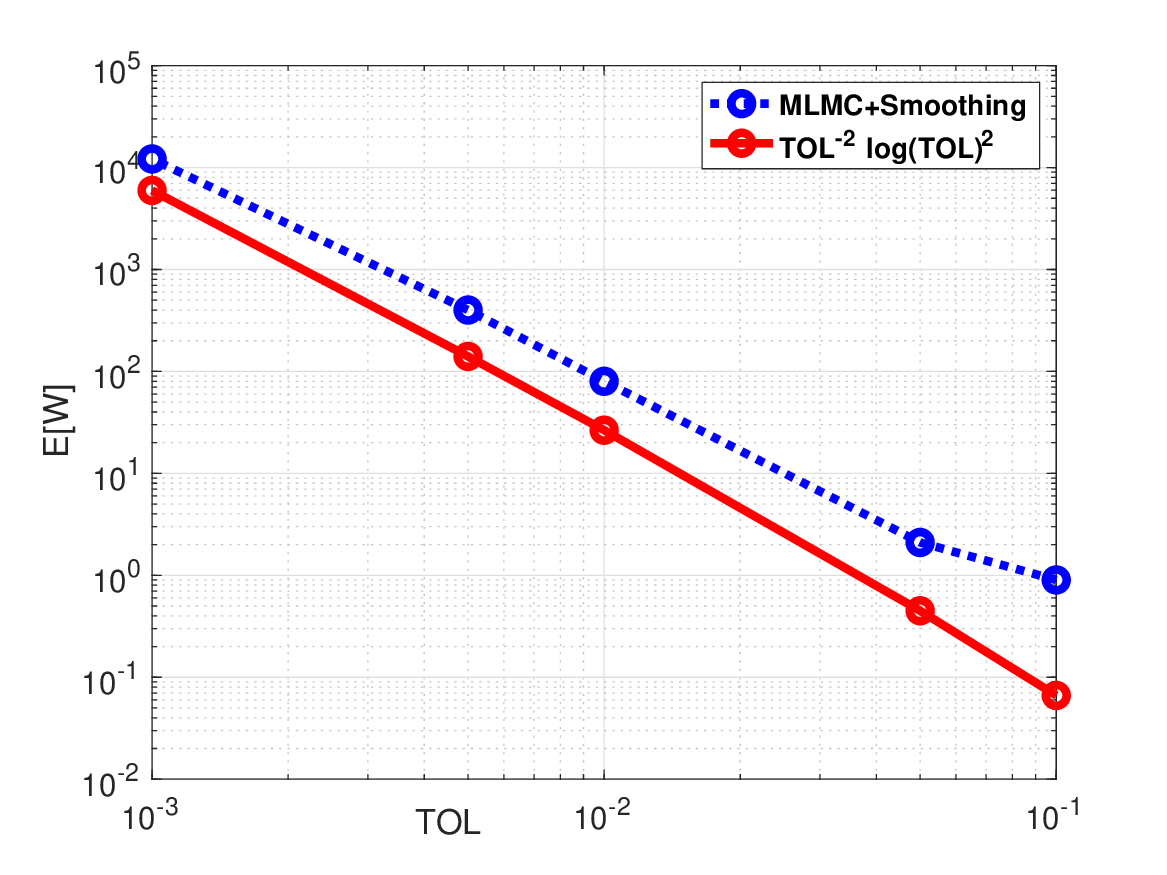}
\caption{Asset price density of Heston: Numerical complexity (expected work, $\expt{W}$  (in seconds), vs tolerance, $\text{TOL}$)
of MLMC with numerical smoothing   for computing the density $\rho_{X(T)}$ at $u=1$, where $X$ follows the Heston dynamics.}
\label{fig:Heston_density_complexity}
\end{figure}

We  compute the density  $\rho_{X(T)}$, given by  \eqref{eq:density_function}, at $u=1$   such that $X$ is a Heston asset \eqref{eq:dynamics Heston}, with parameters: $X_0=1$, $v_0=0.04$, $\mu=0$,  $\rho=-0.9$, $\zeta=1$, $\xi=0.1$, and $\theta=0.0025$.   A reference solution, equal to $2.4475$,  was    obtained  by  applying the fractional Fourier transform to the  characteristic function.   
    Table \ref{table:Summary of our numerical results density Heston.} summarizes  the  results.  Figure \ref{fig:Heston_density_MLMC_with_smoothing_FT_euler}  details the  convergence results for the MLMC estimator combined with the numerical smoothing,  using the FT Euler--Maruyama scheme. This figure verifies that  the kurtosis is bounded and that the  variance decay  rate is of order $1$ for the Euler--Maruyama scheme, resulting  in a  numerical complexity of the MLMC estimator in  the order  of  $\Ordo{\text{TOL}^{-2} \left(\log(\text{TOL})\right)^2} $, as confirmed in Figure \ref{fig:Heston_density_complexity}.  
\begin{small}
\begin{table}[!h]
	\centering
	\begin{tabular}{l*{4}{c}r}
	\toprule[1.5pt]
		Method      &   $\kappa_{L}$ & $\alpha$   &  $\beta$  &  $\gamma$   & Numerical complexity \\
		\hline
     MLMC with numerical smoothing +  (FT Euler--Maruyama) & $9$ & $1$  &  $1$ & $1$ &  $\Ordo{\text{TOL}^{-2} \left(\log(\text{TOL})\right)^2}$ \\
		\bottomrule[1.25pt]
	\end{tabular}
	\caption{Density of Heston: Summary of the MLMC numerical results observed for  computing the density $\rho_{X(T)}$ at  $u=1$, where $X$ follows the Heston dynamics. These results correspond to Figures  \ref{fig:density_heston} and  \ref{fig:Heston_density_complexity}.}
	\label{table:Summary of our numerical results density Heston.}
\end{table}
\end{small}

With the same model parameters, we  compute the joint density $\rho_{X(T), v(T)}$ at $u=1$ and $v=0.04$.   A reference solution was  obtained by the  kernel density estimator. 
 Figure \ref{fig:joint_density_heston} presents the detailed convergence results for the MLMC estimator combined with the numerical smoothing,  using the FT Euler--Maruyama scheme. This figure verifies that  the kurtosis is bounded and that  the variance decay  rate is of order $1$  for  the Euler--Maruyama scheme, resulting  in a  numerical complexity of the MLMC estimator of  the order  of  $\Ordo{\text{TOL}^{-2} \left(\log(\text{TOL})\right)^2}$. 

  
%

\textbf{Acknowledgments}  C. Bayer gratefully acknowledges funding by the Deutsche Forschungsgemeinschaft (DFG, German Research Foundation) under Germany 's Excellence Strategy – The Berlin Mathematics Research Center MATH+ (EXC-2046/1, project ID: 390685689). This publication is based on work supported by the King Abdullah University of Science and Technology (KAUST) Office of Sponsored Research (OSR) under Award No. OSR-2019-CRG8-4033 and the Alexander von Humboldt Foundation. The authors are also very grateful to the anonymous referees for their valuable feedback  that greatly contributed to shape the final version of the paper.
 
%


\bibliographystyle{plain}
\bibliography{smoothing} 

\appendix
\section{Additional  Results for the Proofs of Theorems \ref{corrol: Lipschitz of the integrands} and \ref{corrol: Lipschitz of the integrands_density}}\label{appendix:Theoretical Results for the Proofs of Theorems}
This section states and proves the additional theoretical results for the proofs of Theorems \ref{corrol: Lipschitz of the integrands} and \ref{corrol: Lipschitz of the integrands_density} in Section \ref{sec:Strong Convergence results for MLMC with numerical smoothing}. We use the same notation as in  Section \ref{sec:Strong Convergence results for MLMC with numerical smoothing}. 
\begin{proposition}[Vanishing boundary terms]\label{lemma: boundary_condition_error growth}	Assume that  $a(\cdot)$ and  $b(\cdot)$ in \eqref{eq:SDE_interest}  satisfy Assumption \ref{ass:  uniform boundedness of the drift and diffusion coefficients}, and that Assumption  \ref{ass:boundedness-inverse} holds. Then\footnote{We emphasize that $f(y; B_{\ell})$ in \eqref{eq:BC_limit} is a determinisitc function of $y$.}
	\begin{small}
		\begin{equation}\label{eq:BC_limit}
			\underset{\abs{y} \rightarrow \infty}{\lim}  f(y; B_{\ell}):=	E\left[  \left(e_{\ell}(T; y, B_{\ell}) \; \rho_1(y) \int_{0}^{1}  g (z(\theta,y;B_{\ell}) )  \left(\partial_y z(\theta,y;B_{\ell}) \right)^{-1}   d \theta  \right)^2\right]=0.
		\end{equation}
	\end{small}
\end{proposition}
\begin{proof}\label{proof2: lemma: boundary_condition_error growth}
	We have $g(\cdot)$ is bounded and  by Assumption  \ref{ass:boundedness-inverse}, we have also $ \left(\partial_y z(\theta;y,B_{\ell}) \right)^{-1}$ is bounded in moments (similar to what we showed for \eqref{eq:aux_results} in the proof of Thorem \ref{corrol: Lipschitz of the integrands}). Consequently, we need  to show that $  	\underset{\abs{y} \rightarrow \infty}{\lim}  	E\left[  e^2_{\ell}(T; y, B_{\ell}) \; \right] \rho_1(y) =0$. 
	
	First observe that  $E\left[  e^2_{\ell}(T; y, B_{\ell}) \right] \le  E\left[  \bar{X}^2_{\ell}(T; y, B_{\ell}) \right] + E\left[ \bar{X}^2_{\ell-1}(T; y, B_{\ell}) \right] $. Therefore, to conclude our target result, we just  need to get a bound on $E\left[ \bar{X}^2_{\ell}(T; y, B_{\ell}) \right] $. For that we will  use the discrete version of Gr\"onwall's inequality and  show that, for the time grid at level $\ell$:  $0=t^\ell_{0}<\ldots< t_n^\ell<\dots< t^\ell_{N_{\ell}}$, we have		$E\left[   \bar{X}^2_{\ell}  (t_{n+1}^\ell) \right] \le (1+K \Delta t_{n}^\ell) E\left[   \bar{X}^2_{\ell}  (t_{n}^\ell) \right] +A_n $, to conclude that $  E\left[   \bar{X}^2_{\ell}  (t_{n}^\ell) \right]   \le   \bar{X}^2_{\ell}  (0) e^{K t^\ell_n} + \sum_{i=0}^{n-1} A_i e^{K t_i^\ell}	$.
	
	We recall that 		$\bar{X}_{\ell}  (t_{n+1}^\ell)=  \bar{X}_{\ell}  (t_{n}^\ell)  +  a( \bar{X}_{\ell}  (t_{n}^\ell)) \Delta t_\ell + b( \bar{X}_{\ell}  (t_{n}^\ell)) \left(\frac{y}{\sqrt{T}}   \Delta t^\ell  +\Delta B_{n,\ell}\right)$. Therefore,  using Assumption \ref{ass:  uniform boundedness of the drift and diffusion coefficients}, we obtain
	\begin{small}
		\begin{align}\label{eq: Gronwal discrete condition}
			E\left[   \bar{X}^2_{\ell}  (t_{n+1}^\ell) \right]= &E\left[   \bar{X}^2_{\ell}  (t_{n}^\ell) \right]+ E\left[   b^2 ( \bar{X}_{\ell}  (t_{n}^\ell)) \left(\Delta B_{n,\ell}\right)^2\right]+ E\left[   \left(a ( \bar{X}_{\ell}  (t_{n}^\ell)) +\frac{y}{\sqrt{T}} b ( \bar{X}_{\ell}  (t_{n}^\ell))  \right)^2 \right] \Delta t_\ell^2 \nonumber\\
			&+ 2 E\left[   \bar{X}_{\ell}  (t_{n}^\ell) \left(a ( \bar{X}_{\ell}  (t_{n}^\ell)) +\frac{y}{\sqrt{T}} b ( \bar{X}_{\ell}  (t_{n}^\ell))  \right)  \right] \Delta t_\ell+ 2 E\left[   \bar{X}_{\ell}  (t_{n}^\ell)  b ( \bar{X}_{\ell}  (t_{n}^\ell)) \Delta B_{n,\ell}\right]\nonumber\\
			&+ 2 E\left[   \left(a ( \bar{X}_{\ell}  (t_{n}^\ell)) +\frac{y}{\sqrt{T}} b ( \bar{X}_{\ell}  (t_{n}^\ell))  \right) b ( \bar{X}_{\ell}  (t_{n}^\ell)) \Delta B_{n,\ell}  \right] \Delta t_\ell\nonumber\\
			&\le  E\left[   \bar{X}^2_{\ell}  (t_{n}^\ell) \right]+ C_1^2  \underset{=\Delta t_\ell- \frac{\Delta t_\ell^2}{T}}{\underbrace{E\left[ \left(\Delta B_{n,\ell}\right)^2 \right]}}+ (1+\Delta t_\ell)  \Delta t_\ell (C_2^2 + C_1^2 \frac{y^2}{T}+ 2 C_1 C_2 \frac{y}{\sqrt{T}} ) +\Delta t_\ell   E\left[   \bar{X}^2_{\ell}  (t_{n}^\ell) \right] +  E\left[   \bar{X}^2_{\ell}  (t_{n}^\ell) \right]  \nonumber\\
			&+ C_1^2  \underset{=\Delta t_\ell- \frac{\Delta t_\ell^2}{T}}{\underbrace{E\left[ \left(\Delta B_{n,\ell}\right)^2 \right]}} + 2 (C_1 C_2 +C_1^2 \frac{y}{\sqrt{T}}) \Delta t_\ell  \underset{=0}{\underbrace{E\left[ \Delta B_{n,\ell} \right]}} \nonumber\\
			&=  E\left[   \bar{X}^2_{\ell}  (t_{n}^\ell) \right] \left(2+\Delta t_\ell\right) + \underset{=A_n(y)}{\underbrace{\left( \left(1+\Delta t_\ell\right) \Delta t_\ell  \left(\frac{C_1^2}{T}y^2+ 2  \frac{C_1 C_2}{\sqrt{T}} y\right) +   \Delta t_\ell  \left(2C_1^2 \left(1-\frac{\Delta t_\ell}{T}\right) +C_2^2 \left(1+\Delta t_\ell\right)  \right) \right)}}\nonumber\\ 
			&=2  (1+\frac{\Delta t_{n}^\ell}{2}) E\left[   \bar{X}^2_{\ell}  (t_{n}^\ell)  \right] +A(y).
		\end{align}
	\end{small}
	Using  \eqref{eq: Gronwal discrete condition} and the discrete version of Gr\"onwall's inequality, we conclude that 
	\begin{small}
		\begin{align}\label{eq: bound_final_value _process}
			E\left[   \bar{X}^2_{\ell}  (T) \right] \le  2^{N_\ell}  \bar{X}^2_{\ell}  (0)  e^{T/2} +A(y) \sum_{i=0}^{N_\ell} 2^i e^{t_i^\ell/2}.
		\end{align}
	\end{small}
	Since $A(y)$ is quadratic in $y$, we conclude that, for a given $\Delta t_{\ell}$,  $\underset{\abs{y} \rightarrow \infty}{\lim}  	E\left[  e^2_{\ell}(T; y, B_{\ell}) \; \right] \rho_1(y) =0$.
\end{proof}	
\begin{remark}[Relaxing Assumption \ref{ass:  uniform boundedness of the drift and diffusion coefficients} in the proof of Proposition \ref{lemma: boundary_condition_error growth} and Theorems \ref{corrol: Lipschitz of the integrands} and \ref{corrol: Lipschitz of the integrands_density}]\label{rem:ass_relaxation}
In the  above proof, we showed that  for a given $\Delta t_{\ell}$, we obtain $  	\underset{\abs{y} \rightarrow \infty}{\lim}  	E\left[  e^2_{\ell}(T; y, B_{\ell}) \; \right] \rho_1(y) =0$. Therefore the growth  observed in the bound \eqref{eq: bound_final_value _process} w.r.t $N_{\ell}$ is not problematic for the sake of that proof. However, we emphasise that a better bound can be derived for small $\Delta t_{\ell}$ using the following arguments and sketch of proof: 

The three terms 	$E\left[   \left(a ( \bar{X}_{\ell}  (t_{n}^\ell)) +\frac{y}{\sqrt{T}} b ( \bar{X}_{\ell}  (t_{n}^\ell))  \right) b ( \bar{X}_{\ell}  (t_{n}^\ell)) \Delta B_{n,\ell}  \right]$,  $ E\left[   \bar{X}_{\ell}  (t_{n}^\ell)  b ( \bar{X}_{\ell}  (t_{n}^\ell)) \Delta B_{n,\ell}\right]$, and   $E\left[   b^2 ( \bar{X}_{\ell}  (t_{n}^\ell)) \left(\Delta B_{n,\ell}\right)^2\right]$   in \eqref{eq: Gronwal discrete condition} can be represented as   $E\left[   F(\{\Delta B_{m,\ell}\}_{m=1, m\neq n}^{N_\ell}, \Delta B_{n,\ell}) \left(\Delta B_{n,\ell}\right)^k\right]$, where $k=1,2$ and  $F(\{\Delta B_{m,\ell}\}_{m=1, m\neq n}^{N_\ell},\Delta B_{n,\ell})$ is a function of   $\Delta B_{n,\ell}$ and  the remaing Brownian bridge increments $\{\Delta B_{m,\ell}\}_{m=1, m\neq n}^{N_\ell}$.  For  $\Delta t_{\ell} \rightarrow0$,  applying Taylor expansion for $F(\cdot)$ around  $\Delta B_{n,\ell}$ implies  that 
	\begin{small}
	\begin{align}
		&E\left[   F(\{\Delta B_{m,\ell}\}_{m=1, m\neq n}^{N_\ell}, \Delta B_{n,\ell}) \left(\Delta B_{n,\ell}\right)^k\right]\nonumber\\
		&=   E\left[   \left(F(\{\Delta B_{m,\ell}\}_{m=1, m\neq n}^{N_\ell}, 0)+ F'(\{\Delta B_{m,\ell}\}_{m=1, m\neq n}^{N_\ell},0)\Delta B_{n,\ell}+\text{h.o.t} \right) \left(\Delta B_{n,\ell}\right)^k\right]\nonumber\\
		&=   E\left[   F(\{\Delta B_{m,\ell}\}_{m=1, m\neq n}^{N_\ell}, 0) \right] E\left[\left(\Delta B_{n,\ell}\right)^k\right] + E\left[F'(\{\Delta B_{m,\ell}\}_{m=1, m\neq n}^{N_\ell},0)\right]  E\left[\left(\Delta B_{n,\ell}\right)^{k+1}\right]+\text{h.o.t} \nonumber\\
		&=  E\left[   F^{(2-k)}(\{\Delta B_{m,\ell}\}_{m=1, m\neq n}^{N_\ell}, 0) \right] \underset{=\Delta t_\ell- \frac{\Delta t_\ell^2}{T} \overset{\Delta t_\ell \rightarrow 0}{\longrightarrow} 0}{\underbrace{E\left[\left(\Delta B_{n,\ell}\right)^2\right]}} +\text{h.o.t}, \: k=1,2.
	\end{align}
\end{small}
	In this case we can relax Assumption \ref{ass:  uniform boundedness of the drift and diffusion coefficients} and use  Assumption \ref{ass: globally Lipschitz_drift_diffusion} instead, and  we obtain
	\begin{small}
		\begin{align}\label{eq: Gronwal discrete condition_small_dt}
			E\left[   \bar{X}^2_{\ell}  (t_{n+1}^\ell) \right]= &E\left[   \bar{X}^2_{\ell}  (t_{n}^\ell) \right]+ E\left[   b^2 ( \bar{X}_{\ell}  (t_{n}^\ell)) \left(\Delta B_{n,\ell}\right)^2\right]+ E\left[   \left(a ( \bar{X}_{\ell}  (t_{n}^\ell)) +\frac{y}{\sqrt{T}} b ( \bar{X}_{\ell}  (t_{n}^\ell))  \right)^2 \right] \Delta t_\ell^2 \nonumber\\
			&+ 2 E\left[   \bar{X}_{\ell}  (t_{n}^\ell) \left(a ( \bar{X}_{\ell}  (t_{n}^\ell)) +\frac{y}{\sqrt{T}} b ( \bar{X}_{\ell}  (t_{n}^\ell))  \right)  \right] \Delta t_\ell+ 2 E\left[   \bar{X}_{\ell}  (t_{n}^\ell)  b ( \bar{X}_{\ell}  (t_{n}^\ell)) \Delta B_{n,\ell}\right]\nonumber\\
			&+ 2 E\left[   \left(a ( \bar{X}_{\ell}  (t_{n}^\ell)) +\frac{y}{\sqrt{T}} b ( \bar{X}_{\ell}  (t_{n}^\ell))  \right) b ( \bar{X}_{\ell}  (t_{n}^\ell)) \Delta B_{n,\ell}  \right] \Delta t_\ell\nonumber\\
			&	\underset{\Delta t_\ell \rightarrow 0}{\lessapprox} E\left[   \bar{X}^2_{\ell}  (t_{n}^\ell) \right] \left( 1+ C^2 \Delta t_\ell^2 (1+\frac{y}{\sqrt{T}})^2+ 2 C  \Delta t_\ell (1+\frac{y}{\sqrt{T}})\right) +\text{h.o.t} \nonumber\\
			&=	 E\left[   \bar{X}^2_{\ell}  (t_{n}^\ell) \right] \left( 1+ C^2 \Delta t_\ell^2 (1+\frac{y}{\sqrt{T}})^2+ 2 C  \Delta t_\ell (1+\frac{y}{\sqrt{T}}) \right)+\text{h.o.t} \nonumber\\
			&= (1+K(y) \Delta t_{n}^\ell) E\left[   \bar{X}^2_{\ell}  (t_{n}^\ell) \right] +\text{h.o.t},
		\end{align}
	\end{small}
	with $K(y)= C^2 \frac{\Delta t_\ell^2}{T} y^2+2y \left( C^2 \frac{\Delta t_\ell}{\sqrt{T}} + \frac{C}{\sqrt{T}} \right) \Delta t_\ell+ \Delta t_\ell (C^2 \Delta t_\ell+2 C ) $.
	
	Using  \eqref{eq: Gronwal discrete condition} and the discrete version of Gr\"onwall's inequality, we conclude that 
	\begin{align*}
		E\left[   \bar{X}^2_{\ell}  (T) \right] \le   \bar{X}^2_{\ell}  (0)  e^{K(y) T} ,
	\end{align*}
	and that for sufficiently small $\Delta t_\ell \rightarrow 0$, we obtain  $  	\underset{\abs{y} \rightarrow \infty}{\lim}  	E\left[  e^2_{\ell}(T; y, B_{\ell}) \; \right] \rho_1(y) =0$.
\end{remark}
\begin{lemma}[Moments bounds for  the y-derivative of the error]\label{lemma: boundness of y-derivatives of the weak error}
				Let  $e_{\ell}(t; y, B_{\ell})$ as defined in \eqref{eq:error_eq},  and assume that  $a(\cdot)$ and  $b(\cdot)$ satisfy Assumptions  \ref{ass: globally Lipschitz_drift_diffusion}, \ref{ass:  uniform boundedness of first order derivatives}, \ref{ass:  uniform boundedness of first second order derivatives}, and \ref{ass:  uniform boundedness of the drift and diffusion coefficients}, and that Assumption  \ref{ass:boundedness-derivative} holds. Then, we obtain for $p\ge1$ 
				\begin{equation*}
					\expt{(\partial_y e_{\ell})^{2p} (T)}=\Ordo{\Delta t_{\ell}^{p}}.
				\end{equation*}
			\end{lemma}
				\begin{proof}[Proof of lemma \ref{lemma: boundness of y-derivatives of the weak error}]\label{proof: lemma: boundness of y-derivatives of the weak error}	
					In the following, for ease of  notation, we  denote $e_{\ell}(t; y, B_{\ell})$ by $e_{\ell}(t)$.  	From \eqref{eq: Euler SDE} and \eqref{eq:error_eq} and since  $(\partial_y  e_{\ell})(0)=0$, we obtain\footnote{The transition  related to the diffusion term from the second equality  to the third equality is justified because the integral representation corresponds to finite sums due to construction \eqref{eq: Euler SDE}.}
					\begin{small}
						\begin{align}
									\partial_y  e_{\ell}(t)&= \int_{0}^t  \partial_y\left(a(\bar{X}_{\ell}([s]_{\ell}))-  a(\bar{X}_{\ell-1}([s]_{\ell-1}))\right) ds +\partial_y \left( \int_{0}^{t}   \left(b(\bar{X}_{\ell}([s]_{\ell}))-  b(\bar{X}_{\ell-1}([s]_{\ell-1}))\right) dW_s\right)\nonumber\\
						&	 =   \int_{0}^t  \partial_y\left(a(\bar{X}_{\ell}([s]_\ell))-  a(\bar{X}_{\ell}(s))+ a(\bar{X}_{\ell}(s))- a(\bar{X}_{\ell-1}(s)) +a(\bar{X}_{\ell-1}(s))- a(\bar{X}_{\ell-1}([s]_{\ell-1}))  \right) ds    \nonumber\\
							& +\partial_y \left( \int_{0}^{t}   \left(b(\bar{X}_{\ell}([s]_\ell))-  b(\bar{X}_{\ell}(s))+ b(\bar{X}_{\ell}(s))- b(\bar{X}_{\ell-1}(s)) +b(\bar{X}_{\ell-1}(s))- b(\bar{X}_{\ell-1}([s]_{\ell-1}))    \right) dW_s\right)\nonumber\\
							&= \int_{0}^t  \left(a'(\bar{X}_{\ell}) \partial_y\bar{X}_{\ell}- a'(\bar{X}_{\ell-1})  \partial_y\bar{X}_{\ell-1}\right)(s) \; ds + \int_{0}^{t}   \left(b'(\bar{X}_{\ell}) \partial_y\bar{X}_{\ell}-  b'(\bar{X}_{\ell-1})  \partial_y\bar{X}_{\ell-1}\right)(s)\; dW_s\nonumber\\
							&+ \int_{0}^{t}  \left(b(\bar{X}_{\ell}(s))-  b(\bar{X}_{\ell-1}(s))\right) \frac{ds}{\sqrt{T}}\nonumber\\
							& + \int_{0}^t  \partial_y\left(a(\bar{X}_{\ell}([s]_\ell))-  a(\bar{X}_{\ell}(s))\right) ds  + \int_{0}^t  \partial_y\left(a(\bar{X}_{\ell-1}(s))- a(\bar{X}_{\ell-1}([s]_{\ell-1}))  \right) ds\nonumber\\
								& +  \int_{0}^{t}  \partial_y  \left(b(\bar{X}_{\ell}([s]_\ell))-  b(\bar{X}_{\ell}(s)) +b(\bar{X}_{\ell-1}(s))- b(\bar{X}_{\ell-1}([s]_{\ell-1}))    \right) dW_s\nonumber\\
								& +  \int_{0}^{t}    \left(b(\bar{X}_{\ell}([s]_\ell))-  b(\bar{X}_{\ell}(s)) +b(\bar{X}_{\ell-1}(s))- b(\bar{X}_{\ell-1}([s]_{\ell-1}))    \right)   \frac{ds}{\sqrt{T}}  \nonumber\\
							&= \int_{0}^t  \left(a'(\bar{X}_{\ell}) \partial_ye_{\ell}+ \partial_y\bar{X}_{\ell-1} \left(\int_{0}^{1} a''(\bar{X}_{\ell-1}+\theta e_{\ell}) d\theta \right) e_{\ell}\right)(s)\; ds\nonumber\\
							&+  \int_{0}^t   \left(\int_{0}^{1} b'(\bar{X}_{\ell-1}+\theta e_{\ell}) d\theta \right)(s) e_{\ell}(s) \; \frac{ds}{\sqrt{T}}\nonumber\\
							&+  \int_{0}^t  \left(b'(\bar{X}_{\ell}) \partial_ye_{\ell}+ \partial_y\bar{X}_{\ell-1} \left(\int_{0}^{1} b''(\bar{X}_{\ell-1}+\theta e_{\ell}) d\theta \right)  e_{\ell}\right) (s) \; dW_s\nonumber\\
								& + \int_{0}^t  \partial_y\left(a(\bar{X}_{\ell}([s]_\ell))-  a(\bar{X}_{\ell}(s))\right) ds  + \int_{0}^t  \partial_y\left(a(\bar{X}_{\ell-1}(s))- a(\bar{X}_{\ell-1}([s]_{\ell-1}))  \right) ds\nonumber\\
							& +  \int_{0}^{t}  \partial_y  \left(b(\bar{X}_{\ell}([s]_\ell))-  b(\bar{X}_{\ell}(s)) +b(\bar{X}_{\ell-1}(s))- b(\bar{X}_{\ell-1}([s]_{\ell-1}))    \right) dW_s\nonumber\\
							& +  \int_{0}^{t}    \left(b(\bar{X}_{\ell}([s]_\ell))-  b(\bar{X}_{\ell}(s)) +b(\bar{X}_{\ell-1}(s))- b(\bar{X}_{\ell-1}([s]_{\ell-1}))    \right)   \frac{ds}{\sqrt{T}}.
						\end{align}
					\end{small}
					Therefore, taking expectation, we obtain
					\begin{small}
						\begin{align}\label{eq:proof: lemma: boundness of y-derivatives of the weak error_step1}
							E 	\left[\left(\partial_y  e_{\ell}(t)\right)^{2p}\right]&\le 5^{2p-1}  \underset{(I)}{\underbrace{{E\left[\left(\int_{0}^t  \left(\bar{a}'(\bar{X}_{\ell}) \partial_ye_{\ell}+ \left(\partial_y\bar{X}_{\ell-1} \left(\int_{0}^{1} \bar{a}''(X_{\ell-1}+\theta e_{\ell}) d\theta \right) +   \int_{0}^{1} \frac{1}{\sqrt{T}}\bar{b}'(\bar{X}_{\ell-1}+\theta e_{\ell}) d\theta \right) e_{\ell}\right) ds \right)^{2p}\right]}}} \nonumber\\
							& +5^{2p-1}     \underset{(II)}{\underbrace{{E\left[\left(\int_{0}^t  \left(\bar{b}'(\bar{X}_{\ell})\partial_ye_{\ell}+ \partial_y\bar{X}_{\ell-1} \left(\int_{0}^{1} \bar{b}''(\bar{X}_{\ell-1}+\theta e_{\ell}) d\theta \right) e_{\ell}\right) dWs\right)^{2p}\right]}}}\nonumber\\
							&+5^{2p-1}  	\underset{(III)}{\underbrace{{	E 	\left[ \left( \int_{0}^t  \partial_y\left(a(\bar{X}_{\ell}([s]_\ell))-  a(\bar{X}_{\ell}(s))\right) ds  + \int_{0}^t  \partial_y\left(a(\bar{X}_{\ell-1}(s))- a(\bar{X}_{\ell-1}([s]_{\ell-1}))  \right) ds\right)^{2p}\right]	}  }}  \nonumber\\
								&+5^{2p-1}  	\underset{(IV)}{\underbrace{{	E 	\left[  \left( \int_{0}^{t}  \partial_y  \left(b(\bar{X}_{\ell}([s]_\ell))-  b(\bar{X}_{\ell}(s)) +b(\bar{X}_{\ell-1}(s))- b(\bar{X}_{\ell-1}([s]_{\ell-1}))    \right) dW_s\right)^{2p}\right]	} }}   \nonumber\\
									&+5^{2p-1}  	\underset{(V)}{\underbrace{{	E 	\left[\left( \int_{0}^{t}    \left(b(\bar{X}_{\ell}([s]_\ell))-  b(\bar{X}_{\ell}(s)) +b(\bar{X}_{\ell-1}(s))- b(\bar{X}_{\ell-1}([s]_{\ell-1}))    \right)   \frac{ds}{\sqrt{T}}\right)^{2p} \right]	}}}
						\end{align}
					\end{small}
					The idea now is to  show that    $E \left[\left(\partial_y  e_{\ell}(t)\right)^{2p}\right]\le K \int_{0}^t E \left[\left(\partial_y  e_{\ell}(s)\right)^{2p}\right] ds+ A$, where $0<K<\infty$ and  $A=\Ordo{\Delta t_{\ell}^{p}}$, then,  using Gr\"onwall's inequality we get the result.
					
					Let $p_1,q_1, p_5, q_5 \in (1,+\infty)$ with   $\frac{1}{p_1}+\frac{1}{q_1}=1$ and $\frac{1}{p_5}+\frac{1}{q_5}=1$ such that $p_5 p/p_1 \le 1$. Then using   the  H\"older, Burkholder-Davis-Gundy and   Jensen inequalities, we obtain for (II) in \eqref{eq:proof: lemma: boundness of y-derivatives of the weak error_step1}
					\begin{small}
						\begin{align}\label{eq: proof_derivative_error_Gronwall_struc_termI}
							(II) & \le 2^{p-1}    \left(E\left[\left(\int_{0}^t  \bar{b}'(\bar{X}_{\ell}) \partial_ye_{\ell} dWs \right)^{2p}\right]+ E\left[\left(\int_{0}^t  \partial_y\bar{X}_{\ell-1} \left(\int_{0}^{1} \bar{b}''(\bar{X}_{\ell-1}+\theta e_{\ell}) d\theta \right) e_{\ell} dWs\right)^{2p}\right]\right)\nonumber\\
							&\le K_1 E\left[\left(\int_{0}^t  \left(\bar{b}'(\bar{X}_{\ell}) \partial_ye_{\ell}\right)^2 ds \right)^{p}\right]+ A_1  E\left[\left(\int_{0}^t  \left( \partial_y\bar{X}_{\ell-1} \left(\int_{0}^{1} \bar{b}''(\bar{X}_{\ell-1}+\theta e_{\ell}) d\theta \right) e_{\ell}\right)^2 ds\right)^{p}\right]\nonumber\\
							&\le K_2 \; t^{p-1} E\left[ \int_{0}^t \left( \partial_ye_{\ell}\right)^{2p} ds \right]\nonumber\\
							&+ A_1  E\left[\left(\int_{0}^t  \left( \partial_y\bar{X}_{\ell-1} \left(\int_{0}^{1} \bar{b}''(\bar{X}_{\ell-1}+\theta e_{\ell}) d\theta \right) \right)^{2q_1} ds \right)^{p/q_1} \times  \left(\int_{0}^t \left(e_{\ell}\right)^{2p_1} ds\right)^{p/p_1}\right]\nonumber\\
							&\le K_2 \; t^{p-1}  E\left[\int_{0}^t \left( \partial_ye_{\ell}\right)^{2p} ds \right]\nonumber\\
							&+ A_1  E\left[\left(\int_{0}^t  \left( \partial_y\bar{X}_{\ell-1} \left(\int_{0}^{1} \bar{b}''(\bar{X}_{\ell-1}+\theta e_{\ell}) d\theta \right) \right)^{2q_1} ds \right)^{q_5 p/q_1}\right]^{1/q_5} \times  E\left[\left(\int_{0}^t \left(e_{\ell}\right)^{2p_1} ds\right)^{p_5 p/p_1}\right]^{1/p_5}\nonumber\\ 
							&\le K_2\; t^{p-1}  E\left[\int_{0}^t \left( \partial_ye_{\ell}\right)^{2p} ds \right]\nonumber\\
							&+ A_1 \underset{<\infty} {\underbrace{E\left[\left(\int_{0}^t  \left( \partial_y\bar{X}_{\ell-1} \left(\int_{0}^{1} \bar{b}''(\bar{X}_{\ell-1}+\theta e_{\ell}) d\theta \right) \right)^{2q_1} ds \right)^{q_5 p/q_1}\right]^{1/q_5}}}  \times  \underset{\Ordo{\Delta t_{\ell}^{p}}} {\underbrace{E\left[\left(\int_{0}^t \left( e_{\ell}\right)^{2p_1} ds\right)\right]^{p/p_1}}}
						\end{align}
					\end{small}
					where we  used   \eqref{eq:L_p_moments_estimate}, Assumption \ref{ass:boundedness-derivative},  and that   $  b'(X_{\ell})$ is uniformly bounded due to Assumption  \ref{ass: globally Lipschitz_drift_diffusion}, to get   \eqref{eq: proof_derivative_error_Gronwall_struc_termI}.
					
					For term (I) in \eqref{eq:proof: lemma: boundness of y-derivatives of the weak error_step1},  using  H\"older's  inequality ($p_2,q_2, p_6, q_6 \in (1,+\infty)$ with   $\frac{1}{p_2}+\frac{1}{q_2}=1$ and $\frac{1}{p_6}+\frac{1}{q_6}=1$) and   $2p_6 p/p_2 \le 1$ to use Jensen's  inequality, we obtain
					\begin{small}
						\begin{align}\label{eq: proof_derivative_error_Gronwall_struc_termII}
							(I) & \le 2^{p-1}    E\left[\left( \int_{0}^t  \bar{a}'(\bar{X}_{\ell}) \partial_ye_{\ell} ds \right)^{2p}\right]\nonumber\\
							&+  2^{p-1} E\left[\left( \int_{0}^t   \left(\partial_y\bar{X}_{\ell-1} \left(\int_{0}^{1} \bar{a}''(\bar{X}_{\ell-1}+\theta e_{\ell}) d\theta \right) +   \int_{0}^{1} \frac{1}{\sqrt{T}}\bar{b}'(\bar{X}_{\ell-1}+\theta e_{\ell}) d\theta \right) e_{\ell} ds \right)^{2p}\right]\nonumber\\
							&\le  2^{p-1}  K_3 t^{\frac{2p-1}{2p}}E\left[ \int_{0}^t \left( \partial_ye_{\ell}\right)^{2p} ds \right]\nonumber\\
							&+ 2^{p-1}  E\left[\left(\int_{0}^t   \left(\partial_y\bar{X}_{\ell-1} \left(\int_{0}^{1} \bar{a}''(\bar{X}_{\ell-1}+\theta e_{\ell}) d\theta \right) +   \int_{0}^{1} \frac{1}{\sqrt{T}}\bar{b}'(\bar{X}_{\ell-1}+\theta e_{\ell}) d\theta \right)^{q_2} ds \right)^{2 p/q_2} \times  \left(\int_{0}^t \left( e_{\ell}\right)^{p_2} ds\right)^{2p/p_2}\right]\nonumber\\
							&\le  2^{p-1}  K_3 t^{\frac{2p-1}{2p}}   E\left[\int_{0}^t \left( \partial_ye_{\ell}\right)^{2p} ds \right]\nonumber\\
							&+ 2^{p-1}  \underset{<\infty} {\underbrace{E\left[\left(\int_{0}^t   \left(\partial_y\bar{X}_{\ell-1} \left(\int_{0}^{1} \bar{a}''(\bar{X}_{\ell-1}+\theta e_{\ell}) d\theta \right) +   \int_{0}^{1} \frac{1}{\sqrt{T}}\bar{b}'(\bar{X}_{\ell-1}+\theta e_{\ell}) d\theta \right)^{q_2} ds \right)^{2 p q_6/q_2}\right]^{1/q_6}}} \\\nonumber
							&\times  \underset{\Ordo{\Delta t_{\ell}^{p}}} {\underbrace{E\left[\left(\int_{0}^t \left(e_{\ell}\right)^{p_2} ds\right)^{2p p_6/p_2}\right]^{1/p_6}}}
						\end{align}
					\end{small}
					where we  used  \eqref{eq:L_p_moments_estimate}, Assumption \ref{ass:boundedness-derivative},   and  that  $a'(X_{\ell})$ is uniformly bounded due to Assumption  \ref{ass: globally Lipschitz_drift_diffusion}, to get  \eqref{eq: proof_derivative_error_Gronwall_struc_termII}.
		
			To end the proof, the remaining step is to show that the terms (III), (IV) and (V) in 
				 \eqref{eq:proof: lemma: boundness of y-derivatives of the weak error_step1} are of order $\Ordo{\Delta t_{\ell}^{p}}$. First, observe that for any  $[s]_\ell\le s\le [s]_\ell +\Delta t_{\ell}$,  using \eqref{eq: Euler SDE}  and Assumption \ref{ass:  uniform boundedness of the drift and diffusion coefficients}, we obtain  for any $p\ge 1$
				 \begin{align}\label{eq: moments of increments}
					E\left[\left(	 \bar{X}_{\ell}([s]_\ell)- \bar{X}_{\ell}(s)\right)^{2p}\right] &= E\left[\left(	 a( \bar{X}([s]_\ell)) ([s]_\ell-s) +   b( \bar{X}_{\ell}([s]_\ell)) (W([s]_\ell)-W(s))\right)^{2p}\right]\nonumber\\
					&\le  2^{2p-1} \underset{<\infty \text{ due to Assumption \eqref{ass:  uniform boundedness of the drift and diffusion coefficients}}} {\underbrace{E\left[a( \bar{X}_{\ell}([s]_\ell))^{2p} \right]}} (\Delta t_\ell)^{2p}+  2^{2p-1}   \underset{=\Ordo{\Delta t_{\ell}^{p}} \text{ due to Assumption \eqref{ass:  uniform boundedness of the drift and diffusion coefficients}}} {\underbrace{E\left[ \left(b( \bar{X}_{\ell}([s]_\ell)) (W([s]_\ell)-W(s))\right)^{2p}\right]}}\nonumber\\
					&=\Ordo{\Delta t_{\ell}^p},
				 	\end{align}
				 	and similarly,  using \eqref{eq: Euler SDE}  and Assumptions~\ref{ass:  uniform boundedness of first order derivatives},  \ref{ass:  uniform boundedness of the drift and diffusion coefficients}, and  \ref{ass:boundedness-derivative} (and Lemma \ref{lem:dXdZ}), we obtain
				 	\begin{small}
				 		 \begin{align}\label{eq: moments of _derivative_ increments}
				 		E\left[\left(	\partial_y \left(\bar{X}_{\ell}([s]_\ell)- \bar{X}_{\ell}(s)\right)\right)^{2p}\right] &= E\left[\left(	\partial_y \left( a( \bar{X}([s]_\ell)) ([s]_\ell-s) +   b( \bar{X}_{\ell}([s]_\ell)) (W([s]_\ell)-W(s)) \right) \right)^{2p}\right]\nonumber\\
				 		&= E\left[\left( 	\partial_y a( \bar{X}([s]_\ell)) ([s]_\ell-s) +   	\partial_y b( \bar{X}_{\ell}([s]_\ell)) (W([s]_\ell)-W(s)) +  b( \bar{X}_{\ell}([s]_\ell))  \frac{[s]_\ell-s}{\sqrt{T}} \right)^{2p}\right]\nonumber\\
				 	&\le  3^{2p-1}   \underset{<\infty \text{ due to Assumptions \eqref{ass:  uniform boundedness of the drift and diffusion coefficients}  and  \ref{ass:boundedness-derivative}}}  {\underbrace{E\left[\left(a'( \bar{X}_{\ell}([s]_\ell) ) 	\partial_y  \bar{X}([s]_\ell) \right)^{2p} \right]}}(\Delta t_\ell)^{2p}\nonumber\\
				 	&+  3^{2p-1}   \underset{=\Ordo{\Delta t_{\ell}^{p}} \text{ due to Assumptions \eqref{ass:  uniform boundedness of the drift and diffusion coefficients}  and  \ref{ass:boundedness-derivative} }} {\underbrace{E\left[ \left(b'( \bar{X}_{\ell}([s]_\ell) ) 	\partial_y  \bar{X}([s]_\ell) (W([s]_\ell)-W(s))\right)^{2p}\right]}}  \nonumber\\
				 	&\le  3^{2p-1} T^{-p}  \underset{<\infty \text{ due to Assumption \eqref{ass:  uniform boundedness of first order derivatives}}} {\underbrace{E\left[\left(b( \bar{X}_{\ell}([s]_\ell)  \right)^{2p} \right]}}(\Delta t_\ell)^{2p}\nonumber\\
				 		&=\Ordo{\Delta t_{\ell}^p},
				 	\end{align}
				 	\end{small}
				For the term (III) in \eqref{eq:proof: lemma: boundness of y-derivatives of the weak error_step1}, we  focus on the first integral contribution, and the analysis follows similarly for the second one.  Following similar steps as  in \eqref{eq: proof_derivative_error_Gronwall_struc_termII}, \ie,  using  H\"older's  inequality ($p_2,q_2, p_6, q_6 \in (1,+\infty)$ with   $\frac{1}{p_2}+\frac{1}{q_2}=1$ and $\frac{1}{p_6}+\frac{1}{q_6}=1$) and   $2p_6 p/p_2 \le 1$ to use Jensen's  inequality, we obtain   
				\begin{small}
			\begin{align}\label{eq:term III bound}
		&E\left[\left(		 \int_{0}^t  \partial_y\left(a(\bar{X}_{\ell}([s]_\ell))-  a(\bar{X}_{\ell}(s))\right) ds \right)^{2p} \right]\nonumber\\
		&  \le  2^{2p-1} E\left[\left(  \int_{0}^t  a' (\bar{X}_{\ell}([s]_\ell)) \left(\partial_y\left(\bar{X}_{\ell}([s]_\ell)- \bar{X}_{\ell}(s)\right)   \right) ds \right)^{2p} \right] \nonumber\\
				 &+ 2^{2p-1} E\left[\left(\int_{0}^t  \partial_y \bar{X}_{\ell}(s) \left(a'(\bar{X}_{\ell}([s]_\ell))-a'( \bar{X}_{\ell}(s))   \right)   ds\right)^{2p} \right]\nonumber\\
				 &\le   2^{2p-1}     K_4 t^{\frac{2p-1}{2p}} \underset{\Ordo{\Delta t_{\ell}^{p}}} {\underbrace{E\left[ \int_{0}^t  \left(\partial_y\left(\bar{X}_{\ell}([s]_\ell)- \bar{X}_{\ell}(s)\right)   \right)^{2p} ds \right]}}\nonumber\\
			&+ 2^{p-1}  \underset{<\infty} {\underbrace{E\left[\left(\int_{0}^t   \left(\partial_y\bar{X}_{\ell} \right)^{q_2} ds \right)^{2 p q_6/q_2}\right]^{1/q_6}}} \times   \underset{\Ordo{\Delta t_{\ell}^{p}}} {\underbrace{E\left[\left(\int_{0}^t \left(a'(\bar{X}_{\ell}([s]_\ell))-a'( \bar{X}_{\ell}(s))   \right)^{p_2} ds\right)^{2p p_6/p_2}\right]^{1/p_6}}},
				\end{align}
				\end{small}
				where we used  \eqref{eq: moments of increments},  \eqref{eq: moments of _derivative_ increments}, and  that $a'(\cdot)$ is Lipchitz due to Assumption \ref{ass:  uniform boundedness of first second order derivatives}   to get the bound for the last term in \eqref{eq:term III bound}. 
				 
For the term (IV) in \eqref{eq:proof: lemma: boundness of y-derivatives of the weak error_step1}, we  focus on the first integral contribution, and the analysis follows similarly for the second one.  We  follow similar steps as  in \eqref{eq: proof_derivative_error_Gronwall_struc_termI}.   Let $p_1,q_1, p_5, q_5 \in (1,+\infty)$ with   $\frac{1}{p_1}+\frac{1}{q_1}=1$ and $\frac{1}{p_5}+\frac{1}{q_5}=1$ such that $p_5 p/p_1 \le 1$. Then using   the  H\"older, Burkholder-Davis-Gundy and   Jensen inequalities, we obtain 
\begin{small}
			\begin{align}\label{eq:term IV bound}
				&E\left[\left(		 \int_{0}^t  \partial_y\left(b(\bar{X}_{\ell}([s]_\ell))-  b(\bar{X}_{\ell}(s))\right) d W_s \right)^{2p} \right]\nonumber\\
				&  \le  2^{2p-1} E\left[\left(  \int_{0}^t  b' (\bar{X}_{\ell}([s]_\ell)) \left(\partial_y\left(\bar{X}_{\ell}([s]_\ell))- \bar{X}_{\ell}(s))\right)   \right) dW_s \right)^{2p} \right] \nonumber\\
				&+ 2^{2p-1} E\left[\left(\int_{0}^t  \partial_y \bar{X}_{\ell}(s) \left(b'(\bar{X}_{\ell}([s]_\ell))-b'( \bar{X}_{\ell}(s))   \right)   dW_s\right)^{2p} \right]\nonumber\\
				&\le K_5 \; t^{p-1}  \underset{\Ordo{\Delta t_{\ell}^{p}}} {\underbrace{E\left[\int_{0}^t \left( \partial_y\left(\bar{X}_{\ell}([s]_\ell))- \bar{X}_{\ell}(s))\right)\right)^{2p} ds \right]}}\nonumber\\
			&+ K_6 \underset{<\infty} {\underbrace{E\left[\left(\int_{0}^t  \left( \partial_y\bar{X}_{\ell}  \right)^{2q_1} ds \right)^{q_5 p/q_1}\right]^{1/q_5}}}  \times  \underset{\Ordo{\Delta t_{\ell}^{p}}} {\underbrace{E\left[\left(\int_{0}^t \left(b'(\bar{X}_{\ell}([s]_\ell))-b'( \bar{X}_{\ell}(s)) \right)^{2p_1} ds\right)\right]^{p/p_1}}},
			\end{align}
			\end{small}
				where we used  \eqref{eq: moments of increments},  \eqref{eq: moments of _derivative_ increments}, and  that $b'(\cdot)$ is Lipchitz due to Assumption \ref{ass:  uniform boundedness of first second order derivatives}   to get the bound for the last term in \eqref{eq:term IV bound}. 
				
	For the term (V) in \eqref{eq:proof: lemma: boundness of y-derivatives of the weak error_step1}, we  focus on the first integral contribution, and the analysis follows similarly for the second one. Using H\"older's  inequality,  \eqref{eq: moments of increments}, and    that $b(\cdot)$ is Lipchitz due to Assumption \ref{ass:  uniform boundedness of first second order derivatives}, we obtain 
	\begin{equation}
		E\left[\left(		 \int_{0}^t \left(b(\bar{X}_{\ell}([s]_\ell))-  b(\bar{X}_{\ell}(s))\right) \frac{ds}{\sqrt{T}} \right)^{2p} \right]=\Ordo{\Delta t_{\ell}^p}
	\end{equation}
			\end{proof}
			\begin{remark}[Extending Lemma \ref{lemma: boundness of y-derivatives of the weak error} for higher order derivatives]\label{remark: extending lemma of  boundness of y-derivatives of the weak error}
				It is easy to extend the result of Lemma \ref{lemma: boundness of y-derivatives of the weak error} for higher order terms, that is $	\expt{(\partial^k_y e_{\ell})^{2p}}=\Ordo{\Delta t_{\ell}^{p}} $ for $p\ge1$ and $k \ge 2$. However, we need to further assume  that    $a(\cdot)$ and  $b(\cdot)$ are  of class $C^{k+1}$, besides additional uniform boundedness conditions for the higher order derivatives up to order $k$ as in the proof  of Lemma \ref{lemma: boundness of y-derivatives of the weak error}.
			\end{remark}
\section{Adapting Assumptions 3.2 and 3.3 and Lemma A.1  in \cite{bayer2020numerical} to our Context}\label{appendix:Revisiting Assumptions 3.2 and 3.3}
This section states  assumptions \ref{ass:boundedness-derivative} and  \ref{ass:boundedness-inverse}, and lemma \ref{lem:dXdZ} which are a slightly adapted versions\footnote{We use Brownian bridge construction instead of wavelets.} of  Assumptions 3.2 and 3.3, and Lemma A.1 in \cite{bayer2020numerical}. The   sufficient conditions for the assumptions to be valid are explained in Appendix B in \cite{bayer2020numerical}.

From our construction of the approximate path at level $\ell$  using the Euler--Maruyama scheme based on the Brownian bridge construction, we have    $\bar{X}_\ell(T)$ is a  function of the rdvs $Z_{1}^{\ell}$  (corresponding to the coarsest level of the Brownian bridge $B_{\ell}$) and $\textbf{Z}^{\ell}_{-1} $ (the remaining $N_{\ell}-1$ random variables), \ie,  $\bar{X}_\ell(T):= \bar{X}_\ell(T; \left(Z_{1}^{\ell}, \textbf{Z}^{\ell}_{-1} \right))$. We  write $y \coloneqq z^{\ell}_{1}$ and $\textbf{z}^{\ell}_{-1}$ for the (deterministic) arguments of the function $\bar{X}_\ell(T)$. For convenience, we will denote $\bar{X}_\ell(T)$ by $\bar{X}^{N_\ell}_T$ and  $\bar{X}^{N_\ell}_k$ are the Euler--Maruyama   increments of $\bar{X}^{N_\ell}_T$ for $0\le k \le {N_\ell}$ with  $\bar{X}^{N_\ell}_T= \bar{X}^{N_\ell}_{N_\ell}$.
\begin{assumption}[Adapted version of Assumption 3.2 in \cite{bayer2020numerical}]
	\label{ass:boundedness-derivative}
	There are positive rdvs $C_p$ with finite moments of all orders such that
	\begin{equation*}
		\forall N_{\ell} \in \N,\ \forall k_1, \ldots, k_p \in \{0, \ldots, N_{\ell}-1\}:\ \abs{\f{\pa^p
				\bar{X}^{N_{\ell}}_T}{\pa 	\bar{X}^{N_{\ell}}_{k_1} \cdots \pa 	\bar{X}^{N_{\ell}}_{k_p}}} \le C_p \text{ a.s.}
	\end{equation*}
	In terms of  notation~\ref{notation}, this means that $\f{\pa^p	\bar{X}^{N_{\ell}}_T}{\pa
		\bar{X}^{N_{\ell}}_{k_1} \cdots \pa 	\bar{X}^{N_{\ell}}_{k_p}} = \mathcal{O}(1)$.
\end{assumption}
\begin{assumption}[Adapted version of Assumption 3.3 in \cite{bayer2020numerical}]
	\label{ass:boundedness-inverse}
	For any $p \in \N$ we obtain
	\begin{equation*}
		\left( \f{\pa 	\bar{X}^{N_{\ell}}_T}{\pa y}\left( Z^{\ell}_{1}, \textbf{Z}^{\ell}_{-1}\right) \right)^{-p} = \mathcal{O}(1).
	\end{equation*}
\end{assumption}
\begin{lemma}[Adapted result from Lemma A.1 in \cite{bayer2020numerical}]
	\label{lem:dXdZ}
	If  Assumption \ref{ass:boundedness-derivative} holds, we have the following:
	\begin{equation*}
		\f{\pa 	\bar{X}^{N_{\ell}}_T}{\pa y}(Z^{\ell}_{1}, \textbf{Z}^{\ell}_{-1}) =  \mathcal{O}(1).
	\end{equation*}
\end{lemma}
\begin{proof}
The  proof is similar to the one for Lemma A.1 in Appendix A in 	\cite{bayer2020numerical}.
\end{proof}

\end{document}